\documentclass[fleqn,usenatbib]{mnras}
\pdfoutput=1
\usepackage{newtxtext,newtxmath}

\usepackage[T1]{fontenc}
\usepackage{ae,aecompl}

\usepackage{graphicx}	
\usepackage{amsmath}	
\usepackage{subfig}
\usepackage{float}	
\usepackage{hyperref}
\usepackage{verbatim} 
\usepackage{longtable}
\usepackage{booktabs}
\usepackage[online]{threeparttablex}
\usepackage{siunitx}

\title[FT outbursts in BH LMXBs]{Failed-Transition outbursts in Black hole low-mass X-ray binaries}

\author[K. Alabarta et al.]{
K. Alabarta$^{1,2}$\thanks{E-mail: k.alabarta@soton.ac.uk},
D. Altamirano$^{1}$,
M. M\'endez$^{2}$,
V. A. C\'uneo$^{3,4}$,
F. M. Vincentelli$^{1}$,
\newauthor
N. Castro-Segura$^{1}$,
F. Garc\'ia$^{2}$,
B. Luff$^{1}$ and 
A. Veledina$^{5, 6, 7}$
\\
$^{1}$School of Physics and Astronomy, University of Southampton, Southampton, SO17 1BJ, UK\\
$^{2}$Kapteyn Astronomical Institute, University of Groningen, PO Box 800, NL-9700 AV Groningen, the Netherlands\\
$^{3}$Instituto de Astrof\'isica de Canarias (IAC), V\'ia L\'actea s/n, La Laguna 38205, S/C de Tenerife, Spain \\
$^{4}$Departamento de Astrof\'isica, Universidad de La Laguna, La Laguna, E-38205, S/C de Tenerife, Spain \\
$^{5}$Department of Physics and Astronomy, FI-20014 University of Turku, Finland \\
$^{6}$Nordita, KTH Royal Institute of Technology and Stockholm University, Roslagstullsbacken 23, SE-10691 Stockholm, Sweden \\
$^{7}$Space Research Institute of the Russian Academy of Sciences, Profsoyuznaya Str. 84/32, 117997 Moscow, Russia \\
}

\date{Accepted 2021 July 19. Received 2021 July 19; in original form 2021 February 19}

\pubyear{2020}
\begin{document}
\label{firstpage}
\pagerange{\pageref{firstpage}--\pageref{lastpage}}
\maketitle

\begin{abstract}

Black hole low-mass X-ray binaries (BH~LMXBs) evolve in a similar way during outburst. Based on the X-ray spectrum and variability, this evolution can be divided into three canonical states: low/hard, intermediate and high/soft state. BH~LMXBs evolve from the low/hard to the high/soft state through the intermediate state in some outbursts (here called ``full outbursts''). However, in other cases, BH~LMXBs undergo outbursts in which the source never reaches the high/soft state, here called  ``Failed-Transition outbursts" (FT outbursts). From a sample of 56 BH~LMXBs undergoing 128 outbursts, we find that 36\% of these BH~LMXBs experienced at least one FT outburst, and that FT outbursts represent $\sim$33\% of the outbursts of the sample, showing that these are common events. We compare all the available X-ray data of full and FT outbursts of BH~LMXBs from \textit{RXTE}/PCA, \textit{Swift}/BAT and MAXI and find that FT and full outbursts cannot be distinguished from their X-ray light curves, HIDs or X-ray variability during the initial 10--60 days after the outburst onset. This suggests that both types of outbursts are driven by the same physical process. We also compare the optical and infrared (O/IR) data of FT and full outbursts of GX~339$-$4. We found that this system is generally brighter in O/IR bands before an FT outburst, suggesting that the O/IR flux points to the physical process that later leads to a full or an FT outburst. We discuss our results in the context of models that describe the onset and evolution of outbursts in accreting X-ray binaries.

\end{abstract}




\section{Introduction}

Low mass X-ray binaries (LMXBs) are binary systems in which the primary component is a compact object, either a black hole or a neutron star, and the secondary component is a low-mass star, typically with a mass $<1M_{\odot}$. LMXBs with a black-hole candidate as primary component are known as black hole low-mass X-ray binaries (BH LMXBs). The energy spectra of LMXBs can be described by two main components during outburst: a soft thermal component and a hard power-law component \citep[e.g.][]{Remillard06c,Belloni10}. The soft thermal component is described by a multi-colour disc blackbody model \citep{Mitsuda84} generally peaking at 0.1--1 keV depending on the spectral state \citep[see ][ for a review]{Done07}. This is thought to be produced by a  geometrically thin and optically thick accretion disc \citep{Shakura73}. The hard power-law component is thought to be produced at the corona which, in some scenarios, is a region composed of hot electron plasma with temperatures from tens to hundreds of keV that surrounds the compact object and possibly the accretion disc \citep{Sunyaev79, Sunyaev80}. The emission of this component is due to thermal Comptonisation, by which the low energy photons are inverse Compton scattered by the electrons in the corona \citep{Sunyaev80,Titarchuk94,Zdziarski04,Done07,Burke17}. 

During an outburst, BH LMXBs show different spectral and timing properties \citep[e.g., ][]{VanDerKlis89,Mendez97,VanDerKlis00,Homan05b,Remillard06,Belloni10b,Belloni11,Plant14,Motta16}. Taking into account these properties, different states have been defined. Although different classifications have been proposed, here we use the classification given by \citet{Homan05b}, among others, which defines four spectral states. When the source is in the \textit{low-hard state} (LHS), the X-ray spectrum is dominated by a thermal Comptonisation component that can be described by a hard power-law with a high energy cut-off (around 100 keV) related to the temperature of the electron plasma \citep[see ][ for a review]{Done07}. A weak disc-blackbody component can also be detected in the energy spectrum \citep[e.g., ][]{McConnell02, Zdziarski02, Capitanio09a, Alabarta20}. In the LHS, the Power-Density Spectrum (PDS) is characterised by a strong broadband noise component with a break frequency below 1 Hz and a high fractional rms amplitude \citep[30\%-50\%, e.g.,][]{Belloni05,Remillard06c, Munoz-Darias11, Motta16}. Quasi-periodic oscillations (QPOs) can also be observed in this state with frequencies between 0.01 Hz to 30 Hz \citep[e.g., ][]{Casella04, Belloni05}. The \textit{high-soft state} (HSS) is characterised by an energy spectrum dominated by a soft thermal component due to an optically thick, geometrically thin accretion disc \citep{Shakura73}. A very weak hard power-law component can also be detected \citep[e.g., ][]{Capitanio09a, Alabarta20}. In the HSS, the fractional rms amplitude is less than 5\% and weak quasi periodic oscillations (QPOs) are sometimes detected \citep[e.g.][]{Casella04,Rodriguez04, Munoz-Darias11, Motta11, Sriram13, Motta16}. Between the LHS and the HSS, two intermediate states are defined considering the evolution of the X-ray variability: the \textit{hard-intermediate state} (HIMS) and the \textit{soft-intermediate state} (SIMS) \citep[e.g., ][]{Homan05b, Belloni10b}. The HIMS is characterized by a broadband noise fractional rms amplitude of 10\%$-$30\% \citep{Munoz-Darias11,Motta12}  and sometimes QPOs can be detected \citep[e.g., ][]{Casella04,Belloni05,Belloni14}. The SIMS shows a fractional rms amplitude of a few \% and type-A (in the 6.5$-$8 Hz frequency range) and type-B QPOs (in the 1$-$7 Hz frequency range) can also be found \citep[e.g., ][]{Wijnands99,Homan01,Casella04,Belloni05,Belloni14}. 

\begin{figure}
\centering
\includegraphics[width=\columnwidth]{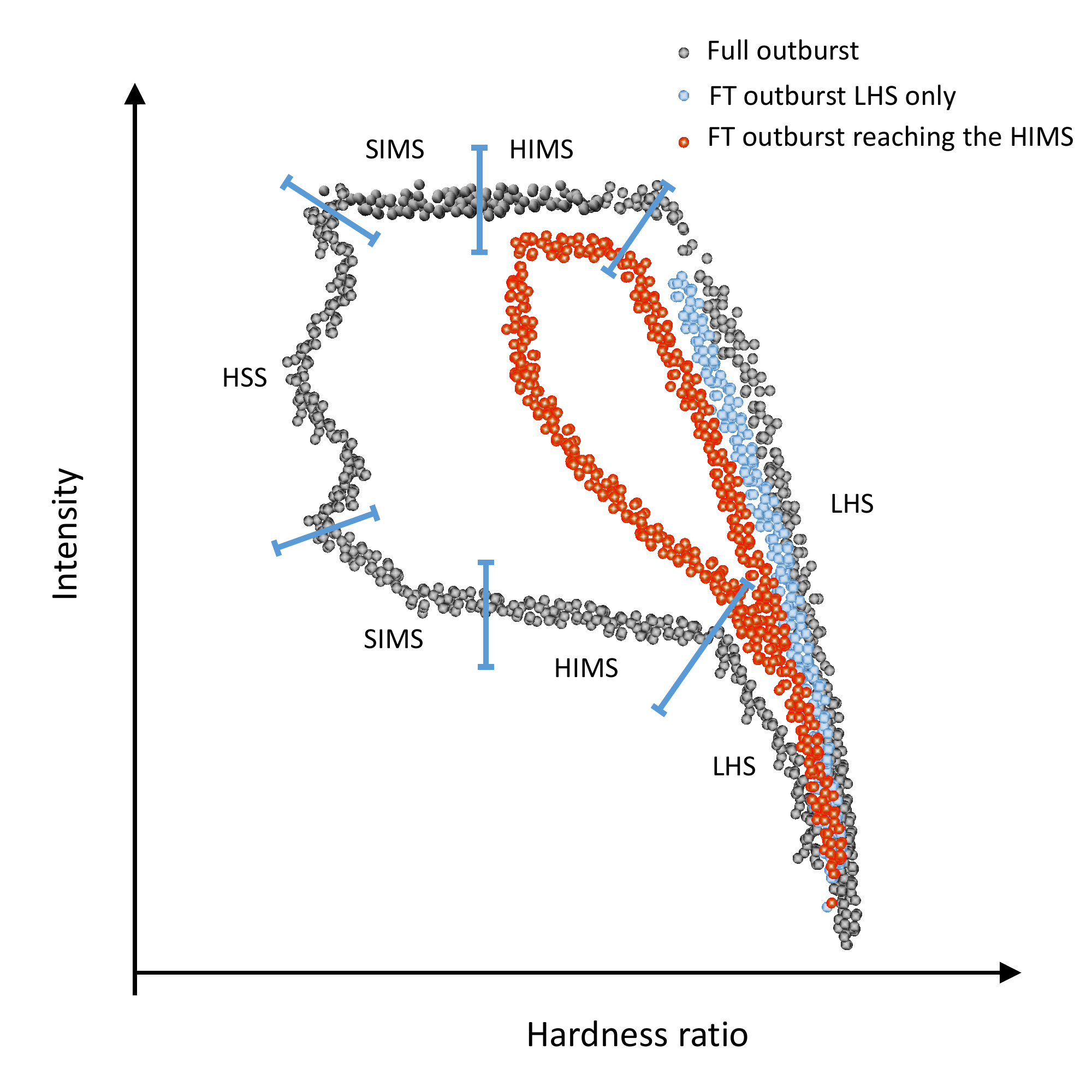}
\caption{Schematic HID of full (grey) and FT (red and blue) outbursts. The red track represents the sources that leave the LHS but do not reach the HSS and the blue track represents the sources that do not leave the LHS.}
\label{fig:diagrams}
\end{figure}

During a so-called canonical outburst, a system evolves through all the different spectral states in the following order: $LHS\rightarrow HIMS \rightarrow SIMS \rightarrow HSS \rightarrow SIMS \rightarrow HIMS \rightarrow LHS$.  This evolution can be well traced using the hardness-intensity diagram \citep[HID; e.g.,][]{Homan01,Remillard06}, the hardness-rms diagram \citep[HRD,][]{Belloni05}, the rms-intensity diagram \citep[RID,][]{Munoz-Darias11} and the power colour-colour diagram \citep[PCC,][]{Heil15a}. 
The grey track in Fig. \ref{fig:diagrams} shows a schematic representation of the track traced by a BH LMXB during an outburst in the HID. This pattern in the HID is known as the q-track. At the beginning, the source is in the LHS, in the right bottom part of the HID. In this phase of the outburst, the source intensity increases without a big change of hardness, following a vertical line on the right of the HID. At some luminosity, the source starts the transition to the HSS moving to the left of the HID, following a horizontal branch without big changes of the source intensity. In this horizontal branch the source enters the HIMS and the SIMS. When the source reaches the HSS state on the top left part of the HID, its intensity decreases at an approximately constant value of the hardness. From then on, the evolution reverses and the source returns to the intermediate states following a bottom horizontal branch of the HID. Finally, at the end of the outburst the source goes through the LHS again. Several sources show this pattern in their outbursts:  e.g. XTE~1550--564, GX~339--4, H1743--322 and GRO~J1655--40 \citep{Homan01,Belloni05,Fender09,Dunn10,Uttley15}. We call ``full outbursts'' to the outbursts that follow the full spectral evolution.

Despite the q-track pattern observed in many BH LMXBs, some sources undergo outbursts that do not show the complete spectral evolution. These kinds of outbursts are known in the literature as \textit{``hard-only state''} outbursts \citep{Tetarenko16}, \textit{``low/hard state''} outbursts \citep{Belloni02}, \textit{``failed outbursts''} \citep[e.g., ][]{Capitanio09a,Curran13} or \textit{``failed state transition outbursts''} \citep[][]{Bassi19}. 
According to the track traced on the HID, the ``failed outbursts'' can be divided in two groups. The first group corresponds to sources that remain in the LHS during the whole outburst \citep[e.g., ][]{Hynes00b, Brocksopp01, Belloni02,Brocksopp04,Curran13}, while the sources belonging to the second group make the transition to the intermediate states without reaching the HSS \citep[e.g., ][]{Zand02b,Capitanio09a,Ferrigno12}. The track followed by these two types of "failed outbursts" are represented with blue and red circles in Fig. \ref{fig:diagrams}, respectively. All ``failed outbursts'', in addition, appear to reach fainter X-ray luminosity peaks than full outbursts \citep{Tetarenko16}. Most of the names presented in the lines above to call this type of outbursts do not completely describe their real nature. \textit{``Failed outbursts''} may cause some confusion by giving the impression that the source did not undergo into outburst. The names of \textit{``hard-only state''} and \textit{``low/hard state''} give the impression that the source never leaves the LHS. However, we have mentioned that some ``failed outbursts'' made the transition to the HIMS. For simplicity, we propose the nomenclature: ``Failed-Transition'' outbursts (hereafter FT outbursts), similar to that used by \citet{Bassi19}. 

Some studies analysed the differences between full and FT outbursts in radio wavelengths. \citet{Corbel13} found the correlation between the X-ray and radio emission. The same correlation was also found only with the hard state emission of the source. \citet{Furst15} used new data to confirm this correlation and they found that the correlation is slightly different for the FT (2008, 2009 and 2013) than for the full outbursts.  \citet{deHaas20} found a flatter radio$-$X-ray correlation using FT outbursts of GX~339--4 than with full outbursts of the same source. They interpreted this flatter correlation as an inefficient coupling between the X-rays and radio emission mechanisms. They also suggested that this correlation could be used to predict the nature of the outburst at its early stages. However, \citet{Williams20} studied the same correlation in the 2018 FT outburst of H~1743--322 and found that full and FT outbursts of this source show a similar correlation.

In this paper we present the observational differences between full and FT outbursts using X-ray and O/IR data during the first days of the outbursts. The aim is to understand why some outbursts behave as full outbursts and some others as FT outbursts and, in particular, whether it is possible to predict the type of outburst during the first days. After an exhaustive search of the X-ray archive and literature, we constrained our analysis of X-ray data to three BH systems that have enough observations during the rising part of the outburst: GX~339--4, H~1743--322 and GRS~1739--278. In addition, we also studied O/IR data of GX~339--4. In section 2 and 3 we describe the source selection and the data analysis, respectively. In section 4.1 we show the fraction of FT outbursts of our sample. In section 4.2 we show the relation between the X-ray peak intensity of the outbursts and the time spent in quiescence between outbursts. In sections 4.3, 4.4 and 4.5 we compare the observational X-ray light curves, HIDs and PCCs between FT and full outbursts. In section 4.6 we analyse the observational differences between both types of outbursts of GX~339--4 in O/IR wavelengths during outburst and quiescence. Finally, in sections 5 and 6 we discuss our results and present our conclusions.

\section{Data \& data analysis}

\subsection{Source selection}

For our study, we selected 46 transient BH LMXBs from the WATCHDOG catalogue \citep{Tetarenko16} which showed at least one full or one FT outburst. In the cases where the outburst was not classified in the literature, we checked the spectral and timing properties of the source to decide whether they were consistent with those of full or FT outbursts, and hence included them in our study. If the spectral and timing properties were not enough to determine the nature of the outburst, the source was excluded. In addition to the 46 BH LMXBs studied by \citet{Tetarenko16}, we considered one more source (MAXI J1828-249) not included in that paper and nine further BH systems that went in outburst from 2016 until now. The final sample used in our study includes 56 sources that meet the above requirements (see Appendix A).

\subsection{X-ray light curves and HIDs}

We used all the available data from the \textit{Rossi X-ray Timing Explorer (RXTE)} Proportional Counter Array \citep[PCA; for instrument information see ][]{Zhang93,Jahoda06} for the selected sources. To calculate tha hardness ratio (HR) we used the 16-s time-resolution Standard 2 mode data. For each of the five PCA detectors (PCUs) we calculated the HR defined as the 16.0--20.0 keV count rate divided by the count rate in the 2.0--6.0 keV band. We also calculated the intensity defined as the count rate in the 2--20 keV energy band. To obtain the count rates in these exact energy ranges, we made a linear interpolation between PCU channels. We then performed deadtime corrections, subtracted the background contribution in each band using the standard bright source background model for the PCA (version 2.1e1) and removed instrumental drop-outs to obtain the HR and intensity for each time interval of 16 s. We also took into account that the \textit{RXTE} gain epoch changed with each new high voltage setting of the PCUs \citep{Jahoda06}. To correct for this effect and the differences in effective area between the PCUs, we normalized our data to the Crab \citep[method introduced by ][]{Kuulkers94}. Following this method, for each PCU we obtained the HR and intensity of the Crab, which is supposed to be constant, and after that we averaged the 16 s Crab HR and intensity for each PCU. Then, we divided them by the average Crab values that are closer in time and in the same \textit{RXTE} gain epoch. We obtained the HR and intensity average over all PCUs and with that the average of HR and intensity for each interval of 16 s. Finally, we averaged the HR and intensity for each ObsID separately. We also compared the \textit{RXTE}/ASM, the 15--50 keV \textit{Swift}/BAT \footnote{https://swift.gsfc.nasa.gov/results/transients/} and the 2--20 keV MAXI light curves \footnote{http://maxi.riken.jp/top/index.html}.

\subsection{Power colour-colour diagrams (PCCs)}

The power ``colour-colour'' diagrams (PCCs) of GX~339--4 analysed here were first presented by \citet{Heil15a}. These authors obtained the so-called power colours, defined as the ratio of the broadband variability in two different frequency bands. They defined two power colours in the 2--16 keV energy band: PC1 as the ratio of the variability in the 0.25--2.0 Hz and 0.0039--0.031 Hz frequency bands; and PC2 as the ratio in the frequency bands 0.031--0.25 Hz and 2.0--16.0 Hz.   

\subsection{Optical and Infrared light curves and colour-colour diagrams of GX 339--4}

The optical and infrared (O/IR) data used in this paper come from the ANDICAM camera \citep[][]{DePoy03} on the Small and Moderate Aperture Research Telescope System \citep[SMARTS; ][]{Subasavage10} and correspond to GX~339--4 \citep[][]{Buxton12}, since this is the only source with enough O/IR data for the purposes of this paper. The SMARTS light curves used in this paper were studied and presented in \citet{Buxton12} and include the 2002--2010 time period. The SMARTS team also provided the O/IR light curves of the 2011--2015 period. The O/IR light curves consist of \textit{V}, \textit{I}, \textit{J} and \textit{H} magnitude. Gaps in the data occur when the source is behind the Sun. We used the colours: \textit{V-H}, \textit{V-I}, \textit{V--J}, \textit{I--J}, \textit{I--H} and \textit{J--H}. From them, we created the colour-magnitude diagrams (CMDs) comparing the colours defined with the O/IR band magnitudes.

\section{Results}

\subsection{Rate of FT outbursts}

Based on the 56 sources that satisfy our selection criteria, showing 128 outbursts in the 1971-2021\footnote{For completeness, we also considered an outburst of 1A~0620--00 occurred in 1917.} time period (see Section 2 and Appendix A), we found that: 

\begin{itemize}
    \item 36\% of the sources showed at least one FT outburst (20 out of 56, see Appendix A for the list of sources). Fig.~\ref{total} shows, for each source that exhibits FT outbursts, the number of outbursts of each type.
    \item 12 of the sources showed only FT outbursts (see Table \ref{tab:table_sources}).
    \item 14\% of the sources show both full and FT outbursts (eight out of 56). 
    \item 33\% of the outbursts were catalogued (see references in Table \ref{tab:table_sources}) as FT outbursts (42 out of 128). Table \ref{tab:table_sources} contains the type of outburst of every outburst of the sources in our sample.
\end{itemize}

\begin{figure*}
\centering
\includegraphics[width=\textwidth]{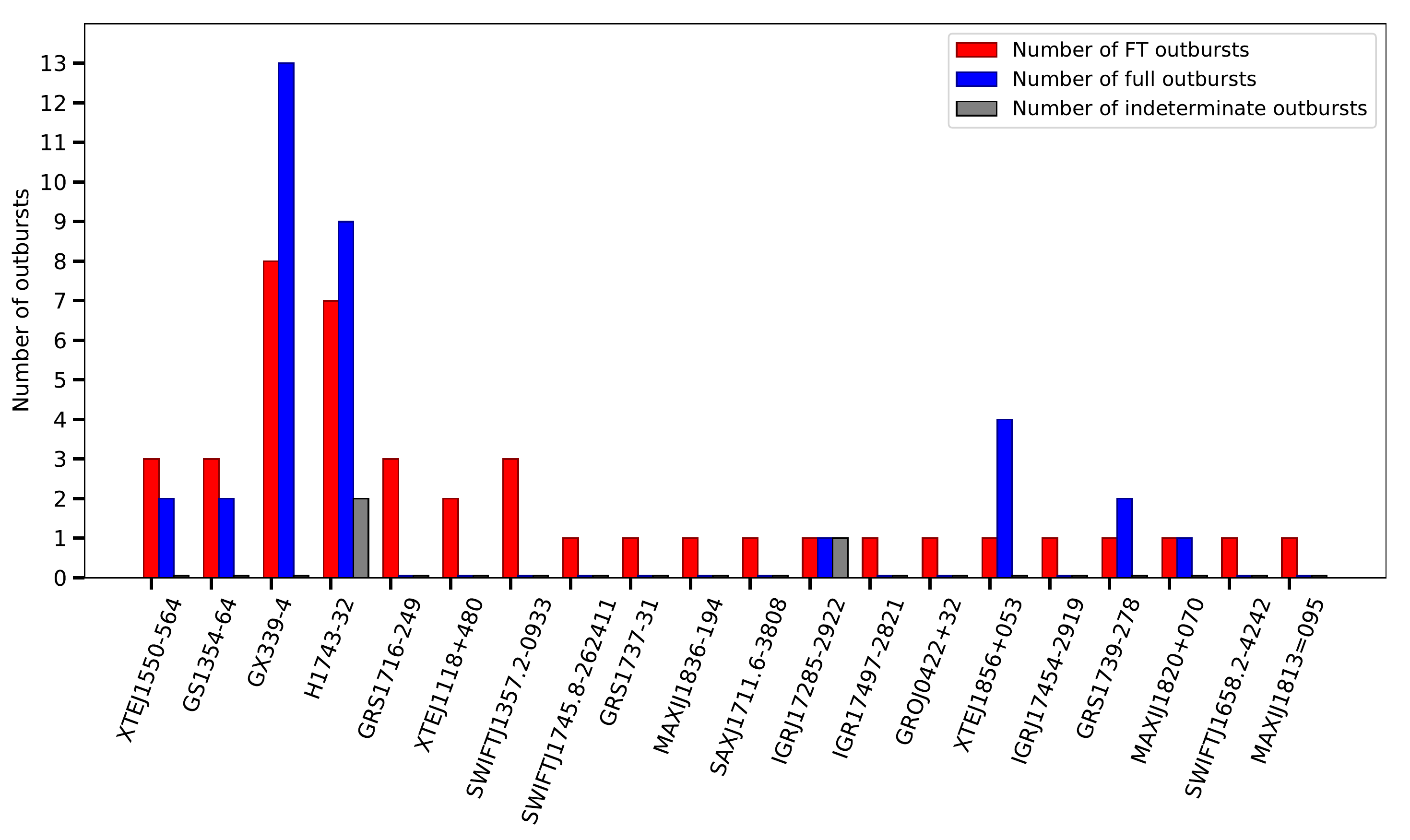}
\caption{Comparison between the number of FT and full outbursts in sources that exhibit both types of outbursts.}
\label{total}
\end{figure*}

\subsection{Quiescence times before full and FT outbursts}

\begin{figure}
\centering
\includegraphics[width=\columnwidth]{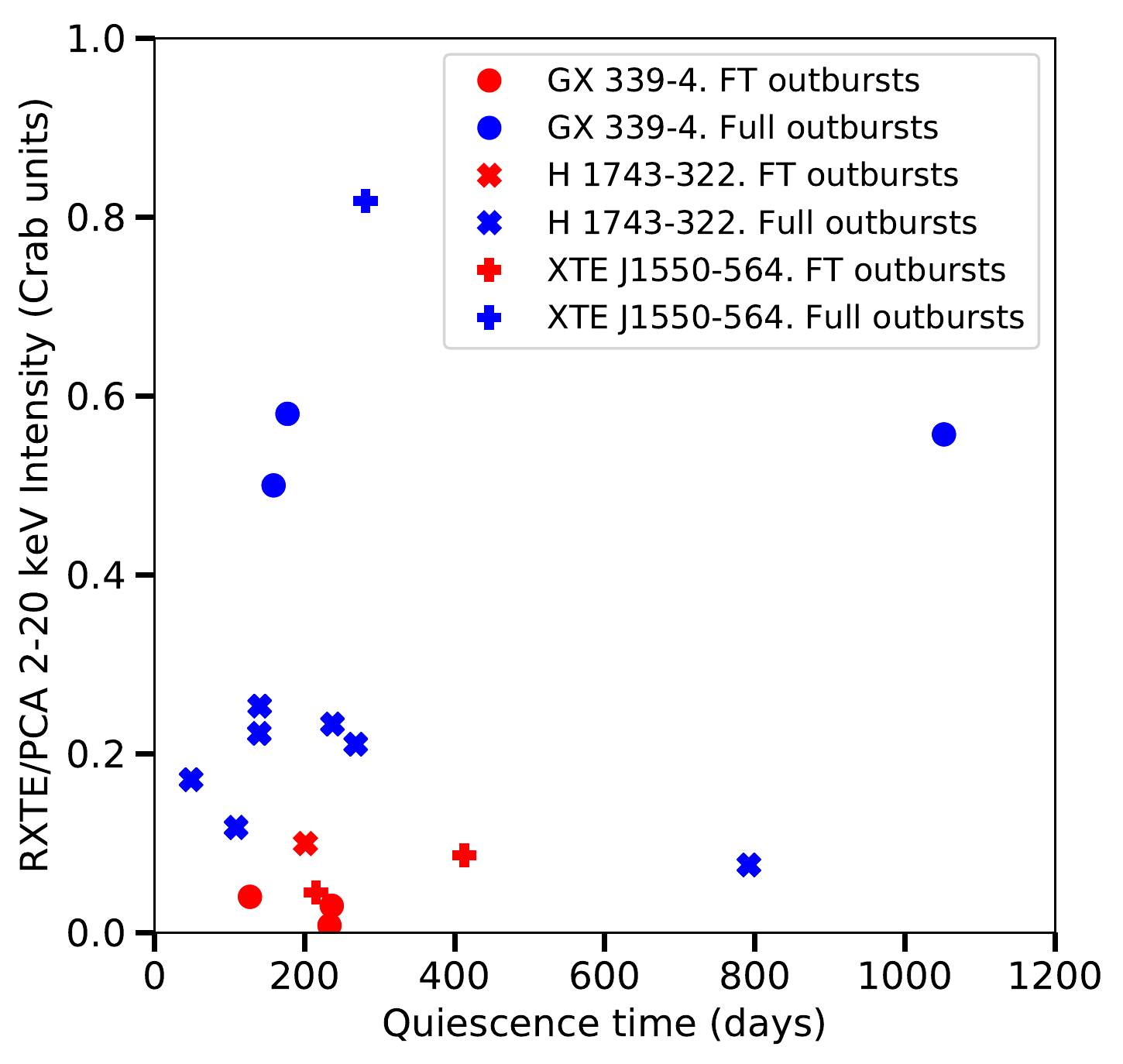}
\caption{\textit{RXTE}/PCA peak intensity vs Quiescence time for the outbursts of GX~339$-$4, H~1743$-$322 and XTE~J1550$-$564.}
\label{fig:waiting_all}
\end{figure}

Full outbursts are brighter in X-rays than FT outbursts \citep{Tetarenko16}. Assuming that the X-ray peak intensity depends on the amount of matter accumulated in the accretion disc and that the mass accretion rate from the companion star is constant, if a source spends more time in quiescence its accretion disc would accumulate more matter and, as a consequence, the following outburst would be brighter. Then, a source would spend more time in quiescence before a full outburst than before an FT outburst. Defining the quiescence time as the time difference between the onset of an outburst and the offset of the previous one in X-rays, we studied the relation between the quiescence time of full and FT outbursts and the \textit{RXTE}/PCA peak intensity of the corresponding outburst. We used data from the sources GX~339$-$4, H~1743$-$322 and XTE~J1550$-$564 since these were the only sources that underwent multiple outbursts of each type and the peak of the outbursts was observed with \textit{RXTE}/PCA. To obtain the quiescence time, we used the begin and end times of the outbursts given by \citet{Tetarenko16}. 
Fig. \ref{fig:waiting_all} shows the relation between the quiescence time and the intensity peak of the outbursts of the sources mentioned before. We found that the three sources spent similar times in quiescence before FT and full outbursts. Only before two full outbursts (the 2002 of GX~339$-$4 and the 2007 outburst of H~1743$-$32) did the source spend a longer time in quiescence compared to the other outbursts.

\subsection{X-ray light curves}

As we mentioned before, full outbursts are brighter than FT outbursts \citep{Tetarenko16}. What causes this difference in the maximum luminosity reached by each type of outburst can also make the light curves evolve in a different way from the beginning of the outburst for both types of outbursts. In order to study that, we compared, for all sources with enough available data, the rising part of the light curves of different FT and full outbursts of the same source. In order to do that, we considered the time of the beginning of the outbursts given in \citep{Tetarenko16} and we then shifted all the outbursts in time such that the start of all outbursts coincide in time. First, we compared the rising parts of both type of outbursts using data from pointed observations with \textit{RXTE}/PCA. Then, we compared both type of outbursts using observations from \textit{Swift}/BAT and MAXI, both all sky monitors. It is important to note that our results can be biased by the method used to determine the time of the beginning of the outbursts. \citet{Tetarenko16} considered the beginning of an outburst as the first 3$\sigma$ detection above the background using different instruments. Because of that, the beginning of an outburst in a specific instrument can differ a few days from the value we used.

\subsubsection{Comparing the rise of FT and full outbursts using X-ray light curves of \textit{RXTE}/PCA}

\begin{figure}
\centering
\includegraphics[width=\columnwidth]{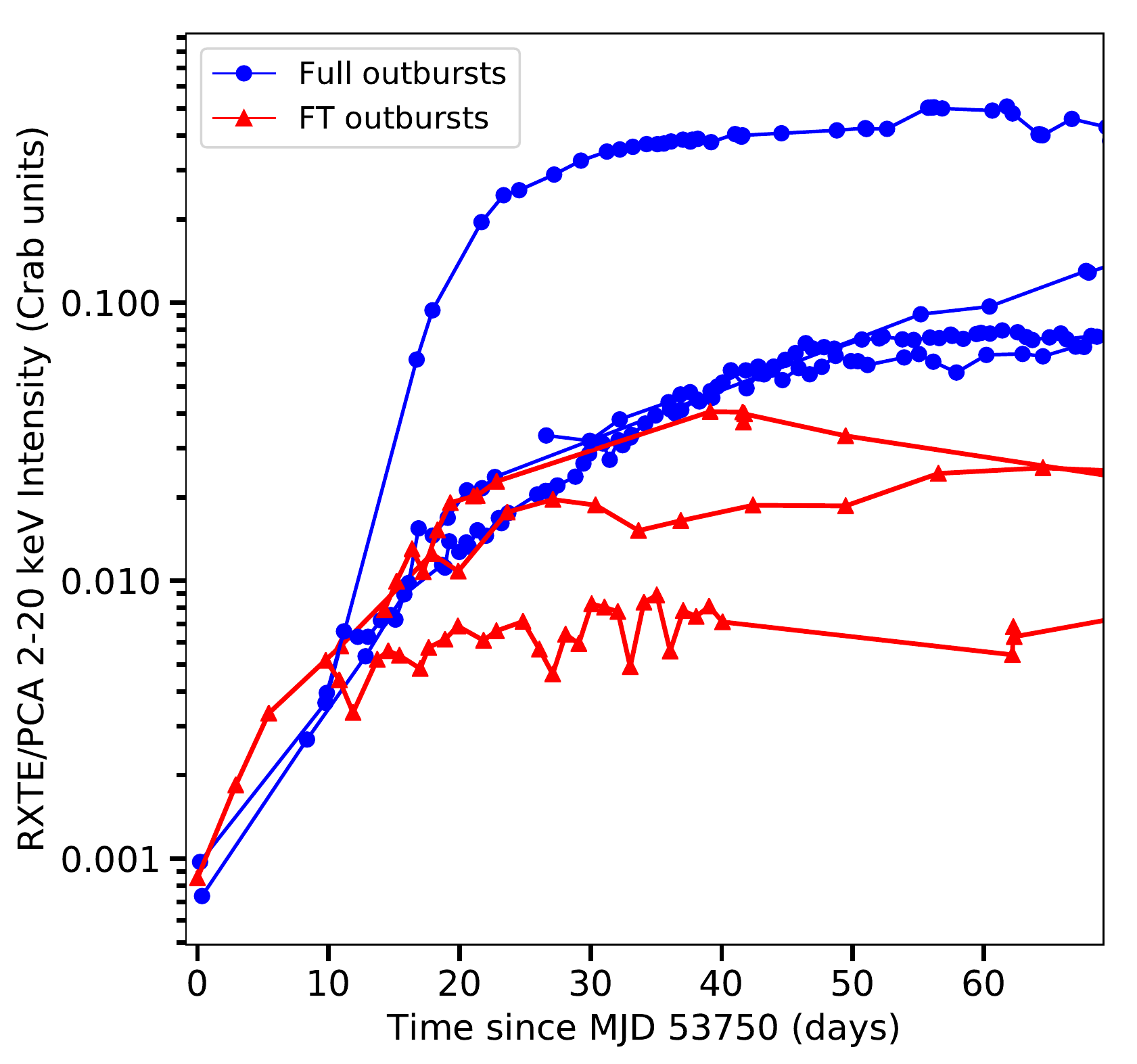}
\caption{\textit{RXTE}/PCA light curves of the outbursts of GX~339$-$4. The beginning of the outbursts are zoomed in.} 
\label{f:PCA339}
\end{figure}

For the comparison of the rising parts of FT and full outbursts using \textit{RXTE}/PCA observations, we focused on GX~339$-$4. This is the only source with multiple outbursts of each type with enough \textit{RXTE}/PCA observations during the rising parts. Fig. \ref{f:PCA339} shows the comparison between \textit{RXTE}/PCA light curves, where we plot four full outbursts in blue circles (2002, 2004, 2006 and 2009) and three FT outbursts in red triangles (2006, 2008 and 2009). We found that during the first $\sim$33 days, the light curves of FT and full outbursts evolved along similar tracks. After $\sim$33 days the flux of full outbursts continued increasing whereas the flux of FT outbursts started to decrease. The only exception is the 2002 full outburst, whose intensity increased faster than the others. 

\subsubsection{Comparing the rise of FT and full outbursts using X-ray light curves of the all sky monitors \textit{Swift}/BAT and MAXI}

\begin{figure*}
\centering
\includegraphics[width=\textwidth]{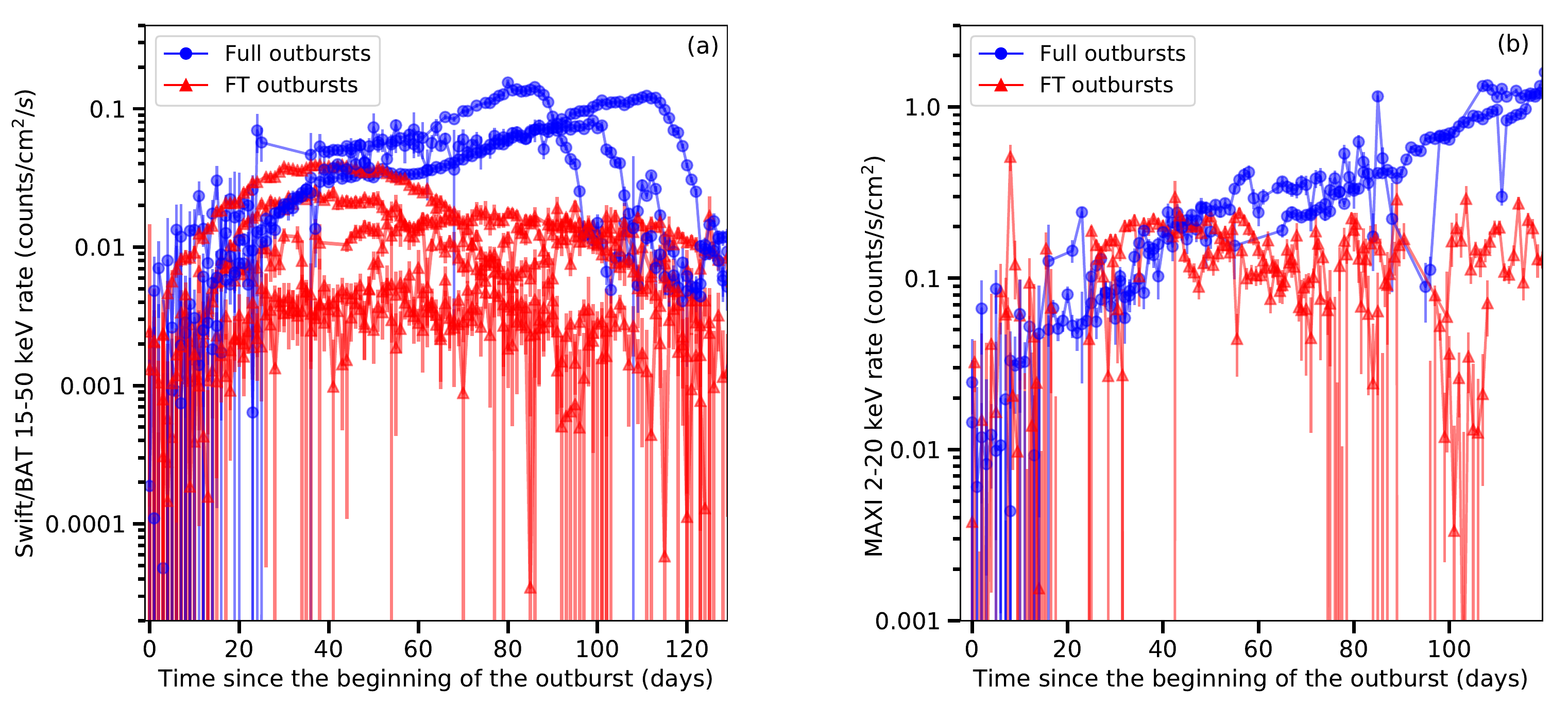}
\caption{15--50 keV \textit{Swift}/BAT (panel (a))and 2--20 keV MAXI (panel(b)) light curves of the outbursts of GX~339$-$4. The beginning of the outbursts are zoomed in.}
\label{f:339risecomp}
\end{figure*}

\begin{figure*}
\centering
\includegraphics[width=\textwidth]{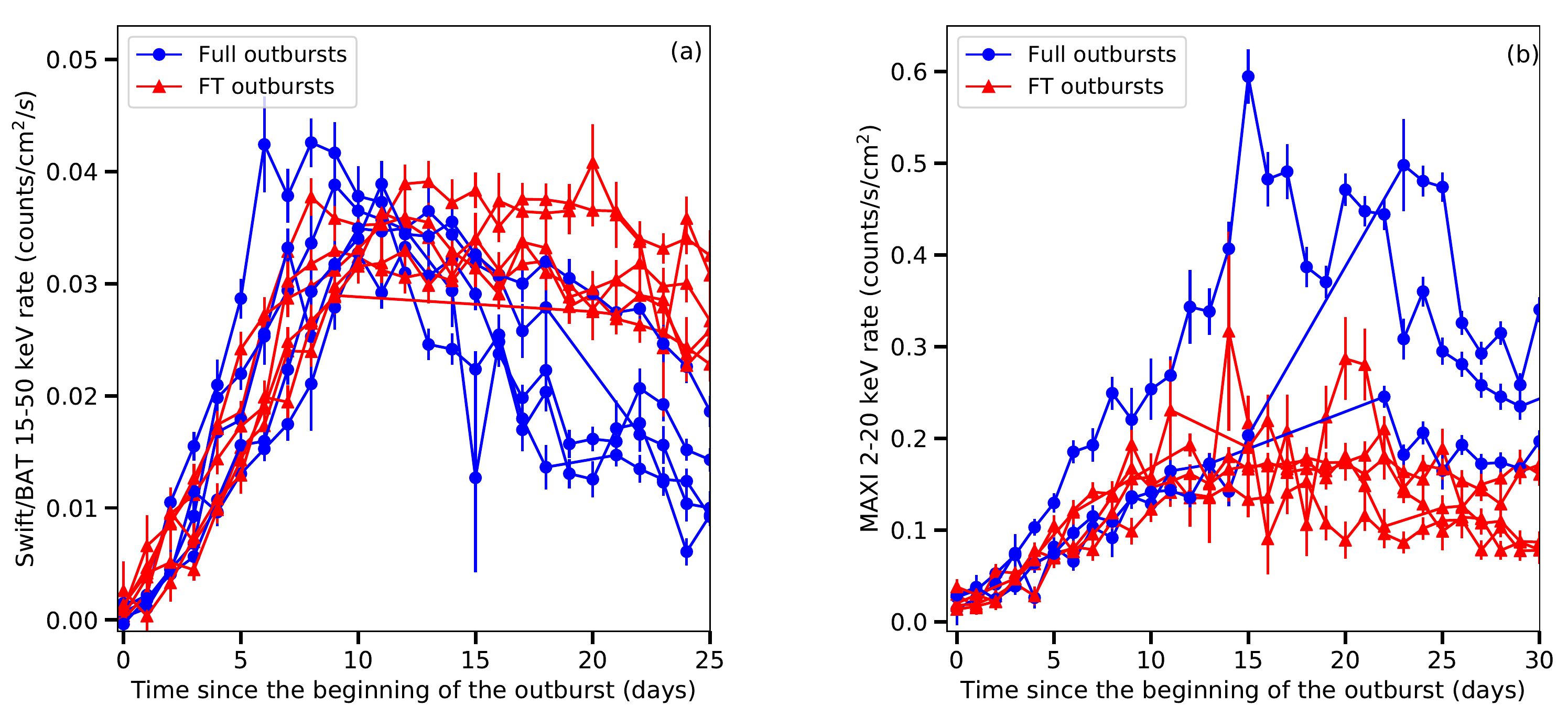}
\caption{15--50 keV \textit{Swift}/BAT (panel(a)) and 2--20 keV MAXI (panel(b)) light curves of the outbursts of H~1743$-$322. The beginning of the outbursts are zoomed in.}
\label{f:BAT1743}
\end{figure*}

\begin{figure*}
\centering
\includegraphics[width=\textwidth]{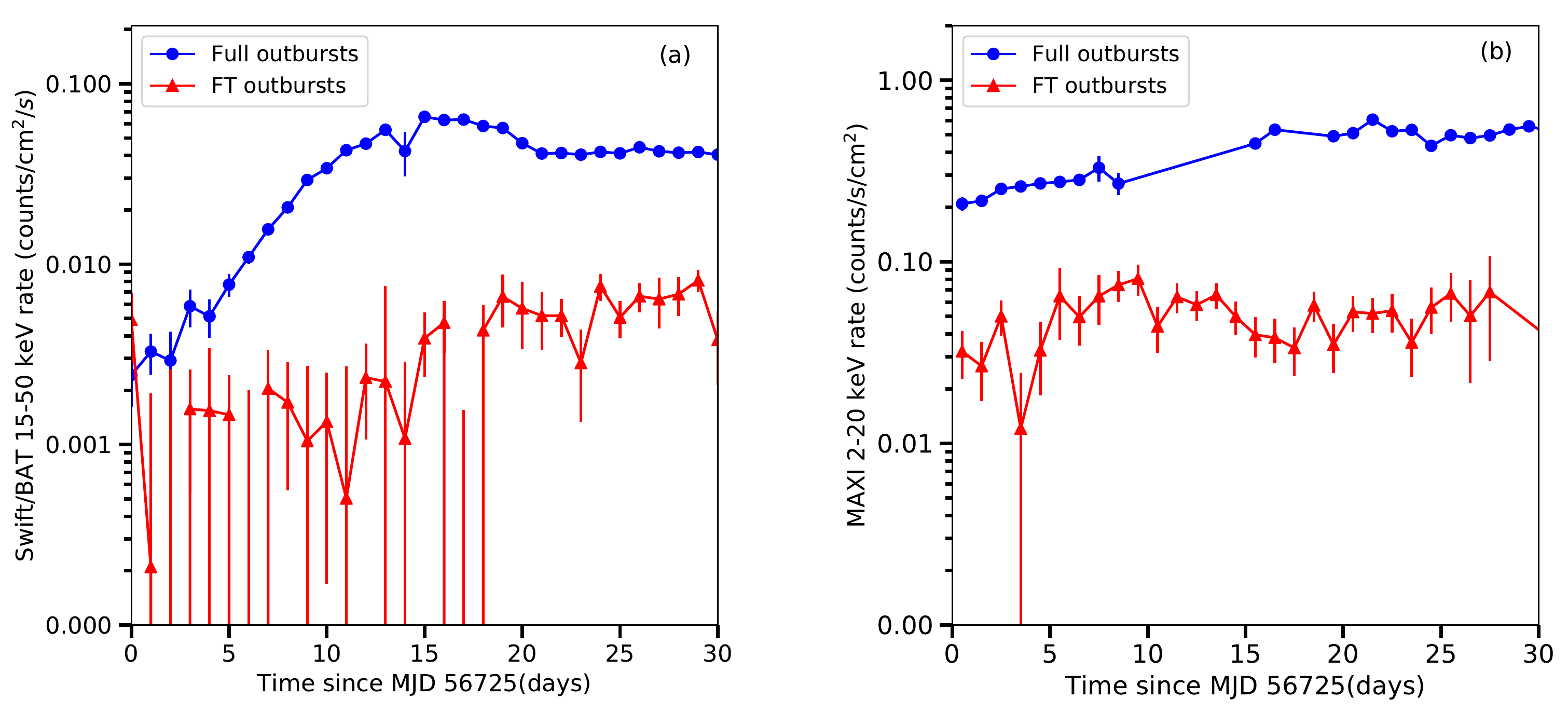}
\caption{15--50 keV \textit{Swift}/BAT (panel(a)) and 2--20 keV MAXI (panel(b)) light curves of the outbursts of GRS~1739$-$278. The beginning of the outbursts are zoomed in.}
\label{f:BAT1739}
\end{figure*}

For the comparison of the rising parts of FT and full outbursts using all sky monitors we used data from \textit{Swift}/BAT and MAXI, given that most of the FT outbursts were too faint to be detected with \textit{RXTE}/ASM. We focused on the sources GX~339$-$4, H~1743$-$322 and GRS~1739$-$278, since the three sources show full and FT outbursts bright enough to be detected by both instruments. As in the previous section, we considered the beginning of the outbursts defined by \citet{Tetarenko16}. However, in the case of H~1743$-$322, the source was first detected at higher energies than the energy band of \textit{MAXI} and \textit{Swift}/BAT, so the light curves are flat during the first days of the outburst. Therefore, we moved the beginning of the outbursts to the beginning of the rise of the outbursts.

Panel (a) of Fig.~\ref{f:339risecomp} shows the \textit{Swift}/BAT light curves of GX~339$-$4, where three full outbursts in blue (2006, 2009 and 2014) and five FT outbursts in red (2006, 2008, 2009, 2013 and 2017) are plotted. The result was the same as with \textit{RXTE}/PCA. During the first $\sim$30 days, the light curves of both types of outbursts followed similar tracks. After that, full outbursts were brighter than FT outbursts. The only exception was the 2013 FT outburst, which started to decay before the 33 days from the beginning. Panel (b) of Fig.~\ref{f:339risecomp} shows the 2--20 keV MAXI light curves of two full outbursts (2009 and 2014) and two FT outbursts (2013 and 2017) of GX~339$-$4. We see that both types of outburst overlap during the first $\sim$50 days.

Panel (a) of Fig.~\ref{f:BAT1743} shows the \textit{Swift}/BAT light curves of six FT outbursts (2008, 2012, 2014, 2015, 2017 and 2018) and five full outbursts (2009, 2010, a second outburst in 2010, hereafter 2010b, 2013 and 2016) of H~1743$-$322. As we found for GX~339$-$4, it is not possible to distinguish between FT and full outbursts at the beginning of the outburst. Panel (b) of Fig.~\ref{f:BAT1743}, on the other hand, shows the corresponding MAXI light curves of H~1743$-$322, including five FT outbursts (2012, 2014, 2015, 2017 and 2018) and four full outbursts (2010, 2010b, 2013 and 2016). The rising part of FT and full outbursts overlap during the first $\sim$20 days, which is the length of the rising part of FT outbursts. The only exception is the 2010 full outburst of H~1743$-$322, whose intensity increased faster than the intensity of the other outbursts.

Panels (a) and (b) of Fig.~\ref{f:BAT1739} show, respectively, the \textit{Swift}/BAT and MAXI light curves of the 2014 (full) and 2016 (FT) outbursts of GRS~1739$-$278. In contrast to GX~339$-$4 and H~1743$-$322, we found that the rising parts of the FT and full outbursts of GRS~1739$-$278 do not overlap during the first 30 days. The 2016 outburst is always fainter than the 2014 full outburst. This is due to the fact that the full outburst started to be observed when its intensity was at higher values than those reached by the FT outburst.

\subsection{Comparing FT outbursts and full outbursts using the HID}

\begin{figure}
\centering
\includegraphics[width=\columnwidth]{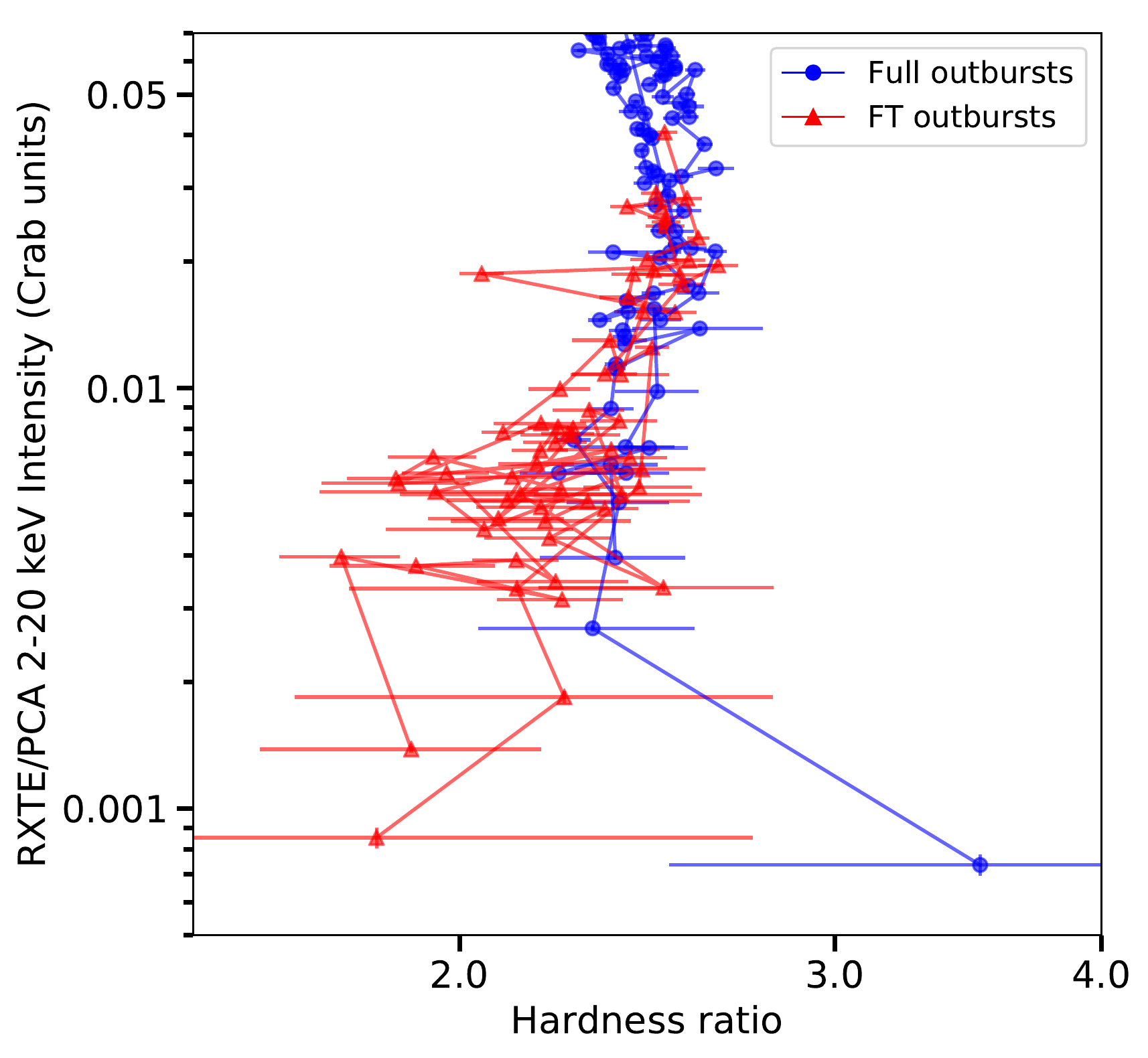}
\caption{HID of 2004 (full), 2006 (FT), a second outburst in 2006 (full, 2006b), 2008 (FT) and 2009 (FT) outbursts of GX 339$-4$. The rising and the end of the decaying parts of the outbursts are zoomed in.}
\label{339risecomphid}
\end{figure}

As we mentioned in Section 1, FT outbursts do not show the full spectral evolution typical of BH LMXBs. This can be observed in the HID since FT outbursts do not follow the full q-track. Given that we observe the same behaviour for full and FT outbursts during the first $\sim$33 days in GX~339$-$4 light curves, we explored whether the track followed along the HID is also similar for both types of outburst. We compared the rise of full and FT outbursts using the \textit{RXTE}/PCA HID of GX~339$-$4. As we mentioned in the previous section, GX~339$-$4 is the only source with enough observations during the rise of the outbursts to compare the track followed in the HID. Fig.~\ref{339risecomphid} shows the rise of both types of outburst in the HID. Two full outbursts in blue (2004 and 2006) and three FT outbursts in red (2006b, 2008 and 2009) are plotted. We did not find significant differences between the rise of full and FT outbursts within error bars. All outbursts followed the same track along the HID.  

\begin{figure*}
\centering
\includegraphics[width=\textwidth]{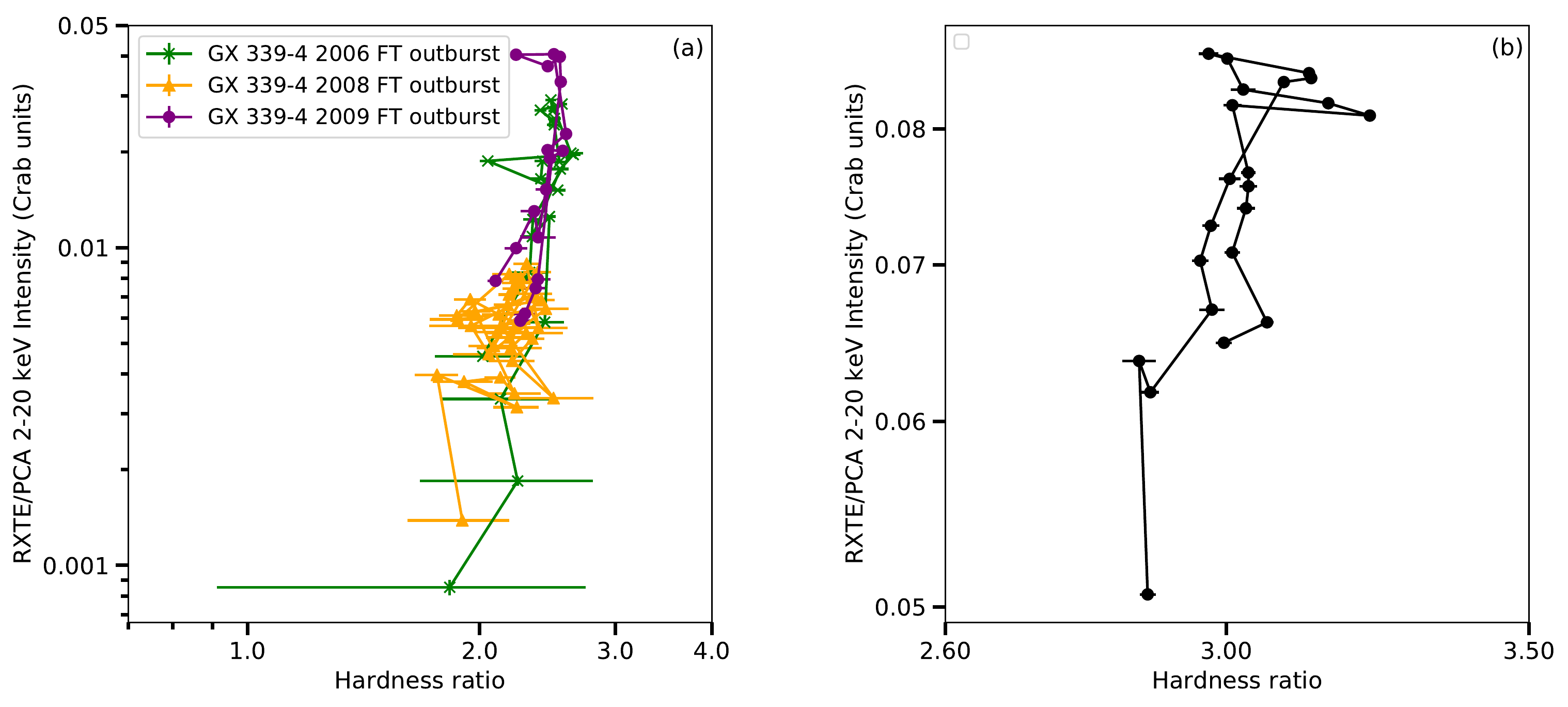}
\caption{\textit{RXTE}/PCA HIDs of the 226, 2008 and 2009 FT outbursts of GX~339$-$4 (panel(a)) and the 2003 FT outburst of XTE~J1550$-$564 (panel(b)). Both panels show the rising and decaying parts of the outbursts.}
\label{f:rdfailed}
\end{figure*}

For completeness, we also compared the rise and the decay of FT and full outbursts in the HID using \textit{RXTE}/PCA data. The only FT outbursts with enough observations during the rise and the decay are the 2006, 2008 and 2009 outbursts of GX~339$-$4 and the 2003 outburst of XTE~J1550$-$564. Panel (a) of Fig. \ref{f:rdfailed} shows the HID of the FT outbursts of GX~339$-$4. We found that the three FT outbursts followed the same track along the HID. In addition, the rise and the decay overlapped in the HID. Panel (b) of Fig. \ref{f:rdfailed} shows the HID of the 2003 FT outburst of XTE~J1550$-$564. The rise and the decay of this outburst followed similar tracks in the HID, as it happens in GX~339$-$4.

\begin{figure*}
\centering
\includegraphics[width=\textwidth]{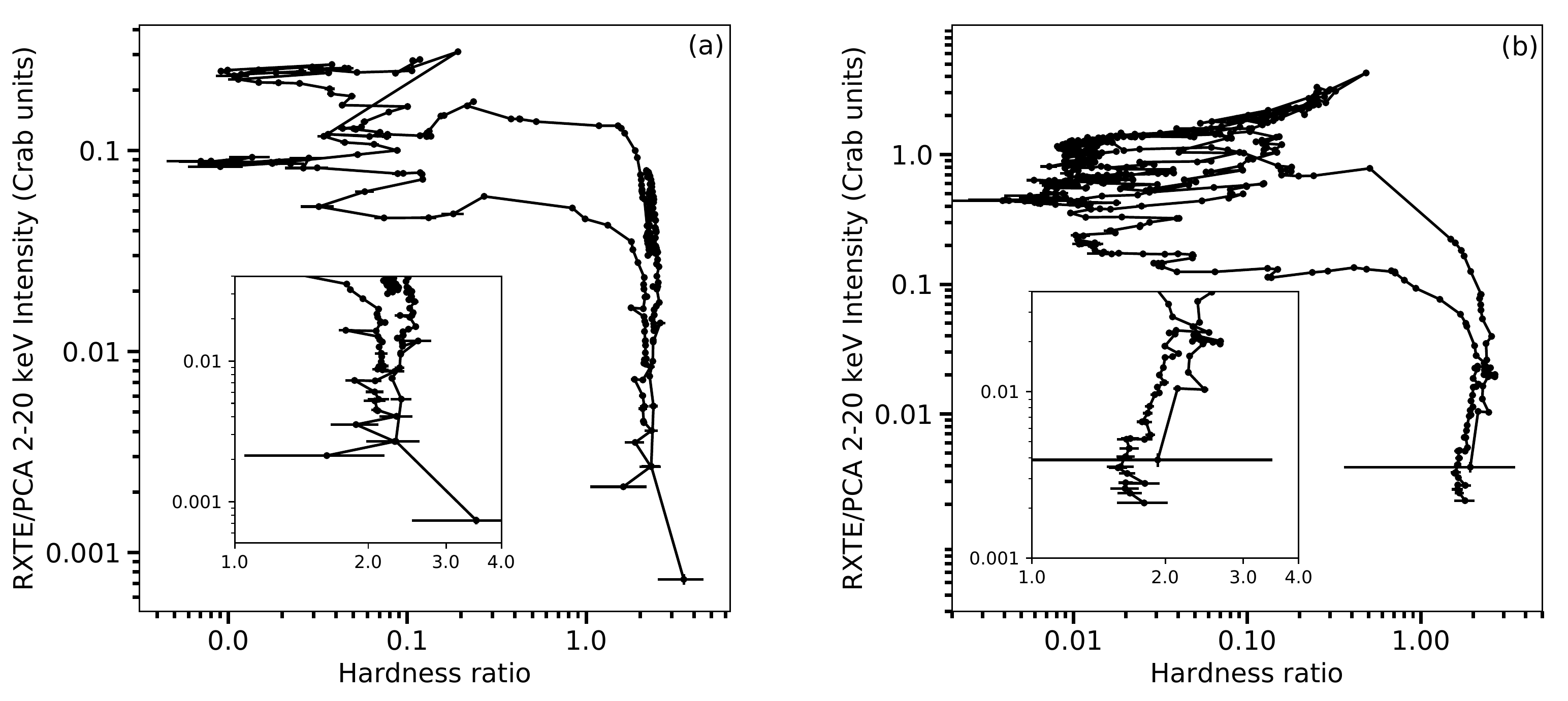}
\caption{\textit{RXTE}/PCA HIDs of the 2004 outburst of GX~339$-$4 (panel(a)) and the 2005 outburst of GRO~1655$-$40 (panel(b)), both full outbursts. The bottom right part of the diagram is zoomed.}
\label{rdfull}
\end{figure*}

As for full outbursts, we compared the rise and the decay of the 2005 outburst of GRO~1655$-$40 and the 2004 outburst of GX~339$-$4 since these are the only full outbursts with enough \textit{RXTE}/PCA observations (Fig.~\ref{rdfull}). We found that the rise and the decay did not follow the same track along the HID, implying that the source is softer in the decay than in the rise of the outbursts, contrary to what we found for FT outbursts. Moreover, both the rise and decay phases of FT outbursts are located in the place of the rising phase of full outbursts.

\subsection{Comparing the power spectra of full and FT outbursts using power colours}

\begin{figure}
\centering
\includegraphics[width=\columnwidth]{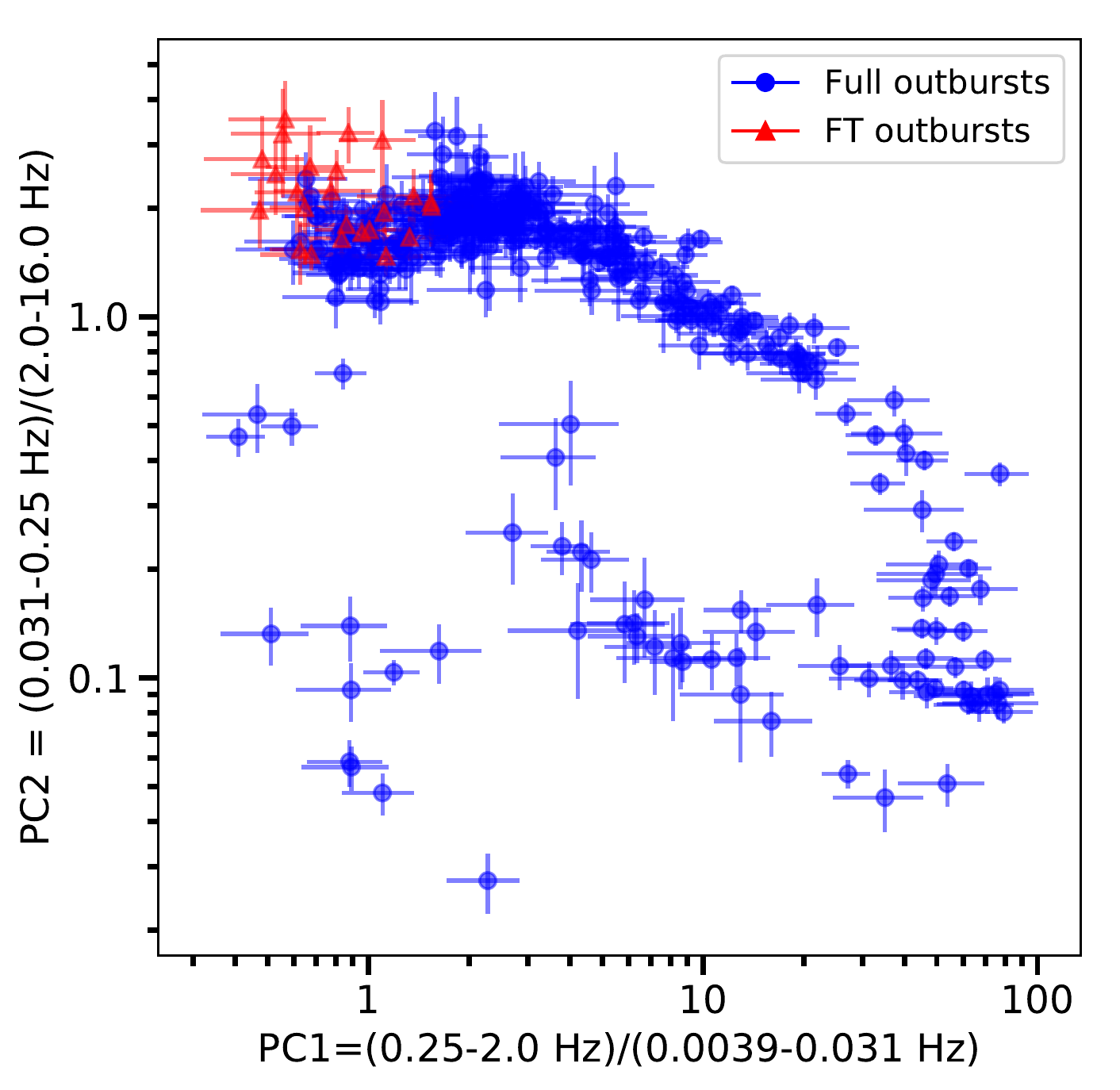}
\caption{Power colours of full (1998, 2002, 2004, 2006b and 2009b) and FT (2006, 2008 and 2009) outbursts of GX~339$-$4.}
\label{fig:power_colors}
\end{figure}

We explored whether the variability properties at the beginning of the outburst can affect the nature of the outburst. Fig. \ref{fig:power_colors} shows the PCC of the rise of full and FT outbursts of GX~339$-$4, obtained with \textit{RXTE}/PCA observations. We found that FT outbursts did not show the whole track along the PCC, but occupy a narrow area in the plot, which is the area corresponding to the LHS \citep{Heil15a}. This result was expected, since FT outbursts of GX~339$-$4 did not leave the LHS. If we only compare the rise of the outbursts, we found that the power colours of both type of outbursts overlap in the PCC.

\subsection{O/IR data of GX~339$-$4}

We explored whether we can distinguish FT outbursts and full outbursts using optical and infrared data. In order to do that we used data from SMARTS. We focused on GX~339$-$4 since this is the most monitored BH~LMXB in O/IR wavelengths. We divided our analysis in two parts. We first compared the rising parts of FT and full outbursts and, after that, we compared the periods of quiescence before FT and full outbursts.

\subsubsection{O/IR data during outburst. Light curves, O/IR correlations and CMDs of FT and full outbursts}

\begin{figure*}
\centering
\includegraphics[width=\textwidth]{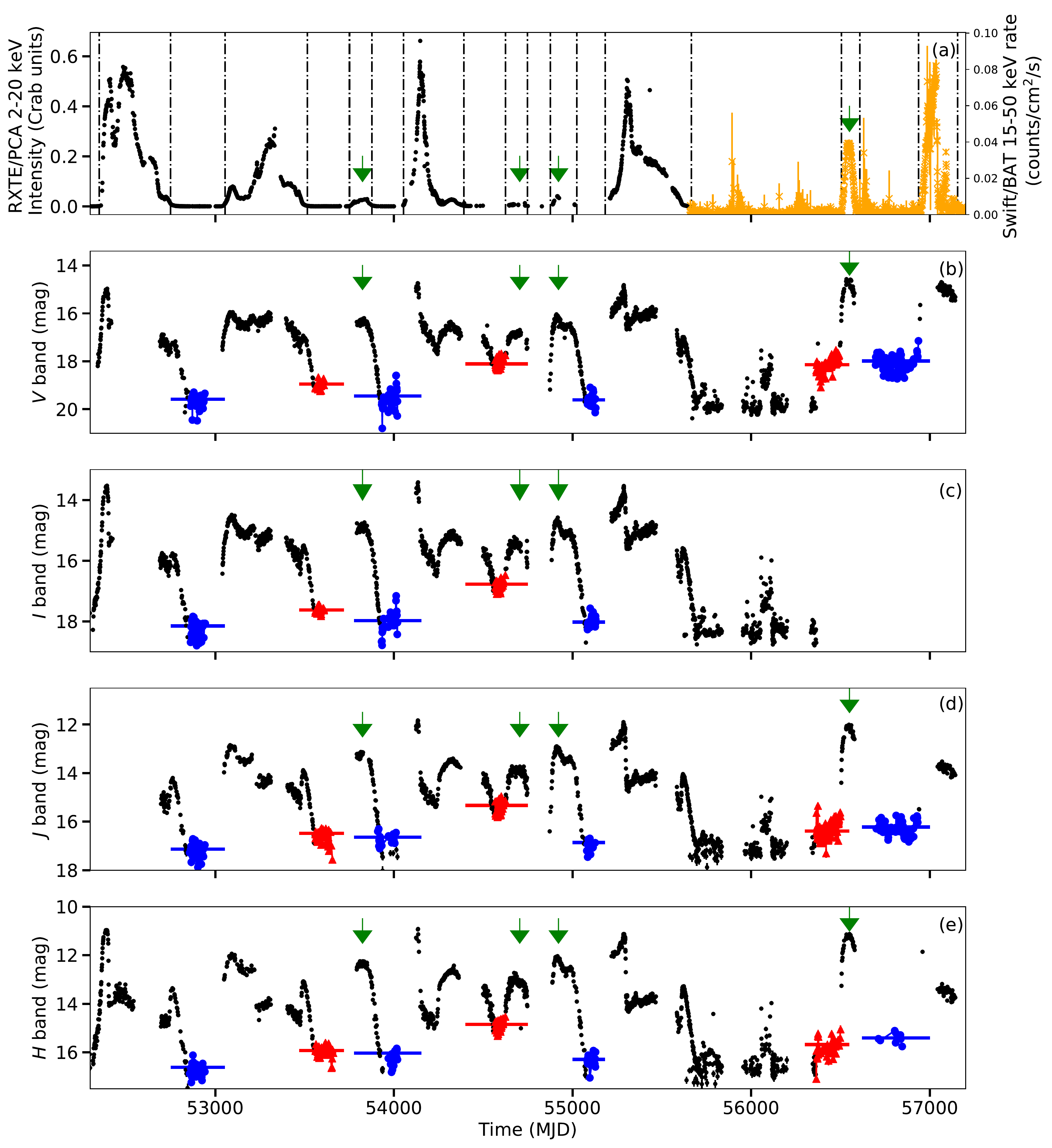}
\caption{Panel (a): 2002--2010 light curve \textit{RXTE}/PCA light curve (black symbols) and 2011--2015 \textit{Swift}/BAT light curve (orange symbols). Panel (b): SMARTS \textit{V}-band light curve. Panel (c): SMARTS \textit{I}-band light curve. Panel (d): SMARTS \textit{J}-band light curve. Panel (e): SMARTS \textit{H}-band light curve. Green arrows mark the FT outbursts. Red and blue symbols represent the intervals where the O/IR emission was at the minimum while the source was in quiescence in X-rays before FT and full outbursts, respectively. Blue and red horizontal lines represent the mean values of these intervals. Black dashed lines represent the onset and the offset of the outbursts in X-rays.}
\label{f:quiescence}
\end{figure*}

Fig. \ref{f:quiescence} shows the \textit{RXTE}/PCA and \textit{Swift}/BAT light curves (panel (a)) and  O/IR light curves (panels (b) to (e)) of GX 339$-$4 of the time period 2002--2015. Five full (2002, 2004, 2006, 2009 and 2014) and four FT outbursts (2006, 2008, 2009 and 2013; marked with green arrows) were observed during this period. 

We found that the O/IR magnitudes of the peak of the 2008 and 2009 FT outbursts were lower than the magnitudes of the peaks of the full outbursts. This result is similar to that reported in \citet{Buxton12} about the 2006 FT outburst. Moreover, the 2008 outburst is fainter than the 2006 and 2009 outbursts in X-rays, the same behaviour observed in O/IR wavelengths. We also found that the 2013 FT outburst peaked at similar magnitude as the full outbursts of GX~339$-$4.  

\begin{figure*}
    \centering
    \includegraphics[width=\textwidth]{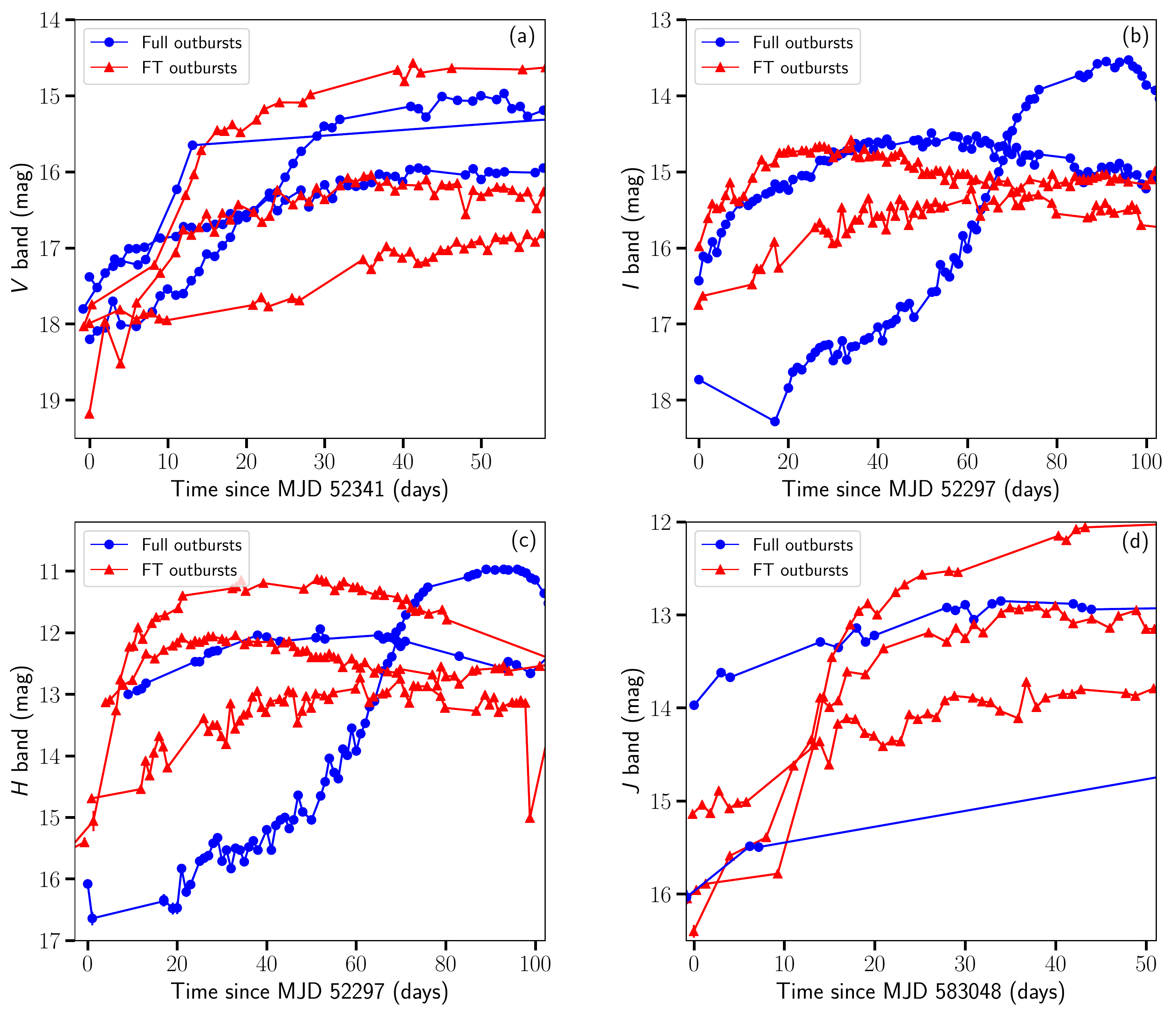}
    \caption{Panel (a): SMARTS V band light curves of the 2002, 2004, 2008, 2009, 2013 and 2014 outbursts of GX~339--4. Panel (b): SMARTS I band light curves of the 2002, 2004, 2008 and 2009. Panel (c): SMARTS J band light curves of the 2002, 2004, 2008, 2009, 2013 and 2014 outbursts. Panel (d): SMARTS V band light curves of the 2002, 2004, 2008, 2009, 2013 and 2014 outbursts. }
    \label{f:oir_rise}
\end{figure*}

We also compared the rising part of FT and full outbursts in O/IR wavelengths as we did in X-rays. Fig. \ref{f:oir_rise} shows the rise of FT and full outbursts in the O/IR light curves, as we did in Section 4.3.1 with the \textit{RXTE}/PCA light curves. We found that, at the beginning of the O/IR outbursts, it is not possible to distinguish between full and FT outburst.

\begin{figure*}
\centering
\includegraphics[width=0.98\textwidth]{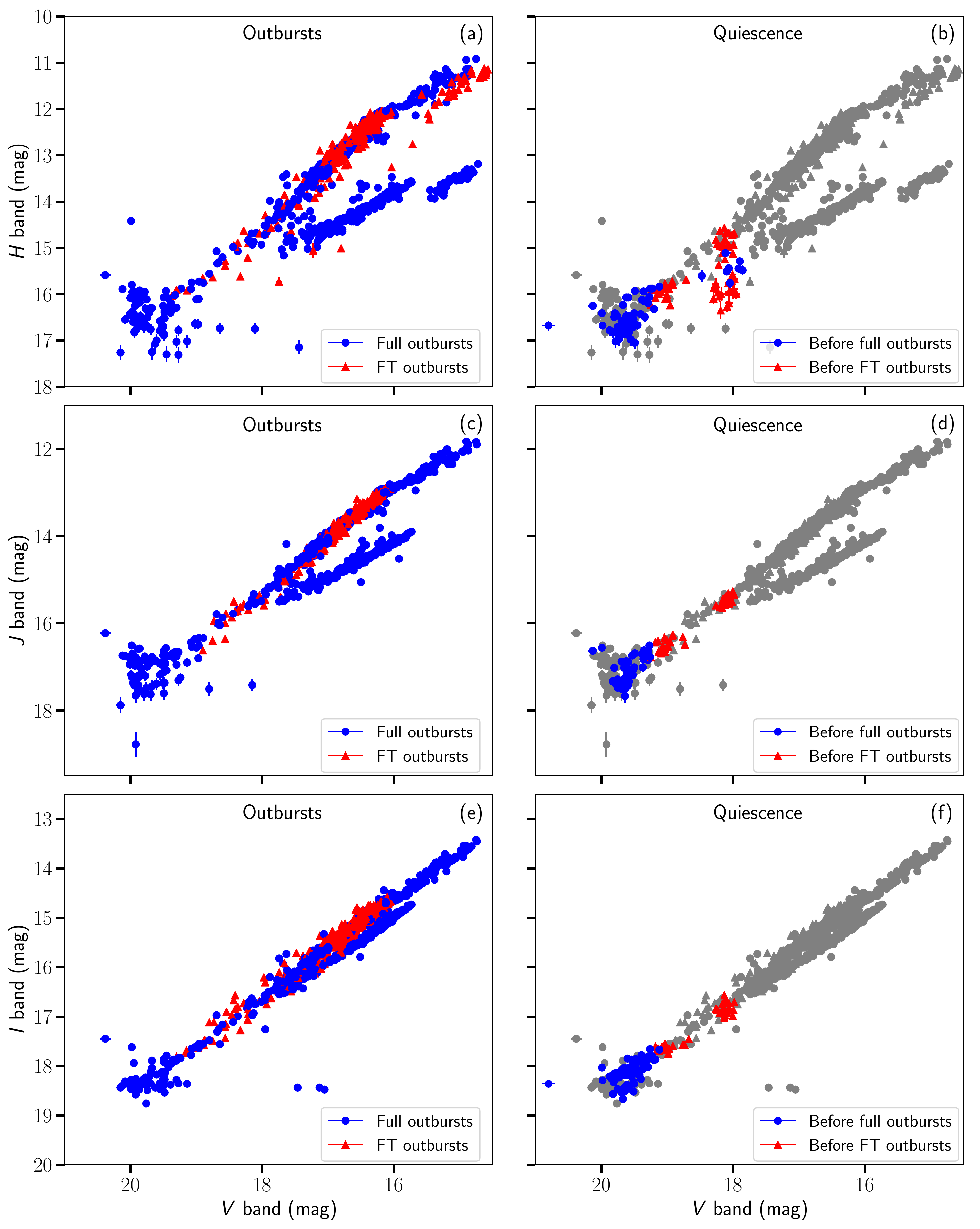}
\caption{Panels (a) and (b): Correlations between the \textit{V}-band and \textit{H}-band magnitudes corresponding to the outbursts from 2004 to 2014 and the periods of quiescence before outbursts, respectively. Panels (c) and (d): Same as panels (a) and (b) for the \textit{V}- and \textit{J}-bands. Panels (e) and (f): Same as panels (a) and (b) for the \textit{V}- and \textit{I}-bands corresponding to the outbursts from 2004 and 2009. Grey points correspond to the data of the outbursts.}
\label{fig:qcorrelation_total}
\end{figure*}

In addition, we investigated the relation between the O/IR bands. Panel (a) of Fig. \ref{fig:qcorrelation_total} shows the \textit{V}-band against \textit{H}-band magnitudes corresponding to the outbursts of the 2004--2014 time period. 
In agreement with \citet{Buxton12}, we observed two different branches in the plot. The top branch is described by a broken power-law (H $\propto$ $V^{\alpha}$) with the break located at $V\sim16$ mag. The slope of the power-law is larger than 1 from the faintest magnitudes to the break, and after the break the slope is less than 1. The bottom branch is well described by a single power-law with a slope less than one. \citet{Buxton12} associated the top and the bottom branches with the LHS and the HSS, respectively. At the beginning of an outburst, GX~339--4 starts at the bottom left part of the diagram. During the LHS, the source follows the top branch until the top right part of the diagram. In the hard-to-soft state transition, the source leaves the top branch part and moves to the bottom branch, when the source reaches the HSS. In the bottom branch, the source evolves backwards to the bottom left side of the diagram. During the soft-to-hard transition the source moves again to the top branch until the end of the outburst. Full outbursts of GX~339--4 follow this track. The FT outbursts of 2006, 2008 and 2009 fall on the top branch, as GX~339--4 was only seen in the LHS. We also found that the outburst of 2013, which is classified as an FT outburst, evolve from the top branch to the bottom branch. 

Plots corresponding to the correlation between the \textit{V}-band and the \textit{J}- and \textit{I}-bands of the outbursts of GX~339$-$4 are on panels (c) and (e) in Fig. \ref{fig:qcorrelation_total}. In both cases we see the two branches. When plotting the \textit{V}- vs the \textit{I}-band magnitudes, we found that both branches are very close to each other and in the \textit{V}- vs the \textit{J}-band the branches are more separated. In conclusion, we see that the two branches in Fig. \ref{fig:qcorrelation_total} become more separated as we compare the V-band to a band with longer wavelengths. Regarding the behaviour of full and FT outbursts, we also see that both types of outbursts fall in the top branch in the three cases.

\begin{figure*}
\centering
\includegraphics[width=\textwidth]{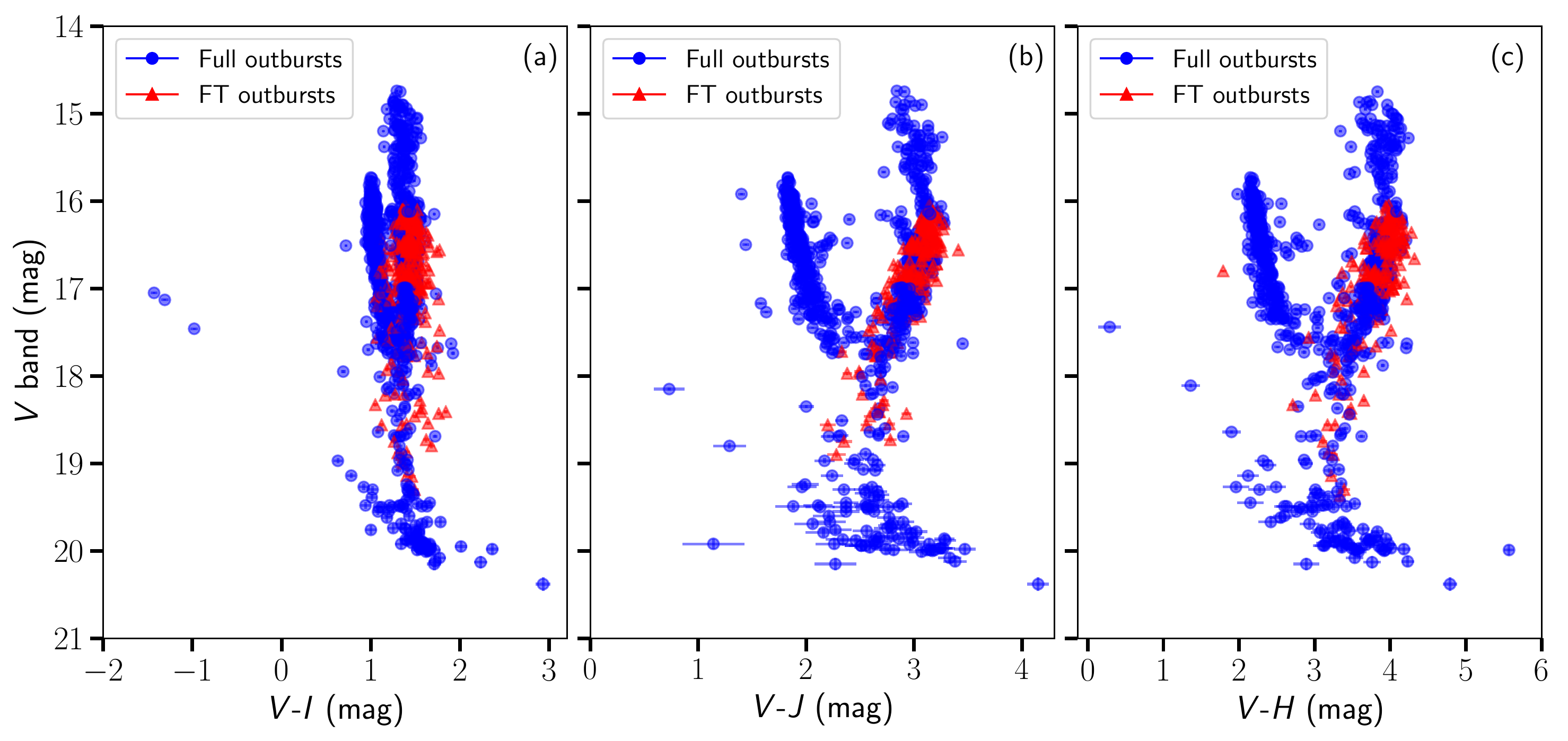}
\caption{Representative colour-magnitude diagrams of GX~339$-$4. Blue symbols correspond to the CMDs of full outbursts and red symbols corresponds to the CMDs of FT outbursts.}
\label{fig:cmd}
\end{figure*}

Following the procedure performed with the HIDs with X-ray data, we compared the track followed by the rise of the full and FT outbursts of GX~339$-$4 in the CMD. Fig.\ref{fig:cmd} shows some representative CMD of the source. We found the same result as with the X-ray HIDs: the rising part of the full and FT outbursts lie on the same region of the CMDs for all the colours. This result is similar to that shown in \citet{Kosenkov20}, in which is reported that both types of outbursts follow the same track along the CMD. In addition to the data shown in \citet{Kosenkov20}, we show the CMDs of the 2013 (FT) and 2014 (full) outbursts of GX~339$-$4.

\subsubsection{O/IR data during quiescence. Light curves, O/IR correlations and CMDs}

GX~339$-$4 was also monitored in O/IR while the system was in quiescence in X-rays. In this paper we define quiescence as the moments where no X-ray emission is detected from the source. This allowed us to study the O/IR light curves of GX~339$-$4 in these periods. When GX~339$-$4 reached the quiescence in X-rays after an outburst, the O/IR magnitudes were still decaying and, at some point, they varied within an interval of $\sim$1 magnitude before starting to rise again. We also found that GX~339$-$4 did not have the same O/IR magnitudes while the source was in quiescence in X-rays before FT and full outbursts. In order to quantify this difference in brightness, we obtained the weighted average of the magnitude of the intervals where the O/IR magnitude varies within an interval of $\sim$1 magnitude that we previously mentioned. Between the end of the 2009 full outburst and the beginning of the 2013 FT outburst, the O/IR light curves show an optical outburst that was not detected in X-rays \citep[][]{Lewis12, Maccarone12}. Just before the 2013 outburst, on MJD 56364, the \textit{V} magnitude increased almost $\sim$2 magnitudes in $\sim $2 day. Because of this behaviour, we decided to calculate the average of the O/IR magnitudes prior to the 2013 FT outburst after MJD 56364. These intervals are plotted with blue circles and red triangles in Fig. \ref{f:quiescence}.
The average values obtained for these intervals are given in Table. \ref{Tab:mean_prior}. We found that GX~339$-$4 was generally brighter in O/IR before undergoing an FT outburst than before a full outburst. In particular, before the 2008 outburst, GX~339$-$4 was always more than 1 magnitude brighter than before all the full outbursts. This can also be explained if the source did not go to quiescence before this outburst. Before the 2006 FT outburst, the source was $\sim$0.5 magnitudes brighter than full outbursts in the \textit{V}- and \textit{I}-bands, but this difference was smaller than before the 2006b and 2009b full outbursts. The only exception occurred before the full outburst of 2014 when the source showed a brightness similar to that reached before FT outbursts in the O/IR bands. 

\begin{table*}
\resizebox{\textwidth}{!}{
\begin{tabular}{cccccc}
\hline 
\textbf{Type of outburst} & \textbf{Year of the outburst} & \textbf{Mean \textit{V} magnitude} & \textbf{Mean \textit{I} magnitude} & \textbf{Mean \textit{J} magnitude} & \textbf{Mean \textit{H} magnitude} \\
 &  & \textbf{(mag)} & \textbf{(mag)} & \textbf{(mag)} & \textbf{(mag)} \\
\hline
FT outbursts & 2006 & $18.996\pm0.005$ & $17.626\pm0.006$ & $16.57\pm0.01$ & $15.94\pm0.01$ \\
 & 2008 & $18.043\pm0.003$ & $16.742\pm0.002$ & $15.399\pm0.005$ & $14.857\pm0.005$ \\
 & 2013 & $18.123\pm0.004$ & - & $16.146\pm0.007$ & $15.66\pm0.02$ \\
\hline
Full outbursts & 2004 & $19.616\pm0.006$ & $17.626\pm0.006$ & $17.15\pm0.02$ & $16.66\pm0.02$ \\
 & 2006b & $19.273\pm0.006$ & $17.926\pm0.005$ & $16.72\pm0.02$ & $16.14\pm0.02$ \\
 & 2009b & $19.543\pm0.004$ & $17.998\pm0.006$ & $16.97\pm0.03$ & $16.24\pm0.02$ \\
 & 2014 & $17.978\pm0.005$ & - & $16.182\pm0.009$ & $15.40\pm0.03$ \\
\hline
\end{tabular}}
\caption{Mean magnitude in the \textit{V}, \textit{I}, \textit{J} and \textit{H} bands of the regions prior to outbursts.}
\label{Tab:mean_prior}
\end{table*}

We show the relation between the \textit{V}- and \textit{H}- bands during quiescence before FT and full outbursts in panel (b) of Fig. \ref{fig:qcorrelation_total}. We found that the periods of quiescence before full and FT outbursts that occurred before 2011 fall on the upper branch. The periods of quiescence prior to the outbursts of 2013 and 2014, on the other hand, fall on the lower branch. The same happens in the \textit{V}- vs. \textit{I}- bands and \textit{V}- vs. \textit{J}-bands plots (panels (d) and (f)) of Fig. \ref{fig:qcorrelation_total}, respectively.)

\begin{figure*}
\centering
\includegraphics[width=\textwidth]{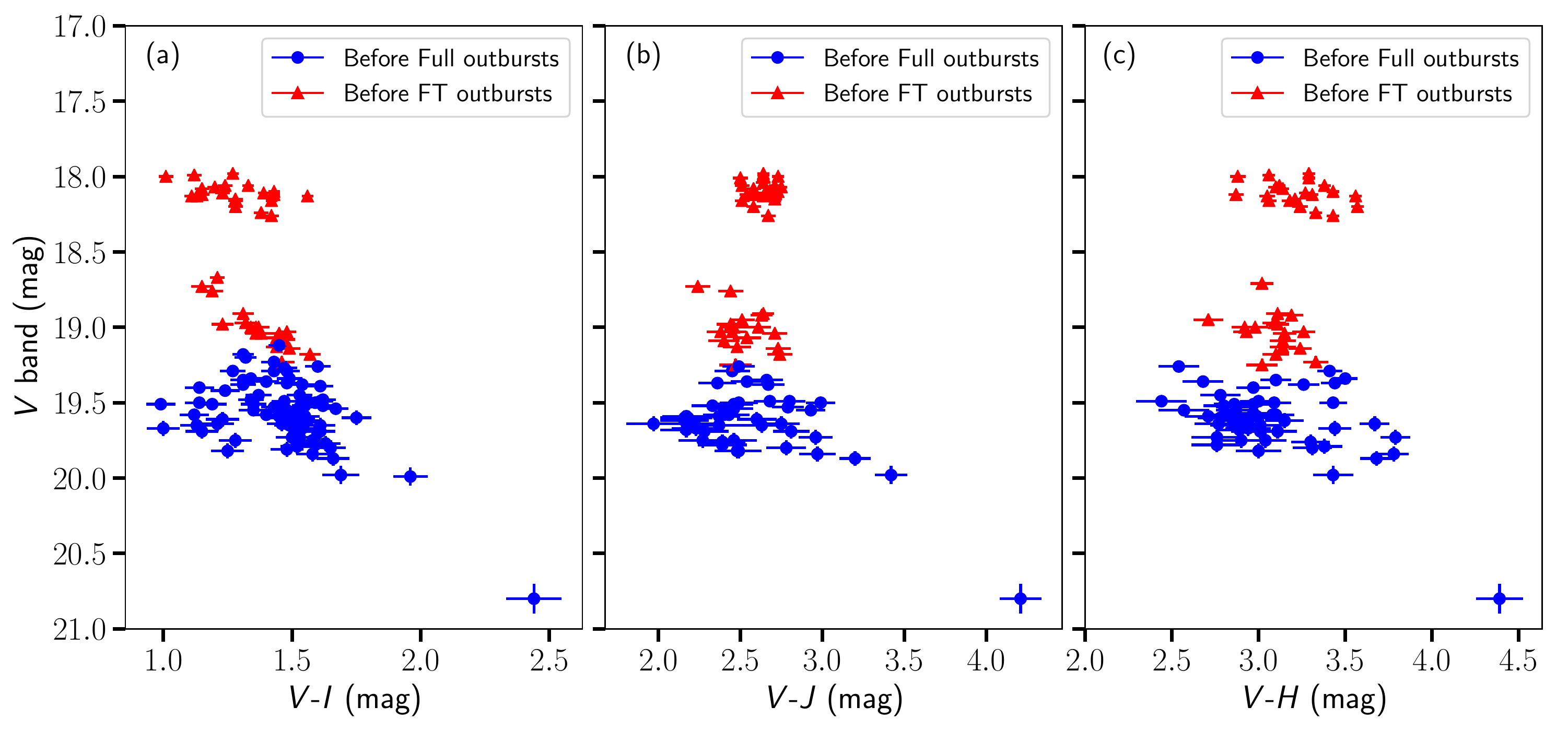}
\caption{Representative colour-magnitude diagrams of GX~339$-$4 during quiescence. Blue symbols correspond to the CMDs of quiescence intervals prior to full outbursts and red symbols corresponds to the CMDs of quiescence intervals prior to FT outbursts.}
\label{fig:cmd_q}
\end{figure*}

Fig.\ref{fig:cmd_q} shows the CMDs of the O/IR intervals corresponding to quiescence periods prior to FT outbursts (red triangles) and full outbursts (blue circles). We found that both intervals show the same values in terms of colours. The significant difference between them is their \textit{V} magnitude (see also Section 4.4.2), as can be appreciated in Fig. \ref{f:quiescence}.

We also estimated the temperature of the accretion disc during quiescence. \citet{Kosenkov20} plots the line corresponding to a blackbody emission in Figs. 4--11 (solid orange line). The intervals of quiescence before full outbursts (with $V\sim19.25-20$ mag) correspond to a blackbody temperature of 8--10 kK. The temperatures of the quiescence intervals before the 2006, 2009 and 2013 outbursts range from $\sim$10 kK to $\sim$12.5 kK. The quiescence before the 2008 FT outburst is more brighter and its disc temperature is $\sim$15 kK, approximately.

\section{Discussion}

We present the first systematic search for, and study of, the X-ray and O/IR data of FT and full outbursts. For all sources in our sample, we collected all the available data from \textit{RXTE}/PCA, MAXI and \textit{Swift}/BAT to study the outbursts in X-rays. We found that 33\% of the outbursts of our sample are FT outbursts. We found that only three sources had enough data during the rise to compare both FT and full outbursts: GX~339$-$4, H~1743$-$322 and GRS~1739$-$278. When we compared the X-ray light curves of both types of outbursts, we found that during the first days of the outbursts of GX~339$-$4 and H~1743$-$322, the X-ray light curves of FT and full outbursts follow similar tracks. In contrast, the X-ray \textit{Swift}/BAT and MAXI light curves of the 2014 (full) and the 2016 (FT) outbursts of GRS~1739$-$278 show that the 2014 outburst was always brighter than the 2016 FT outburst. Regarding the evolution along the HID and the PCC, both types of outbursts of GX~339$-$4 follow the same track. For GX~339$-$4, the source in our sample showing a large number of full and FT outbursts, we were also able to combine the X-ray data with O/IR observations with the SMARTS telescope (the O/IR data for other sources is too sparse to be able to do detailed analysis). FT and full outbursts of GX~339$-$4 in the O/IR light curves and CMDs are similar. The only significant difference is that GX~339$-$4 was brighter in O/IR before FT outbursts than before full outbursts, while the source was in quiescence in X-rays.

\subsection{Fraction of FT outbursts}

We found that the 36\% of the black hole X-ray binaries included in this study showed FT outbursts, and that 33\% of the outbursts included in our sample were identified as FT outbursts. The percentage of FT outbursts that we found is similar to that obtained by \cite{Tetarenko16}. The reason for the small difference between the two results is that we did not include some of the sources of \cite{Tetarenko16} and we updated the list of BH systems showing full or FT outbursts until 2021. The fact that three or four out of ten outbursts is an FT one tells us that FT outbursts are common events in BH LMXBs, as \citet{Tetarenko16} pointed out.

\subsection{FT and full outbursts in X-rays}

Comparing the rising part of the outbursts of GX~339$-$4 and H~1743$-$322 we found that the X-ray light curves of FT and full outbursts overlap during the first 20--50 days, depending on the source and the instrument. After that, full outbursts continue brightening while FT outbursts faint. 
We also found that both types of outburst of GX~339$-4$ follow the same track along the HID during the rise of the outbursts, making impossible to predict their evolution during the first days. 
In addition, the behaviour of both type of outbursts in the PCC diagram is also similar.  
These three results suggest that the physical processes involved at the beginning of both types of outbursts are the same. It has been proposed that a LMXB makes a state transition if the source reaches a certain mass accretion rate \citep[e.g., ][]{Esin97}. Based on this hypothesis, the only difference between FT and full outbursts is that FT outbursts do not reach the mass accretion rate level needed to trigger the hard-to-soft state transition. \citet{Tetarenko16} found that the luminosity at which the transition to the soft state occurs in the 2004 full outburst of GX~339$-$4 is $\sim0.11\;L_{\rm edd}$, and that the luminosity of FT outbursts is always equal or lower than this value. The question is why FT outbursts reach lower peak X-ray luminosities than full outbursts. If we assume that the accretion rate from the secondary star is constant, one possible explanation is that a longer quiescence time between outbursts could lead to a brighter outburst due to a larger accumulation of matter in the accretion disc. However, we found that there is no correlation between the quiescence time between outbursts and their X-ray peak intensity. This is similar to the result of \citet{Campana13} for the NS LMXB Aql~X$-$1. In addition, the fact that some sources have only undergone FT outbursts (see \ref{tab:table_sources}) suggests that the nature of the outburst, being either an FT or a full outburst, does not depend on the quiescence time between outbursts. Another possible explanation for the lower peak flux in FT outbursts is that the outer disc is not irradiated enough to be involved in the outburst \citep[e.g., different outbursts of Aql X--1][]{Campana13, Gungor14}. \citet{Yu04} and \citet{Yu07} also suggested that the total mass of the accretion disc of BH LMXBs influences the peak flux of the LHS of an outburst. They found a positive correlation between the peak flux of the LHS before the hard-to-soft transition and the peak flux of the HSS in full outbursts of XTE~J1550$-$564, Aql~X$-$1 and 4U~1705$-$44. They speculated that the hard-to-soft transition occurs when the inner disc dominates the emission over the contribution of the outer disc. This is consistent with the mass of the accretion disc driving the state transitions, since at the beginning of an outburst the mass of the disc is concentrated far from the compact object. Based on these ideas, the main difference between full and FT outbursts would be the mass distribution of the accretion disc, being the outer parts of the accretion disc more massive than the inner parts during a FT outburst. This, however, does not allow us to predict during the first days of an outburst whether it will be an FT or a full outburst.

\subsection{FT and full outbursts in O/IR}

Comparing the rising part of the O/IR light curves of the outbursts of GX~339$-$4 we found the same result as in X-rays: we cannot distinguish between full and FT outbursts during the first days. Similarly, we found that both types of outbursts lie in the same regions of the O/IR CMDs. \citet{Kosenkov20} studied the 2002--2010 outbursts of GX~339$-$4 and found that both types of outbursts follow the same tracks in O/IR CMDs. In addition, we studied the CMDs of the outbursts of GX~339$-$4 in 2013 (FT) and 2014 (full), reaching the same conclusion: both outbursts lie in the same region of the CMD. Similarly to \citet{Buxton12}, we also found that FT outbursts of GX~339$-4$ in 2006, 2008 and 2009 fall on the top branch of the correlation between O/IR bands, corresponding to the LHS of BH LMXBs. It was also expected, since we know that the FT outbursts of GX~339$-$4 never left the LHS. The only exception is the FT outburst of 2013. Some observations of this outburst fall on the bottom branch, corresponding to the HSS. One possible explanation for this is that the source started a hard-to-soft transition during this outburst without reaching the HSS. Another possible explanation is that the source started an O/IR transition but not an X-ray transition \citep[][]{Kalemci13}. 

These three results suggest that, as in X-rays, the beginning of both FT and full outbursts are driven by the same physical mechanisms. However, we found that GX~339$-$4 was brighter before FT outbursts than before full outbursts in O/IR wavelengths. The only exception to this is the full outburst of 2014, in which the O/IR magnitudes are similar to the intervals before FT outbursts.

The O/IR emission is thought to be produced at the outer parts of the accretion disc irradiated by the X-ray emission from the central object. Subsequent studies suggested that the O/IR emission comes from two components: the accretion disc and a non-thermal component that can be a hot flow, a corona, or a jet, or a combination of all of them \citep[e.g., ][]{Fabian82, Motch83, Kanbach01, Gandhi08, Casella10, Veledina11}. We found that the main difference between FT and full outbursts is that, in quiescence, GX~339$-$4 is brighter before FT outbursts than before full outbursts. We can consider that the O/IR emission comes from the corona and/or the outer parts of the accretion disc. If we assume that the O/IR emission is mostly originated at the outer parts of the disc, this suggests that the outer part of the disc is more massive before FT outbursts than before full outbursts. This is in agreement with the conclusion achieved from our study of X-ray data, which is that the outer part of the disc is more massive in FT outbursts than in full outbursts. On the other hand, the emission can be mostly originated in the corona, independently of the mass of the accretion disc. 

The O/IR emission could also come from the jet. The detection of single frequency radio detections by \citet{Gallo06} and \citet{Gallo14} during quiescence for the systems A0620$-$00 and XTE~J1118$+$480 suggested the presence of a radio jet in quiescence. Later, \citet{Dincer18} found an inverted radio spectrum for A0620$-$00, reinforcing this idea. Nevertheless, it is difficult to determine the origin of the O/IR emission. Although emission from the accretion disc has been detected in quiescence \citep[e.g., ][]{McClintock95, Froning11}, some studies suggest that the emission comes only from the jet \citep[][]{Dincer18}. \citet{Poutanen14} showed that at low luminosities non-thermal emission from a hot accretion flow could explain the O/IR emission of XTE~J1550$-$564. Similarly, \citet{Kosenkov20} concluded that the O/IR emission of GX~339$-$4 could be originated by two non-thermal components: a jet and the hot flow.  

In addition to that, \citet{Dincer18} found that A0620$-$00 shows a significant increase in its X-ray flux during quiescence. This has also been observed in other systems, as  GS~1354$-$64 \citep[][]{Koljonen16} and Muscae 1991 \citep[][]{Wu16}. This can be interpreted as the changes in the outer disc and the accretion flow due to the build up of the disc prior to the onset of a new outburst \citep[][]{Dubus01, Dincer18}. This increase is not clearly seen in GX~339$-$4 in O/IR wavelengths. This could suggest that the accretion rate of GX~339$-$4 while the source is in quiescence is constant. However, following this idea the mass accretion rate would be higher in quiescence before FT outbursts than before full outbursts. This conclusion is also supported by the estimations of the temperature of the disc in quiescence. Under the assumption that the radius of the disc is constant in quiescence, the mass accretion rate is proportional to $T^{4}$. Since the temperature of the disc is higher before the onset of FT outbursts, the mass accretion rate is higher, too. 

The evolution of the X-ray power-law index of A0620$-$00 during quiescence was also studied by \citet{Dincer18}. The authors found that the photon index does not change with the increasing of the flux. This result suggested that the emission was produced in the same region. However, they suggested that the size of the emission region could be changing. We can link this conclusion with our results. The difference in O/IR magnitude between quiescence prior to FT and full outbursts could be due to a difference in the size of the emitting region. Further studies of O/IR observations while the source is in quiescence before an outburst will show whether this interpretation is correct.

\subsection{Inside-out and outside-in outbursts}

In the standard disc instability model (DIM) the outbursts are triggered by an instability in the accretion disc \citep[for a review, see][]{Lasota01}. The instability propagates throughout the disc via two heating fronts: one of them moving inwards and the other moving outwards through the accretion disc \citep[e.g., ][]{Smak84c, Menou99}. Depending on the radius at which the outburst is triggered, two types of outburst are distinguished: outside-in and inside-out outbursts. In outside-in outbursts, the ignition occurs close to the outer edge of the accretion disc, and the outwards heating front reaches the outer part of the disc faster than the inwards heating front reaches the inner parts. This type of outburst produces strongly asymmetric light curves \citep[][]{Smak84c}. In inside-out outbursts, the ignition radius is small and the inwards heating front reaches the inner disc radius faster than the outwards heating front reaches the outer part of the disc. This type of outburst produces approximately symmetric light curves \citep[][]{Smak84c}. It can happen that the outward heating front fails to propagate to the outer part of the disc because it finds regions of higher densities. Because of that, a cooling front would be developed and it will cool the material, shooting down the outbursts \citep{Hellier01}. Depending on whether the instability reaches the outer parts of the disc, two subtypes of inside-out outbursts can be distinguished: subtype Ba and subtype Bb \citep[][]{Smak84c}. In subtype Ba outbursts, the instability reaches the outer edge of the disc, while in subtype Bb outbursts the instability does not reach the outer part of the disc. Because of that, the light curve amplitude and duration of subtype Bb outbursts are shorter. The O/IR light curves of the FT outbursts of GX~339$-$4 presented in Fig. \ref{f:quiescence} are more similar to that corresponding to the inside-out outbursts \citep[see Fig. 3 in][]{Smak84c}. The light curves corresponding to full outbursts, on the other hand, are more asymmetric, what could mean that the instability that triggers the outburst occurs in the outer part of the accretion disc. If this interpretation is correct, this would mean that both types of outbursts can appear in the same source. If the binary parameters are the same, outside-in outbursts occur at high mass accretion rates, while inside-out outbursts occur at low accretion rates. The fact that the two types are observed in the same source suggests that the mass accretion rate changes from one outburst to another. 

\subsection{FT outbursts and reflares}

Reflares, also known as rebrightenings, rebursts and mini-outbursts, \citep[see e.g., ][]{Callanan95, Chen97, Jonker12} are renewed episodes of increased luminosity experienced by some X-ray binaries during the decay of the main outburst, and before returning to quiescence.
FT outbursts might sometimes be confused with reflares because both events reach X-ray peak intensities lower than that of main full outbursts. 
However, they are essentially different. FT outbursts are isolated events similarly to full outbursts. 

As discussed in the introduction, FT outbursts never make the transition to the HSS \citep[e.g.,][for XTE~J1650--500 and MAXI~J1659--152, respectively]{Tomsick04, Homan13}. Instead, during reflares, systems can both stay in the LHS, or go through the different spectral states \citep[e.g., ][]{Yan17, Cuneo20, Zhang20}. 
Recent studies of the reflares of MAXI~J1348$-$630 show that they follow a slightly softer track in the HID than the rising part of the main outburst \citep[see Fig. 2 of][]{Zhang20}. While this is a single case, it is a significant difference with the typical phenomenology observed in FT outbursts.

It has been proposed that reflares and main outbursts are driven by the same physical processes \citep[e.g., ][]{Yan17, Cuneo20} and that they depend on general accretion physics rather than to the nature of the compact object \citep[e.g., ][]{Patruno16}. The similarities observed in the spectral behaviour of FT and full outbursts, reflares and main outbursts, imply that all these events might be driven by the same mechanisms. However, although it has been widely discussed in the literature \citep[see, e.g. ][ and references therein]{Cuneo20}, it is still unknown what powers the renewed activity, either during the decay of an outburst or from quiescence, and which is the mechanism driving and regulating the state transitions.

\section{Summary and conclusions}

We present the first systematic search for, and study of, the X-ray and O/IR data of FT and full outbursts. We searched in the literature and X-ray archives for data of full and FT outbursts and found that only three sources show enough data to compare the rising part of the outburst: GX~339$-$4, H~1743$-$322 and GRS~1739$-$278. We also looked for O/IR data in the literature and found that the only source with both full and FT outbursts with enough observations during the rise is GX~339$-$4. We then focused our study on the analysis of the sources with enough observations during the rise of the outbursts: three sources in X-rays (GX~339$-$4, H~1743$-$322 and GRS~1739$-$278) and one source in O/IR wavelengths (GX~339$-$4). Our results shed light on the behaviour of full and FT outbursts. The summary of our conclusions is the following:

\begin{enumerate}
    \item During the first days of the outbursts of the sources studied, X-ray light curves of FT and full outbursts follow similar tracks. In addition, the X-rays HID and PCC tracks are the same. Since the quiescence time before FT and full outbursts is similar, two possible explanations are possible. The first possibility is that the outer part of the accretion disc is more massive than the inner part in a FT outburst, preventing a hard-to-soft state transition. The second option is that the outer part of the disc is not irradiated enough to have a significant effect in the evolution of the outburst.
    \item We found that the O/IR light curves of FT and full outbursts of GX~339$-$4 cannot be distinguished during the rising part of them. However, we found that this system is brighter before the onset of an FT outburst than before the onset of a full outburst. This could be explained by the fact that the mass accretion rate is higher before the onset of FT outbursts. However, the difference between full and FT outburst could also be the size of the O/IR emission region. 
    \item According to the shape of the O/IR light curves of full and FT outbursts of GX~339$-$4, the latter would be inside-out outbursts: i.e., the instability is triggered at a radius close to the inner edge of the accretion disc and it cannot reach the outer part of the disc. This could explain the fact that FT outbursts are shorter in time and less bright than full outbursts. 
\end{enumerate}

In conclusion, it is not possible to predict whether an outburst will be a full or an FT during the first days of the outbursts only using X-ray data. The key to predict the nature of the outbursts might be the O/IR magnitude of the source while the source is in quiescence, since we have found that GX~339$-$4 is brighter in O/IR wavelengths before the onset of an FT than before the onset of a full outburst. Although it remains unknown the detail of the mechanisms triggering full and FT outbursts, it is evident that they are different manifestations of one unique accretion-related process. Further analysis of the different types of outbursts of X-ray binaries at O/IR wavelengths during quiescence will help to determine whether the type of outburst depends on the accreted mass, the disc geometry, or some other property that still needs to be considered.

\section*{Acknowledgements}

This research has made use of \textit{RXTE} data provided by the High Energy Astrophysics Science Archive Research Center (HEASARC), a service of the Astrophysics Science Division at NASA/GSFC and the High Energy Astrophysics Division of the Smithsonian Astrophysical Observatory. This research has made use of \textit{MAXI} light curves provided by RIKEN, JAXA, and the \textit{MAXI} team and has also made use of \textit{Swift}/BAT transient monitor results provided by the \textit{Swift}/BAT team. This paper has also made use of SMARTS optical and infrared data. KA acknowledges support from a UGC-UKIERI Phase 3 Thematic Partnership (UGC-UKIERI-2017-18-006; PI: P. Gandhi). D.A. acknowledges support from the Royal Society. MM and FG acknowledge support from the research programme Athena with project number 184.034.002, which is (partly) financed by the Dutch Research Council (NWO). VAC acknowledges support from the Spanish \textit{Ministerio de Ciencia e Innovaci\'on} under grant AYA2017-83216-P. AV acknowledges support from the Academy of Finland grant 309308. FMV acknowledges support from STFC under grant ST/R000638/1. We acknowledge Dr. Phil Uttley for providing the data for the power colour-colour diagrams. We acknowledge Prof. Charles Baylin and Dr. Tolga Dincer for providing SMARTS data. We also acknowledge Dr. Liang Zhang for the discussion about the reflares of MAXI~J1348$-$630. 

\section*{Data availability}

The \textit{RXTE} data underlying this article are publicly available in the High Energy Astrophysics Science Archive Research Center (HEASARC) at \url{https://heasarc.gsfc.nasa.gov/db-perl/W3Browse/w3browse.pl}. The \textit{MAXI} data underlying this article are publicly available in \url{http://maxi.riken.jp/top/index.html}. The \textit{Swift}/BAT data underlying this article are publicly available in \url{https://swift.gsfc.nasa.gov/results/transients/}. The SMARTS data underlying this article until 2011 was published on \citet{Buxton12} and the data covering from 2011 was provided privately by the SMARTS team. 





\bibliographystyle{mnras}
\bibliography{bib_all}



\appendix

\section{Table with all the outbursts of the different sources included in this paper}

\onecolumn

\begin{ThreePartTable}
\footnotesize\setlength{\tabcolsep}{2.6pt}
    \sisetup{table-format=-1.4}
\begin{longtable}[c]{cccc}
\caption{Spectral fitting parameters. All fluxes are unabsorbed, in units of erg cm$^{-2}$ s$^{-1}$, and in the 0.5--10 keV band.}\\
\hline
Source name & Type of outbursts & Outbursts & References \\
\hline
\endfirsthead
\multicolumn{4}{c}%
{\tablename\ \thetable\ -- \textit{Continued from previous page}} \\
\hline
Source name & Type of outbursts & Outbursts & References \\
\hline
\endhead
\hline \multicolumn{4}{r}{\textit{}} \\
\endfoot
\hline
\endlastfoot
\hline
XTE~J1550$-$564 & Full & 1998, 2000 & 1-6; 7-12 \\
 & FT & 2001, 2002, 2003 & 13, 14; 15, 16; 17-19   \\
\hline
GX~339$-$4 & Full & 1973, 1981, 1988, 1991, 1992, 1993, 1996, 2002, 2004, 2006b & 20, 21; 22; 23, 21; 24-26, 21;\\ 
 & & 2009b, 2014, 2018 & 24-26; 25, 26; 21, 26-28;\\
 & & & 29, 30; 30-31; 30, 32, 33; \\
 & & & 34; 35-39; 550-555 \\
 & FT & 1994, 1995, 1995, 2006, 2008, 2009, 2013, 2017 & 21, 26; 21, 26; 21, 26; 32;  \\
 & & & 32; 32; 40-42, 43-47, 549, 550 \\
\hline
H~1743$-$322 & Full & 2003, 2004, 2005, 2007, 2009, 2010, 2010b, 2013, 2016 & 29, 48-51; 52; 52; 73, 556; 53-55 \\
 & & &   56-58; 59-61; 59, 61; 62, 63 \\
 & FT & 2008, 2012, 2014, 2015, 2016b, 2017, 2018 & 56, 64; 71, 72, 556; 65-67; 67-69; \\
 & & & 74, 556; 75, 556; 556 \\
 & Indeterminate & 2009b, 2011 & 59; 70 \\
\hline
XTE~1118$+$480 & FT & 2000, 2004 & 76-83; 79, 84  \\
\hline
Swift~J1745.8$-$262411 & FT & 2012 & 85-90 \\
\hline
GRS~1716$-$249 & FT & 1993, 1995, 2016 & 91-93; 91, 94; 235-243 \\
\hline
GS~1354$-$64 & Full & 1967, 1987 & 93, 95-97; 93, 96, 97 \\
 & FT & 1972, 1997, 2015 & 93, 96-98; 93, 99, 100; 101-106 \\
\hline
Swift~J1357.2$-$0933 & FT & 2011, 2017, 2019 & 107, 422-433; 434-441; 442-449 \\
\hline
GRS~1737$-$31 & FT & 1997 & 93, 108-111 \\
\hline
MAXI~J1836$-$194 & FT & 2011 & 112-114 \\
\hline
SAX~J1711.6$-$3808 & FT & 2001 & 115, 116 \\
\hline
IGR~J17285$-$2922 & Full & 2004 & 117, 118 \\
 & FT & 2010 & 119 \\
 & Indeterminate & 2019 & 525-527 \\
\hline
IGR~J17497$-$2821 & FT & 2006 & 120-122 \\
\hline
GRO~J0422$+$32 & FT & 1992 & 123-125 \\
\hline
XTE~J1856$+$053 & Full & 1996, 1996b, 2007, 2015 & 126, 128, 129;  126, 128-131;  \\
 & & & 126, 131 \\
 & FT & 2015 & 127, 132, 133 \\ 
\hline
IGR~J17454$-$2919 & FT & 2014 & 134, 135  \\
\hline
XTE~J1748$-$288 & Full & 1998 & 136, 137 \\
\hline
4U~1543$-$475 & Full & 1971, 1983, 2002 & 93, 138; 93, 139; 93, 140 \\
 & Indeterminate & 1992 & 93, 141 \\
\hline
GRO~1655$-$40 & Full & 1994, 1996, 2005 & 142; 143-145; 146-149 \\
\hline
MAXI~J1659$-$152 & Full & 2011 & 150-152 \\
\hline
XTE~J1752$-$223 & Full & 2009 & 153-155 \\
\hline
XTE~J1859$+$226 & Full & 1999 & 156-159 \\
\hline
1A~0620$-$00 & Indeterminate & 1917 & 160 \\
 & Full & 1975 & 161, 162 \\
\hline
GS~1124$-$684 & Full & 1992 & 163, 164 \\
\hline
XTE~J1908$+$094 & Full & 2002, 2003, 2013 & 165-167; 166; 168-172 \\
\hline
XTE~J1817$-$330 & Full & 2006 & 173-175 \\
\hline
GRS~1009$-$45 & Full & 1993 & 176 \\
\hline
H~1705$-$250 & Full & 1975 & 177-179 \\
\hline
XTE~J1818$-$245 & Full & 2005 & 182-184 \\
\hline
MAXI~J1305$-$704 & Full & 2012 & 185-187 \\
\hline
MAXI~J1543$-$564 & Full & 2012 & 188, 189\\
\hline
SLX~1746$-$331 & Full & 2003, 2007, 2010 & 190-192; 193, 194; 195\\
 & Indeterminate & 1985, 1990 & 196, 197 \\
\hline
Swift~J1539.2$-$6227 & Full & 2008 & 198, 199\\
\hline
XTE~J1650$-$500 & Full & 2001 & 200-205\\
\hline
XTE~J1652$-$453 & Full & 2009 & 206-209\\
\hline
XTE~J1720$-$318 & Full & 2003 & 210\\
\hline
GRS~1739$-$278 & Full & 1996, 2014 & 211-213; 214-216 \\
 & FT & 2016 & 251-254 \\
\hline
GRS~1730$-$312 & Full & 1994 & 211-213; 217 \\
\hline
KS~1732$-$312 & Full & 1988 & 211-213; 218, 219 \\
\hline
SWIFT~J1842.5$-$1124 & Full & 2008 & 220-222 \\
 & Indeterminate & 2010 & 199 \\
\hline
XTE~J1755$-$324 & Full & 1997 & 223, 224 \\
\hline
XTE~J2012$+$381 & Full & 1998 & 225-227 \\
\hline
Swift~J1910.2$-$0546 & Full & 2012 & 228-230 \\
\hline
EXO~1846$-$031 & Full & 1985, 2019 & 231; 458-470 \\
\hline
XTE~J1719$-$291 & Full & 2008 & 232 \\
\hline
XTE~J1817$-$330 & Full & 2006 & 233, 234 \\
\hline
SWIFT~J174540.7$-$290015 & Full & 2016 & 244-250 \\
\hline
MAXI~J1820$+$070 & Full & 2018 & 255-353 \\
 & FT & 2021 & 354, 355 \\
\hline
MAXI~J1535$-$571 & Full & 2018 & 356-397 \\
\hline
SWIFT~J1658.2$-$4242 & FT & 2018 & 398-411 \\
\hline
MAXI~J1727$-$203 & Full & 2018 & 412-421 \\
\hline
MAXI~J1813$-$095 & FT & 2018 & 450-457 \\
\hline
MAXI~J1348$-$630 & Full & 2019 & 471-504 \\
\hline
MAXI~J0637$-$430 & Full & 2019 & 505-516 \\
\hline
MAXI~J1631$-$479 & Full & 2018 & 517-524 \\
\hline
MAXI~J1828$-$249 & Full & 2013 & 528-539 \\
\label{tab:table_sources}
\end{longtable}
\end{ThreePartTable}
\begin{ThreePartTable}
       \begin{tablenotes}
            \setlength\labelsep{4pt}
            \footnotesize   
  \item \textbf{References}: (1) \citet{Cui99}, (2) \citet{Remillard99a}, (3) \citet{Sobczak00}, (4) \citet{Homan01}, (5) \citet{Remillard02},  (6) \citet{Kubota04}, (7) \citet{Jain01}, (8) \citet{Tomsick01a}, (9) \citet{Corbel01}, (10) \citet{Miller01a}, (11) \citet{Kalemci01}, (12) \citet{Rodriguez04}, (13) \citet{Tomsick01b}, (14) \citet{Curran13}, (15) \citet{Tomsick01b}, (16) \citet{Curran13}, (17) \citet{Aref'ev04}, (18) \citet{Sturner05}, (19) \citet{Chaty11}, (20) \citet{Markert73}, (21) \citet{Kong02}, (22) \citet{Motch85}, (23) \citet{Miyamoto91}, (24) \citet{Harmon94b}, (25) \citet{Trudolyubov98}, (26) \citet{Rubin98}, (27) \citet{Belloni99b}, (28) \citet{Zdziarski04}, (29) \citet{Homan05}, (30) \citet{Plant14}, (31) \citet{Belloni06}, (32) \citet{Buxton12}, (33) \citet{Motta09}, (34) \citet{Debnath15}, (35) \citet{Bernardini15}, (36) \citet{Yan14}, (37) \citet{Pawar15}, (38) \citet{Zhang15b}, (39) \citet{Yan15}, (40) \citet{Buxton13b}, (41) \citet{Belloni13}, (42) \citet{Pawar13b}, (43) \citet{Gandhi17}, (44) \citet{Gandhi17b}, (45) \citet{Remillard17}, (46) \citet{Garcia17}, (47) \citet{Paice18}, \citet{Garcia19}, (48) \citet{Capitanio05}, (49) \citet{Miller06c}, (50) \citet{Kalemci06}, (51) \citet{McClintock09}, (52) \citet{Capitanio2006b}, (53) \citet{Capitanio10}, (54) \citet{Zhou13}, (55) \citet{Jonker10}, (56) \citet{Motta10}, (57) \citet{Chen10}, (58) \citet{Miller-Jones12}, (59) \citet{Zhou13}, (60) \citet{Coriat11}, (61) \citet{Debnath13a}, (62) \citet{Lin16}, (63) \citet{Chand20}, (64) \citet{Capitanio09a}, (65) \citet{Ducci14}, (66) \citet{Stiele16b}, (67) \citet{Tetarenko16}, (68) \citet{Zhang15a}, (69) \citet{Neilsen15}, (70) \citet{Negoro12}, (71) \citet{Shidatsu12}, (72) \citet{Shidatsu14}, (73) \citet{Nakahira13}, (74) \citet{Shidatsu16}, (75) \citet{Zhang17}, (76) \citet{Remillard00}, (77) \citet{Hynes00b}, (78) \citet{McClintock01b}, (79) \citet{Brocksopp10b}, (80) \citet{Frontera01}, (81) \citet{Revnivtsev00b}, (82) \citet{Wren00}, (83) \citet{Uemura00a}, (84) \citet{Zurita06}, (85) \citet{Belloni02}, (86) \citet{Grebenev12}, (87) \citet{Vovk12}, (88) \citet{Sbarufatti13}, (89) \citet{Tomsick12}, (90) \citet{Takagi15}, (91) \citet{Revnivtsev98b}, (92) \citet{Hooft99b}, (93) \citet{Brocksopp04}, (94) \citet{Hjellming96b}, (95)\citet{Francey71}, (96) \citet{Makino87}, (97) \citet{Kitamoto90}, (98) \citet{Markert79}, (99) \citet{Brocksopp01}, (100) \citet{Revnivtsev00a}, (101) \citet{Miller15a}, (102) \citet{Miller15b}, (103) \citet{Russell15}, (104) \citet{Coriat15}, (105) \citet{Koljonen15}, (106) \citet{Stiele16}, (107) \citet{Padilla13a}, (108) \citet{Sunyaev97}, (109) \citet{Cui97a}, (110) \citet{Heise97}, (111) \citet{Trudolyubov99}, (112) \citet{Negoro11b}, (113) \citet{Ferrigno11}, (114) \citet{Reis12}, (115) \citet{Zand01}, (116) \citet{Zand02b},  (117) \citet{Walter04}, (118) \citet{Barlow05}, (119) \citet{Sidoli11}, (120) \citet{Soldi06}, (121) \citet{Walter07}, (122) \citet{Paizis09}, (123) \citet{Callanan95}, (124) \citet{Callanan96}, (125) \citet{Shrader97}, (126) \citet{Sala08}, (127) \citet{Sanna15}, (128) \citet{Marshall96}, (129) \citet{Remillard99}, (130) \citet{Barret96a}, (131) \citet{Tanaka14}, (132) \citet{Suzuki15}, (133) \citet{Rushton15}, (134) \citet{Tendulkar14}, (135) \citet{Paizis15}, (136) \citet{Revnivtsev00c}, (137) \citet{Brocksopp07}, (138) \citet{Matilsky72}, (139) \citet{Kitamoto84},  (140) \citet{Park04}, (141) \citet{Harmon92}, (142) \citet{Zhang97}, (143) \citet{Sobczak99b}, (144) \citet{Remillard99b}, (145) \citet{Tomsick99}, (146) \citet{Brocksopp06}, (147) \citet{Saito06}, (148) \citet{Shaposhnikov07}, (149) \citet{Motta12}, (150) \citet{Kennea11b}, (151) \citet{Yamaoka12}, (152) \citet{Kuulkers13}, (153) \citet{Munoz-Darias10}, (154) \citet{Stiele11}, (155) \citet{Reis11}, (156) \citet{Casella04}, (157) \citet{Farinelli13}, (158) \citet{Brocksopp02}, (159) \citet{Cui00}, (160) \citet{Eachus76}, (161) \citet{Elvis75}, (162) \citet{Ricketts75}, (163) \citet{Kitamoto92}, (164) \citet{Ebisawa94}, (165) \citet{Zandt02c}, (166) \citet{Gogus04}, (167) \citet{Jonker04}, (168) \citet{Coriat13b}, (169) \citet{Krimm13a}, (170) \citet{Krimm13b}, (171) \citet{Miller-Jones13a}, (172) \citet{Zhang15}, (173) \citet{Sala07}, (174) \citet{Gierlinski09}, (175) \citet{Roy11}, (176) \citet{Kaniovski93}, (177) \citet{Griffiths78}, (178) \citet{Watson78}, (179) \citet{Wilson&Rothschild83}, (180) \citet{Tsunemi89}, (181) \citet{Rodriguez20}, (182) \citet{Levine05a}, (183) \citet{CadolleBel09}, (184) \citet{ZuritaHeras11}, (185) \citet{Kennea12b}, (186) \citet{Suwa12}, (187) \citet{Morihana13}, (188) \citet{Munoz11a}, (189) \citet{Stiele12}, (190) \citet{Markwardt03b}, (191) \citet{Remillard-Levine03}, (192) \citet{Homan&Wijnands03}, (193) \citet{Markwardt&Swank07}, (194) \citet{Kuulkers08}, (195) \citet{Ozawa11}, (196) \citet{Motch98}, (197) \citet{Skinner90}, (198) \citet{Krimm11b}, (199) \citet{Krimm13c}, (200) \citet{Markwardt01}, (201) \citet{Homan03a}, (202) \citet{Tomsick03}, (203) \citet{Kalemci03}, (204) \citet{Rossi04}, (205) \citet{Tomsick04}, (206) \citet{Markwardt&Beardmore09}, (207) \citet{Coriat09}, (208) \citet{Han11}, (209) \citet{Hiemstra11}, (210) \citet{CadolleBel04}, (211) \citet{Vargas97}, (212) \citet{Borozdin98}, (213) \citet{Borozdin00}, (214)\citet{Krimm14b}, (215) \citet{Filippova14}, (216) \citet{Mereminskiy19}, (217) \citet{Vargas96}, (218) \citet{Makino88}, (219) \citet{Yamauchi04}, (220) \citet{Markwardt08a}, (221) \citet{Racusin08}, (222) \citet{Zhao16}, (223) \citet{Revnivtsev98a}, (224) \citet{Goldoni99}, (225) \citet{White98}, (226) \citet{Vasiliev00}, (227) \citet{Campana02}, (228) \citet{Kimura12}, (229) \citet{Nakahira12b}, (230) \citet{King12a}, (231) \citet{Parmar93}, (232) \citet{Armas-Padilla11}, (233) \citet{Sala07}, (234) \citet{Roy11}, (235) \citet{Negoro16}, (236) \citet{Masumitsu16}, (237) \citet{Armas-Padilla17}, (238) \citet{Bassi17}, (239) \citet{DelSanto17a}, (240) \citet{DelSanto17b}, (241) \citet{Bassi19}, (242) \citet{Rout20}, (243) \citet{Chatterjee21}, (244) \citet{Reynolds16}, (245) \citet{Esposito16}, (246) \citet{Maan16}, (247) \citet{Baganoff16}, (248) \citet{Bower16}, (249) \citet{Degenaar16}, (250) \citet{Ponti16}, (251) \citet{Mereminskiy16}, (252) \citet{Yan17a}, (253) \citet{Mereminskiy17}, (254) \citet{Parikh18b}, (255) \citet{Tucker18}, (256) \citet{Kawamuro18}, (257) \citet{Kennea18_1820}, (258) \citet{Denisenko18}, (259) \citet{Baglio18}, (260) \citet{Bright18}, (261) \citet{Parikh18b}, (262) \citet{Littlefield18}, (263) \citet{Uttley18}, (264) \citet{Garnavich18}, (265) \citet{DelSanto18}, (266) \citet{Paice18_1820}, (267) \citet{Gandhi18}, (268) \citet{Trushkin18}, (269) \citet{Tetarenko18}, (270) \citet{Casella18}, (271) \citet{Mandal18}, (272) \citet{Bozzo18}, (273) \citet{Munoz-Darias18}, (274) \citet{Mereminskiy18_1820}, (275) \citet{Russell18_1820}, (276) \citet{Trushkin18a}, (277) \citet{Polisensky18}, (278) \citet{Homan18}, (279) \citet{Buisson18}, (280) \citet{Yu18a}, (281) \citet{Richmond18}, (282) \citet{Homan18a}, (283) \citet{Homan18b}, (284) \citet{Bright18jul}, (285) \citet{Tetarenko18a}, (286) \citet{Yamanaka18}, (287) \citet{Parikh18b}, (288) \citet{Broderick18}, (289) \citet{Broderick18a}, (290) \citet{Zampieri18}, (291) \citet{Zhang18}, (292) \citet{Negoro18a}, (293) \citet{Bright18a}, (294) \citet{Motta18}, (295) \citet{Homan18c}, (296) \citet{Baglio18a}, (297) \citet{Shidatsu18}, (298) \citet{Russell191820}, (299) \citet{Veledina19}, (300) \citet{Ulowetz19}, (301) \citet{Bahramian19}, (302) \citet{Williams19}, (303) \citet{Baglio19}, (304) \citet{Munari19}, (305) \citet{Vozza19}, (306) \citet{Shidatsu19}, (307) \citet{Tomsick19}, (308) \citet{Zampieri19}, (309) \citet{Gandhi19}, (310) \citet{Munoz-Darias19}, (311) \citet{Hoang19}, (312) \citet{Isogai19}, (313) \citet{Kara19}, (314) \citet{Hambsch19}, (315) \citet{Xu19}, (316) \citet{Bright19}, (317) \citet{Hankins19}, (318) \citet{Bharali19}, (319) \citet{Torres19}, (320) \citet{Kajava19}, (321) \citet{Buisson19}, (322) \citet{Paice19}, (323) \citet{Mudambi20}, (324) \citet{Adachi20}, (325) \citet{Sasaki20}, (326) \citet{Stiele20}, (327) \citet{Homan20}, (328) \citet{Atri20}, (329) \citet{Bright20}, (330) \citet{Xu20}, (331) \citet{Torres20}, (332) \citet{Fabian20}, (333) \citet{Espinasse20}, (334) \citet{Wang20}, (335) \citet{Markoff20}, (336) \citet{Davis20}, (337) \citet{Kosenkov20a}, (338) \citet{Sanchez-Sierras20}, (339) \citet{Chakraborty20}, (340) \citet{Zhao20}, (341) \citet{Shaw21}, (342) \citet{Buisson21}, (343) \citet{Ma21}, (344) \citet{You21}, (345) \citet{DeMarco21}, (346) \citet{Ghosh21}, (347) \citet{Dzielak21}, (348) \citet{Zdziarski21}, (349) \citet{Rodi21}, (350) \citet{Wang21}, (351) \citet{Tetarenko21}, (352) \citet{Guan21}, (353) \citet{Zdziarski21a}, (354) \citet{Baglio21}, (355) \citet{Homan21}, (356) \citet{Barthelmy17}, (357) \citet{Negoro17}, (358) \citet{Kennea17}, (359) \citet{Scaringi17}, (360) \citet{Negoro17a}, (361) \citet{Russell17}, (362) \citet{Dincer17}, (363) \citet{Nakahira17}, (364) \citet{Kennea17a}, (365) \citet{Palmer17}, (366) \citet{Mereminskiy17a}, (367) \citet{Tetarenko17}, (368) \citet{Shidatsu17a}, (369) \citet{Russell17a}, (370) \citet{Xu18}, (371) \citet{Shidatsu17}, (372) \citet{Negoro18_1535}, (373) \citet{Russell18}, (374) \citet{Parikh18}, (375) \citet{Negoro18_1535_2}, (376) \citet{Miller18}, (377) \citet{Mereminskiy18}, (378) \citet{Lepingwell18}, (379) \citet{Stevens18}, (380) \citet{Huang18}, (381) \citet{Nakahira18}, (382) \citet{Baglio18_1535}, (383) \citet{Stiele18}, (384) \citet{Tao18}, (385) \citet{Shang19}, (386) \citet{Parikh19a}, (387) \citet{Parikh19}, (388) \citet{Sreehari19}, (389) \citet{Sridhar19}, (390) \citet{Bhargava19}, (391) \citet{Chauhan19}, (392) \citet{Russell19a}, (393) \citet{Kong20}, (394) \citet{Chatterjee20}, (395) \citet{Cuneo20}, (396) \citet{Russell20}, (397) \citet{Vincentelli21}, (398) \citet{Barthelmy18}, (399) \citet{Grebenev18}, (400) \citet{Mereminskiy18_1658}, (401) \citet{Lien18}, (402) \citet{Grinberg18}, (403) \citet{Xu18_1658}, (404) \citet{Russell18_1658}, (405) \citet{Russell18_1658a}, (406) \citet{Ducci18}, (407) \citet{Xu18_1658a}, (408) \citet{Beri18a}, (409) \citet{Xu19_1658}, (410) \citet{Jithesh19}, (411) \citet{Bogensberger20}, (412) \citet{Yoneyama18}, (413) \citet{Ludlam18}, (414) \citet{Rau18}, (415) \citet{Thorstensen18}, (416) \citet{Negoro18}, (417) \citet{Kennea18}, (418) \citet{Paice18_1727}, (419) \citet{Bahramian18}, (420) \citet{Tomsick18}, (421) \citet{Alabarta20}, (422) \citet{Krimm11c}, (423) \citet{Rau11b}, (424) \citet{Krimm11a}, (425) \citet{Torres11}, (426) \citet{Milisavljevic11}, (427) \citet{Sivakoff11}, (428) \citet{Padilla13b}, (429) \citet{Shahbaz13}, (430) \citet{Armas-Padilla14}, (431) \citet{Weng15}, (432) \citet{Mata-Sanchez15}, (433) \citet{Plotkin16}, (434) \citet{Drake17}, (435) \citet{Sivakoff17}, (436) \citet{Dincer17a}, (437) \citet{Stiele18_1357}, (438) \citet{Russell18_1357}, (439) \citet{Beri19_1357}, (440) \citet{Paice19_1357}, (441) \citet{Jimenez-Ibarra19_1357}, (442) \citet{vanVelzen19}, (443) \citet{Gandhi19_1357}, (444) \citet{Russell19_1357}, (445) \citet{Neustroev19}, (446) \citet{Beri19_1357atel}, (447) \citet{Rao19}, (448) \citet{Jimenez-Ibarra19}, (449) \citet{Pirbhoy19}, (450) \citet{Beardmore18}, (451) \citet{Mazaeva18}, (452) \citet{Kawase18}, (453) \citet{Kennea18_1813}, (454) \citet{Rau18_1813}, (455) \citet{Russell18_1813}, (456) \citet{Fuerst18}, (457) \citet{Armas-Padilla19}, (458) \citet{Negoro19_1846}, (459) \citet{Mereminskiy19_1846}, (460) \citet{Bult19_1846}, (461) \citet{Miller-Jones19}, (462) \citet{Bellm19}, (463) \citet{Williams19a}, (464) \citet{Miller19}, (465) \citet{Yang19a}, (466) \citet{Yang19b}, (467) \citet{Kim19}, (468) \citet{Draghis20}, (469) \citet{Liu20}, (470) \citet{Wang21_1820}, (471) \citet{Yatabe19}, (472) \citet{Denisenko19}, (473) \citet{Kennea19}, (474) \citet{Russell19_1348}, (475) \citet{Sanna19}, (476) \citet{Wang21}, (477) \citet{Russell19}, (478) \citet{Cangemini19a}, (479) \citet{Cangemini19b}, (480) \citet{Chen19}, (481) \citet{Bassi19b}, (482) \citet{Charles19}, (483) \citet{Baglio19_1348}, (484) \citet{Carotenuto19}, (485) \citet{Jana19}, (486) \citet{Russell19atel}, (487) \citet{Negoro19}, (488) \citet{AlYazeedi19}, (489) \citet{Pirbhoy20}, (490) \citet{Zhang20a}, (491) \citet{Wang21}, (492) \citet{Oeda20}, (493) \citet{Baglio20}, (494) \citet{Tominaga20}, (495) \citet{Jana20}, (496) \citet{Belloni20}, (497) \citet{Negoro20}, (498) \citet{Baglio20a}, (499) \citet{Zhang20}, (500) \citet{Lamer21}, (501) \citet{Chauhan21}, (502) \citet{Weng21}, (503) \citet{Garcia21}, (504) \citet{Carotenuto21}, (505) \citet{Negoro19_0637}, (506) \citet{Kennea19_0637}, (507) \citet{Strader19}, (508) \citet{Tomsick19_0637}, (509) \citet{Russell19_0637}, (510) \citet{Kravtsov19}, (511) \citet{Murata19}, (512) \citet{Knigge19}, (513) \citet{Baglio20_0637}, (514) \citet{Johar20}, (515) \citet{Tomsick20}, (516) \citet{Tetarenko21_0637}, (517) \citet{Miyasaka18}, (518) \citet{Russell19_1631}, (519) \citet{Onori19}, (520) \citet{Negoro19_1631}, (521) \citet{Eijnden19}, (522) \citet{Kong19}, (523) \citet{Fiocchi20}, (524) \citet{Xu20_1631}, (525) \citet{Ducci19}, (526) \citet{Armas-Padilla19_17285}, (527) \citet{Eijnden19_17285}, (528) \citet{Nakahira13}, (529) \citet{Filippova13}, (530) \citet{Kennea13}, (531) \citet{Kennea13a}, (532) \citet{Negoro13}, (533) \citet{Miller-Jones13b}, (534) \citet{Krivonos13}, (535) \citet{Tomsick14a}, (536) \citet{Corbel14}, (537) \citet{Filippova14a}, (538) \citet{Grebenev13}, (539) \citet{Oda19}, (540) \citet{Cummings12a}, (541) \citet{Rau12b}, (542) \citet{Vovk12}, (543) \citet{Tomsick12}, (544) \citet{Grebenev12}, (545) \citet{Hynes12}, (546) \citet{Belloni12}, (547) \citet{Russell13_174510}, (548) \citet{DelSanto16}, (549) \citet{Garcia19}, (550) \citet{Bhowmick21}, (551) \citet{Tremou18}, (552) \citet{Garcia18}, (553) \citet{Paice19_339}, (554) \citet{Rao19_339}, (555) \citet{Paice20}, (556) \citet{Grebenev20}.
        \end{tablenotes}
\end{ThreePartTable}%

\section{O/IR intervals selected to study FT and full outburst during quiescence}

\begin{figure*}
\centering
\subfloat[2004 outburst]{
\label{f:2004qv}
\includegraphics[width=0.3\textwidth]{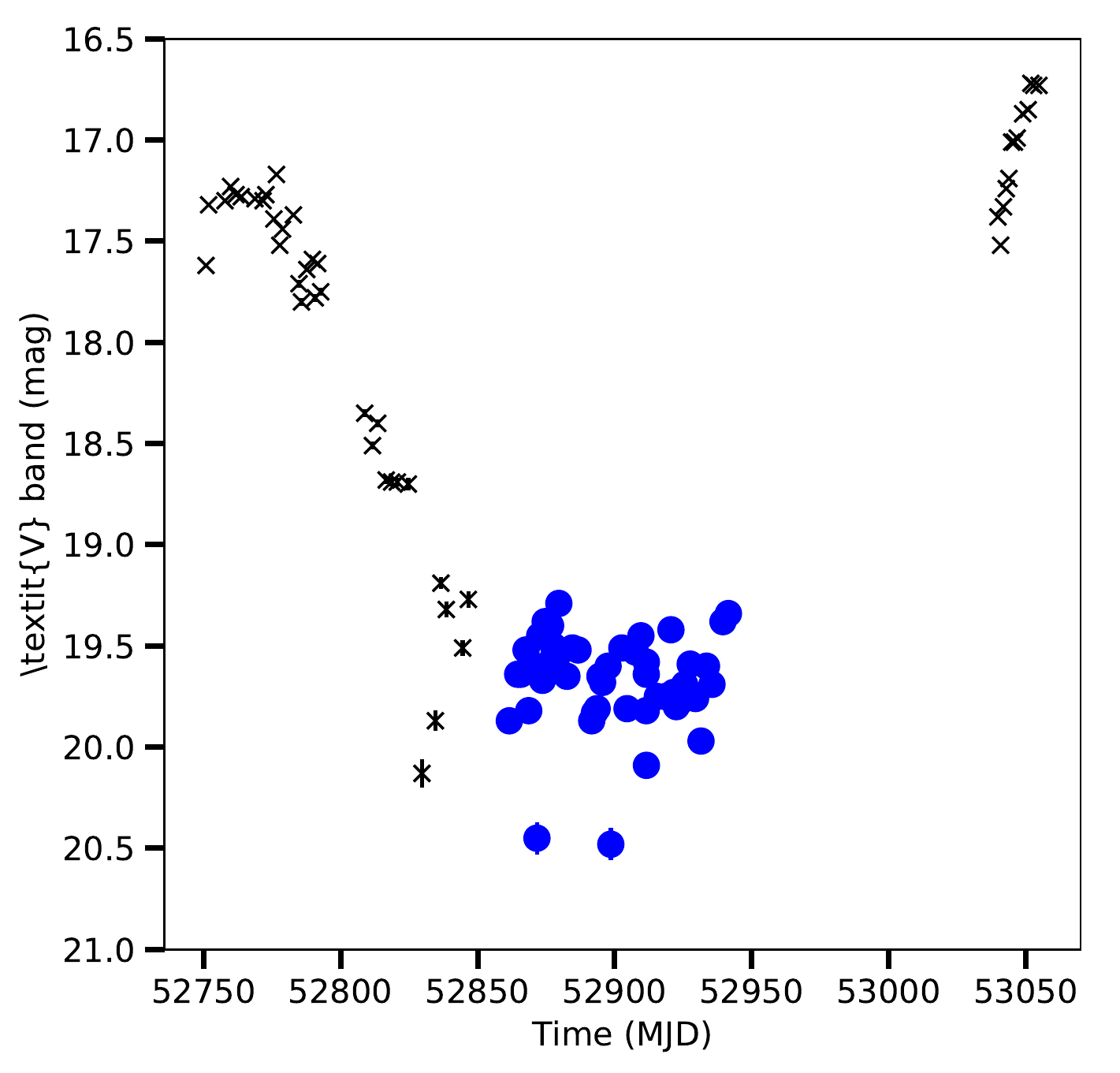}}
\subfloat[2006 outburst]{
\label{f:2006qv}
\includegraphics[width=0.3\textwidth]{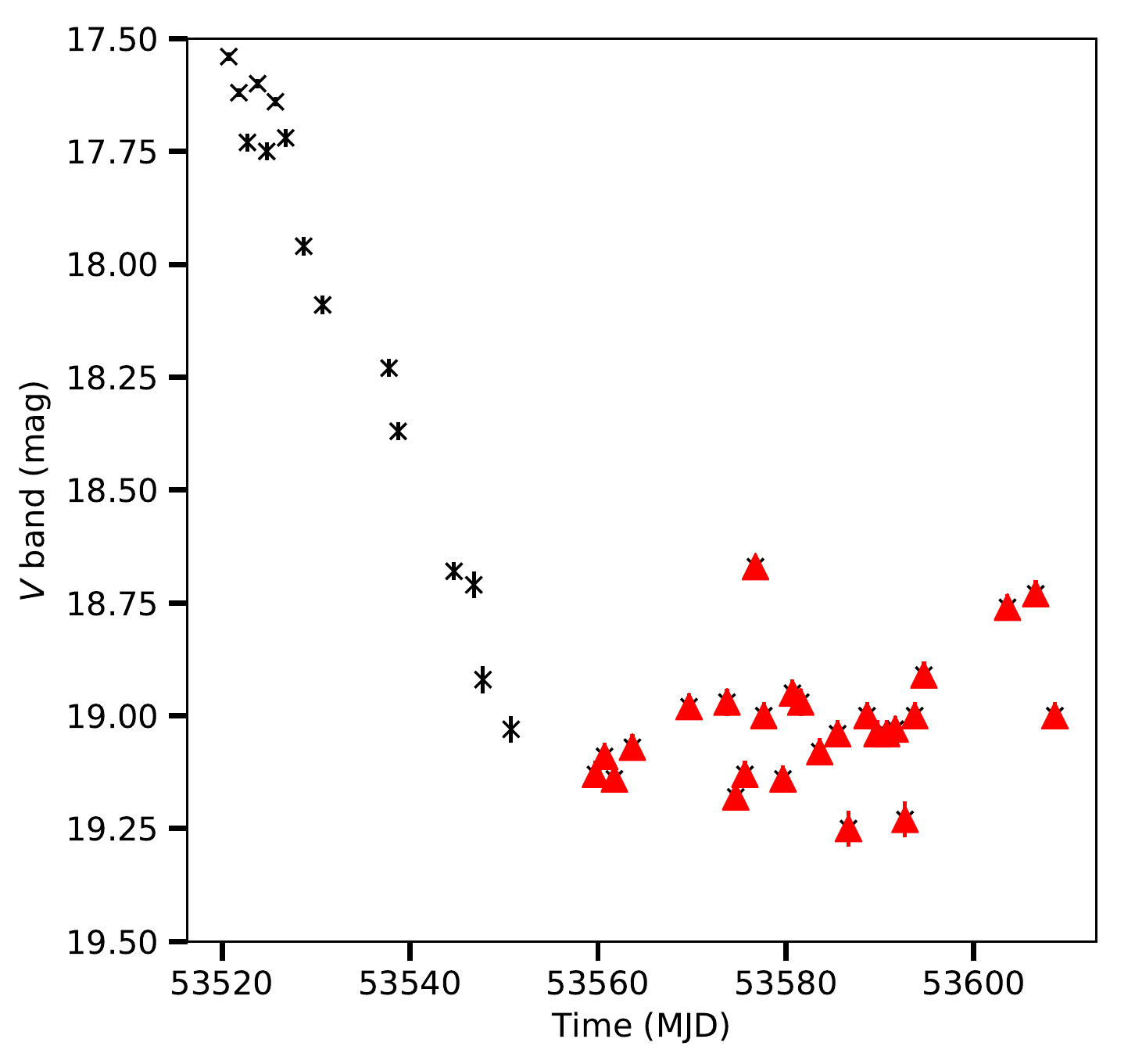}}
\subfloat[2006b outburst]{
\label{f:2006bqv}
\includegraphics[width=0.3\textwidth]{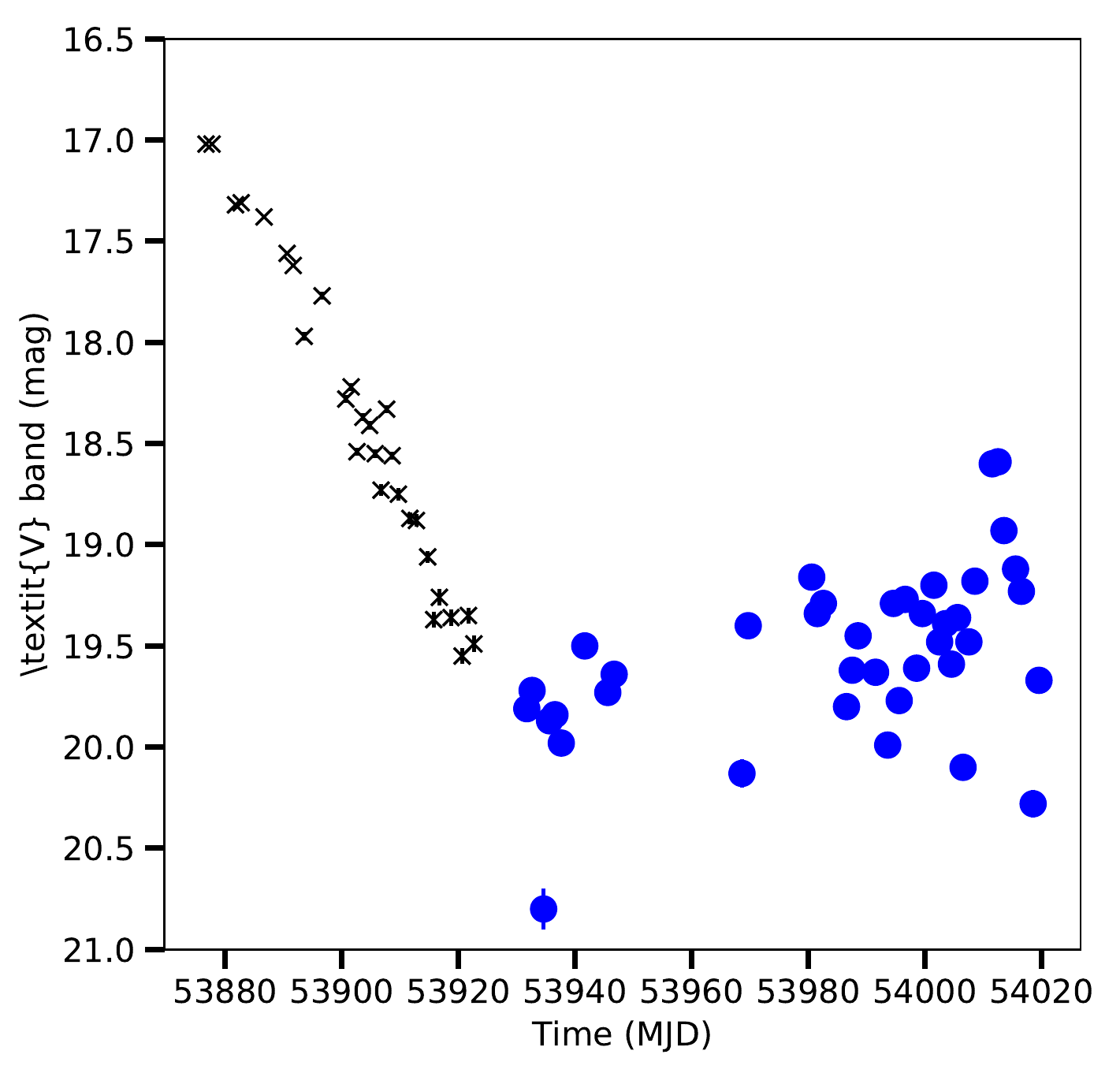}}\\
\subfloat[2008 outburst]{
\label{f:2008qv}
\includegraphics[width=0.3\textwidth]{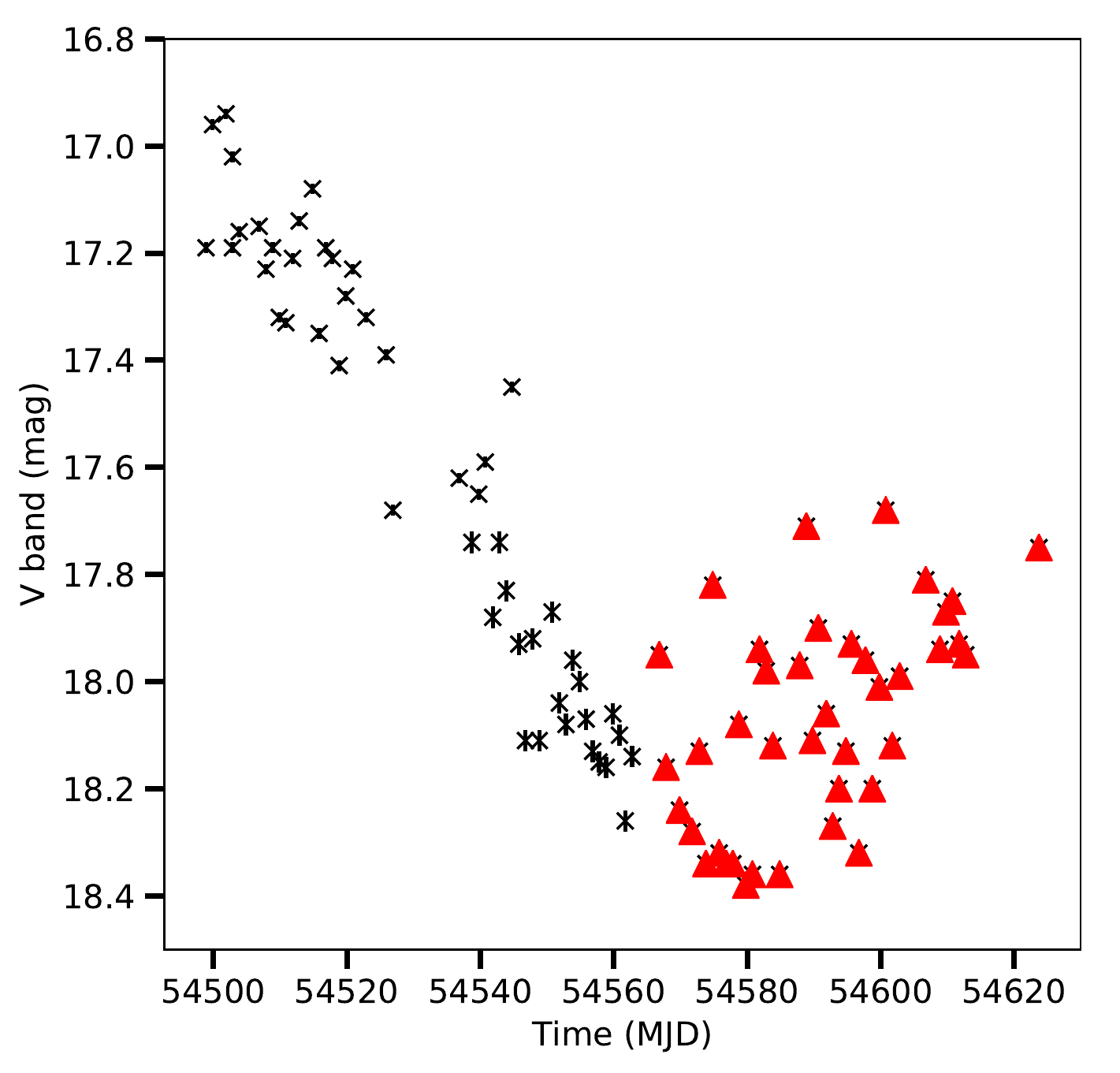}}
\subfloat[2009b outburst]{
\label{f:2009bqv}
\includegraphics[width=0.3\textwidth]{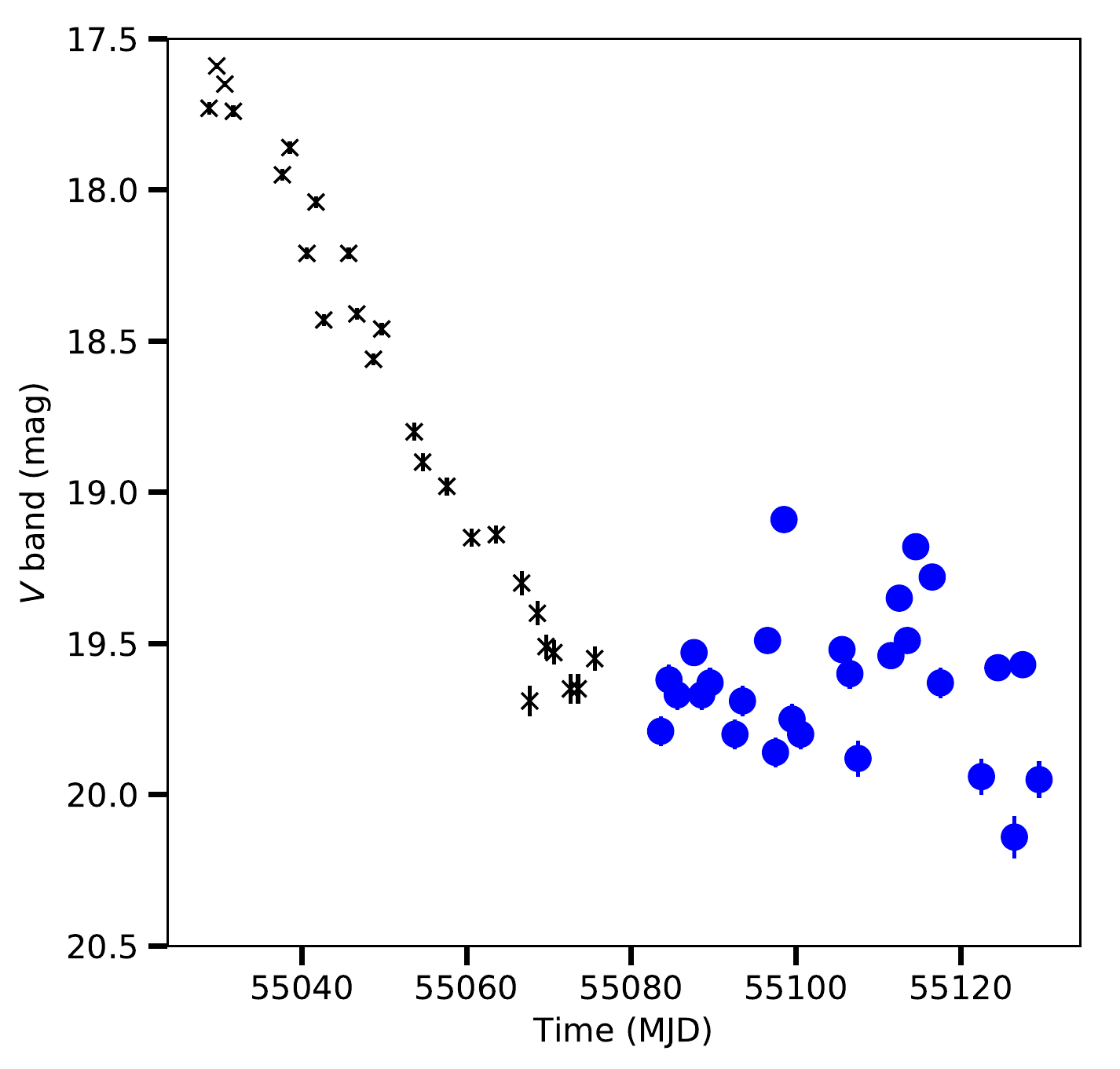}}
\subfloat[2013 outburst]{
\label{f:2013qv}
\includegraphics[width=0.3\textwidth]{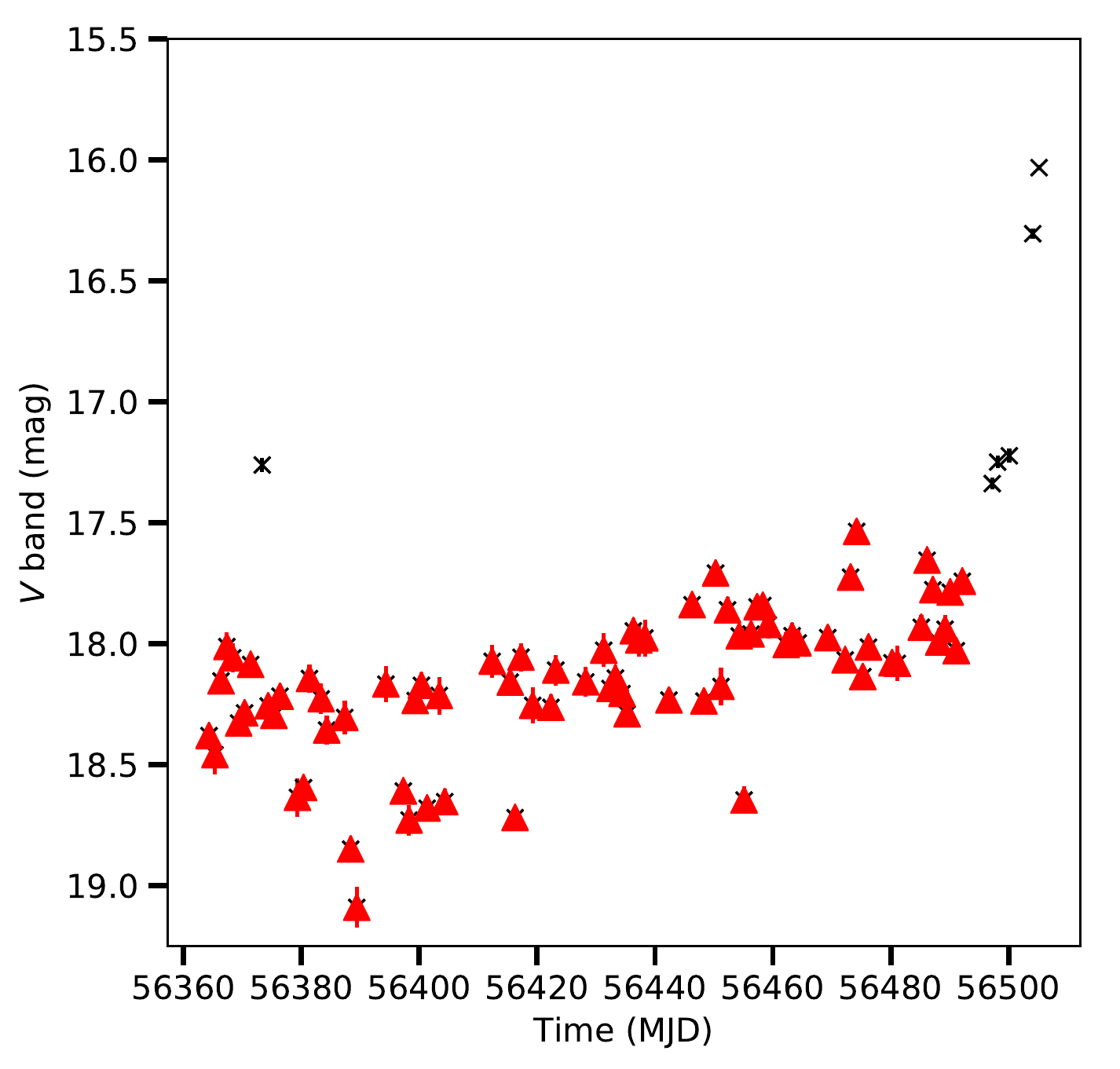}}\\
\subfloat[2014 outburst]{
\label{f:2014qv}
\includegraphics[width=0.3\textwidth]{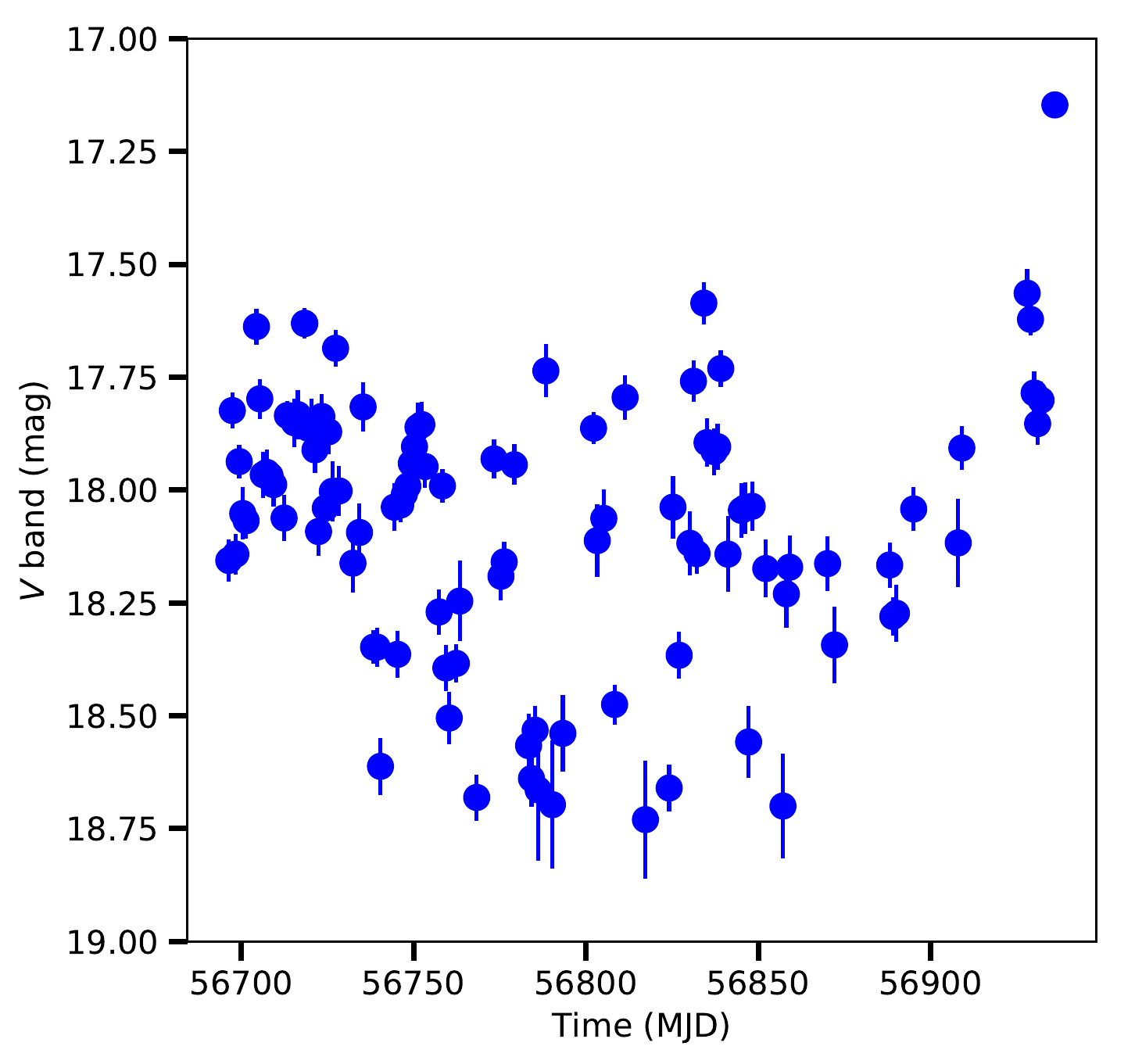}}
\caption{V-band light curve of GX 339$-$4 of the intervals when the source was in quiescence in X-rays. Blue circles and red triangles correspond to the intervals used to compared the magnitude of the V-band before full and FT outbursts, respectively.}
\label{fig:priorv}
\end{figure*}

\begin{figure*}
\centering
\subfloat[2004 outburst]{
\label{f:2004qi}
\includegraphics[width=0.3\textwidth]{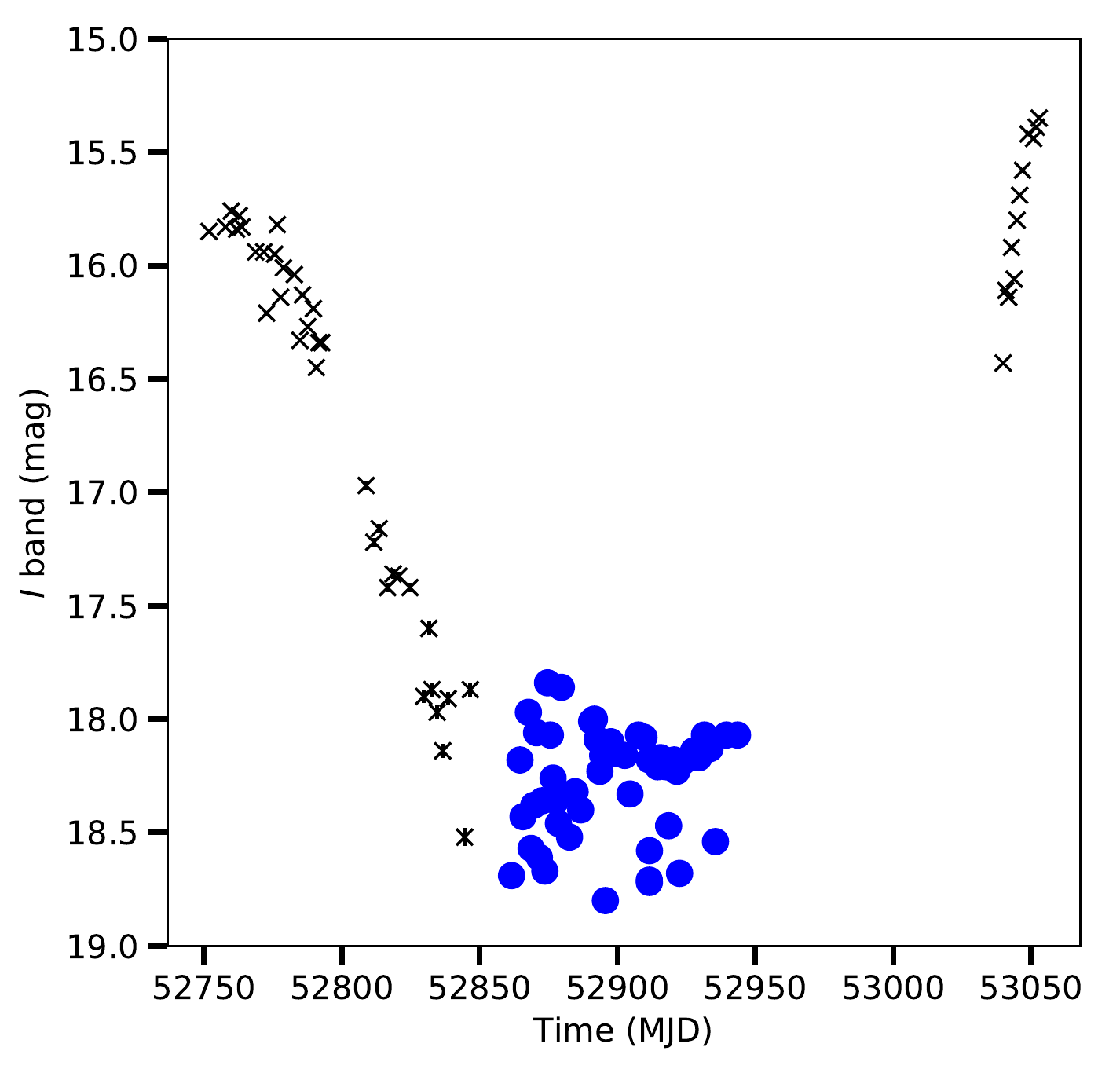}}
\subfloat[2006 outburst]{
\label{f:2006qi}
\includegraphics[width=0.3\textwidth]{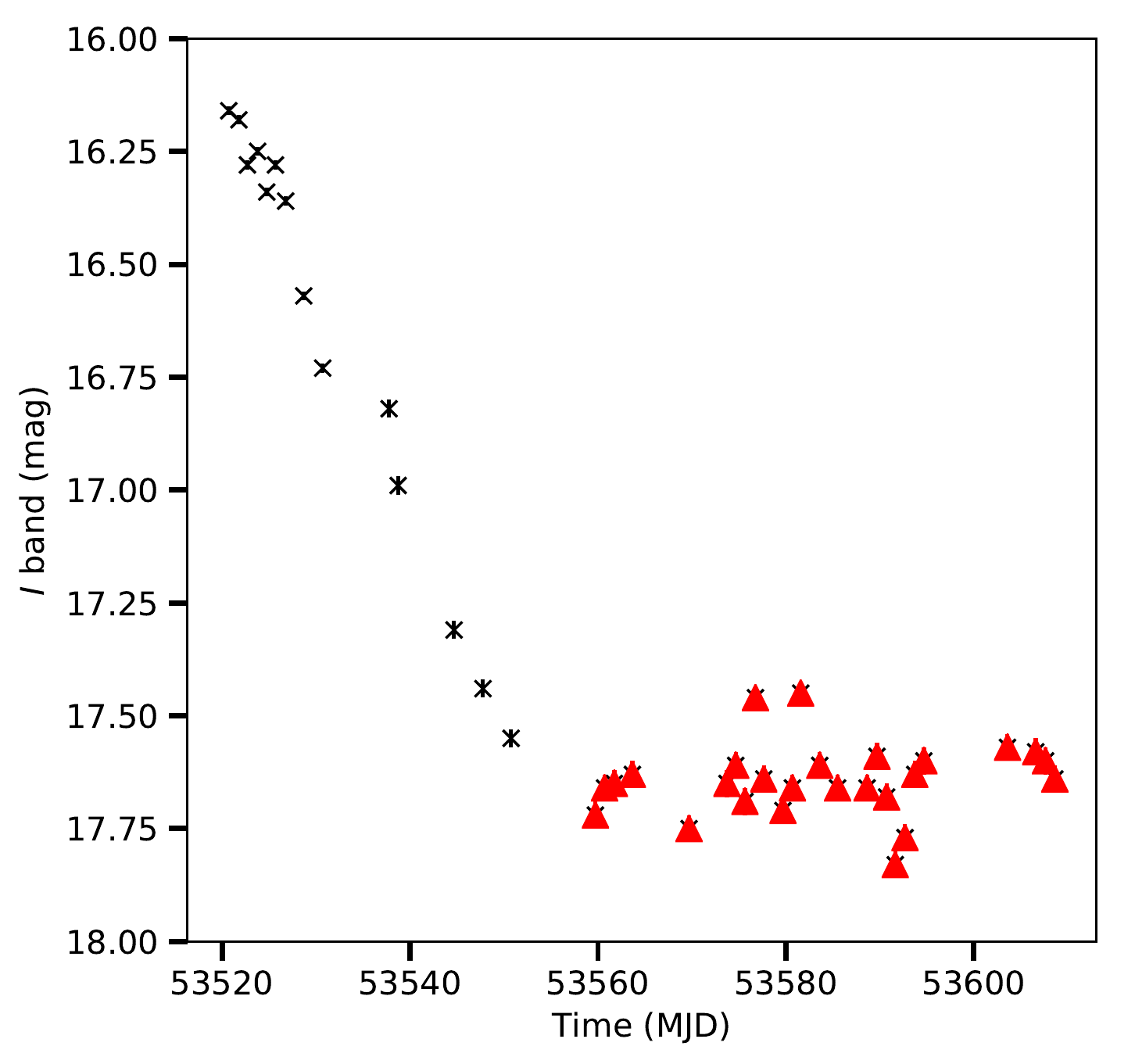}}
\subfloat[2006b outburst]{
\label{f:2006bqi}
\includegraphics[width=0.3\textwidth]{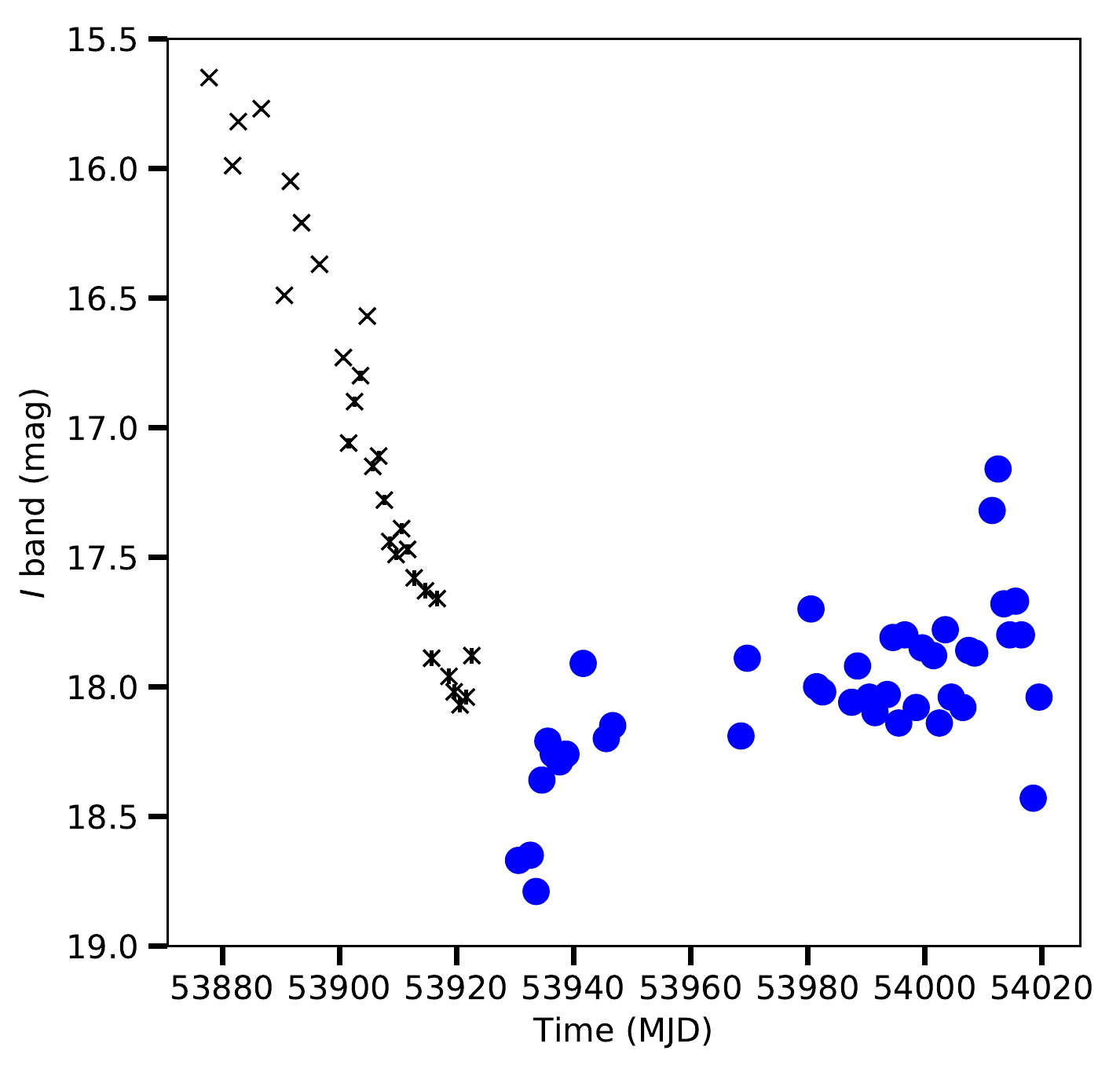}}\\
\subfloat[2008 outburst]{
\label{f:2008qi}
\includegraphics[width=0.3\textwidth]{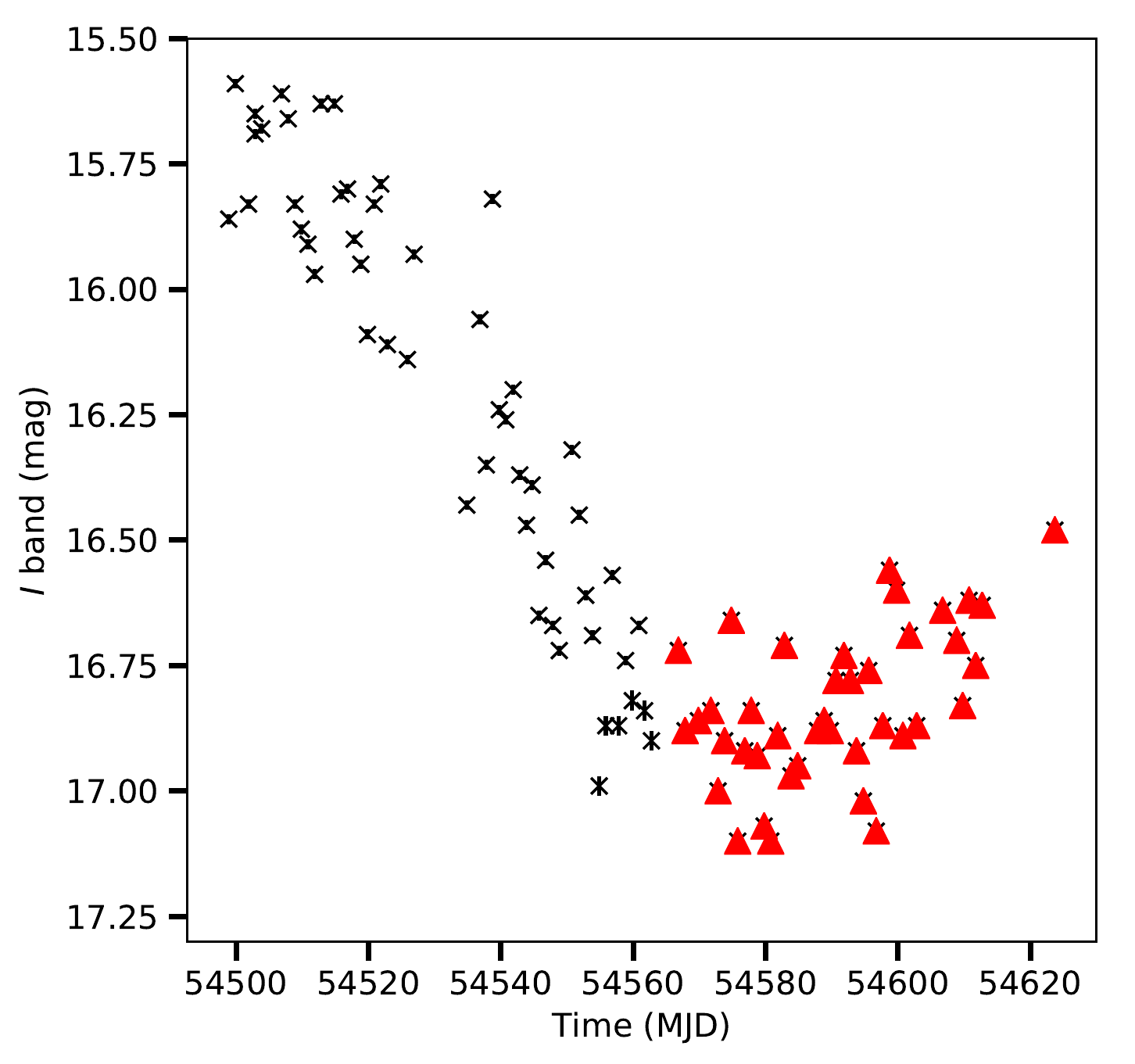}}
\subfloat[2009b outburst]{
\label{f:2009bqi}
\includegraphics[width=0.3\textwidth]{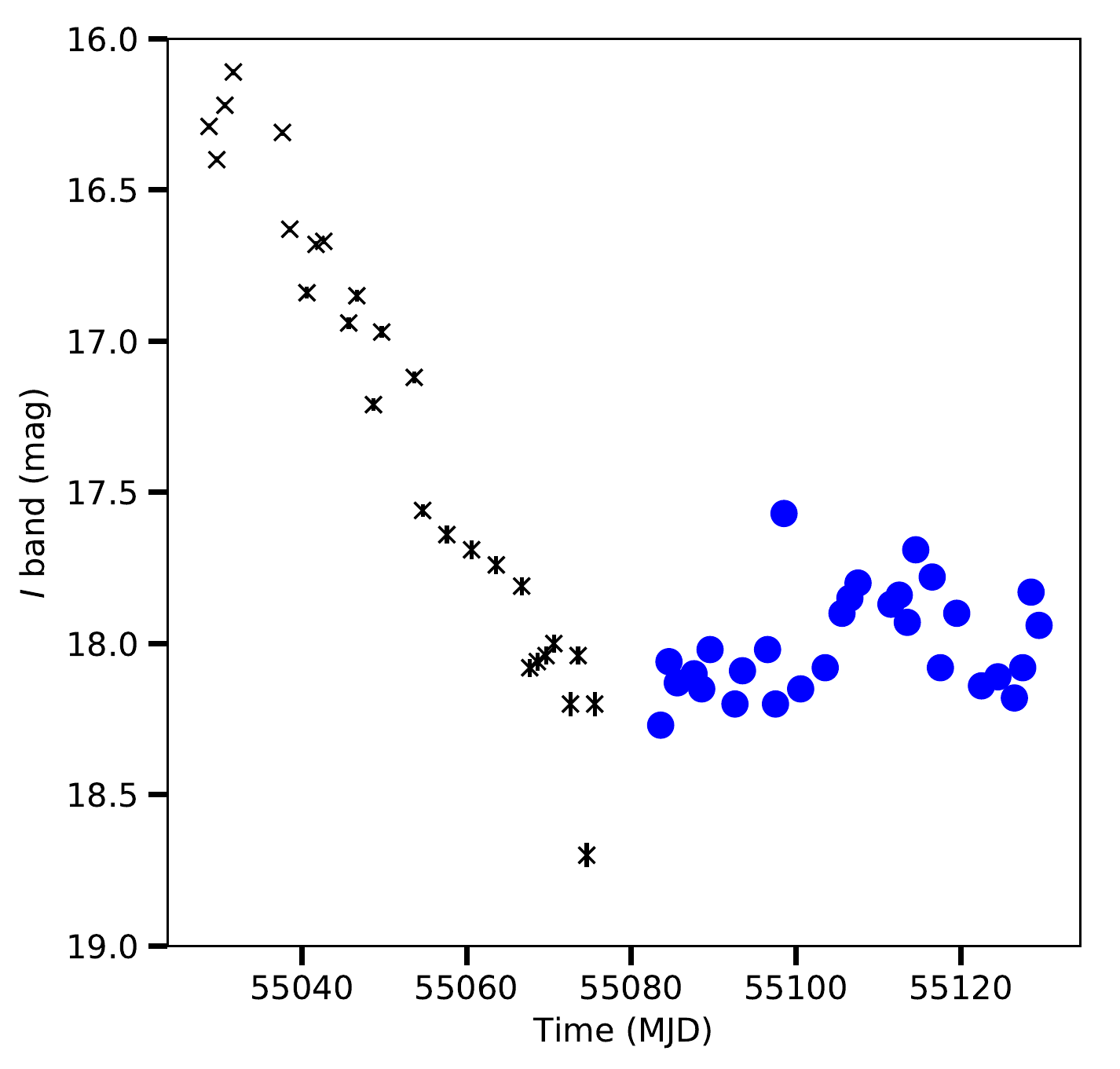}}
\caption{Same as Fig. \ref{fig:priorv} for the I-band.}
\label{fig:priori}
\end{figure*}

\begin{figure*}
\centering
\subfloat[2004 outburst]{
\label{f:2004qj}
\includegraphics[width=0.3\textwidth]{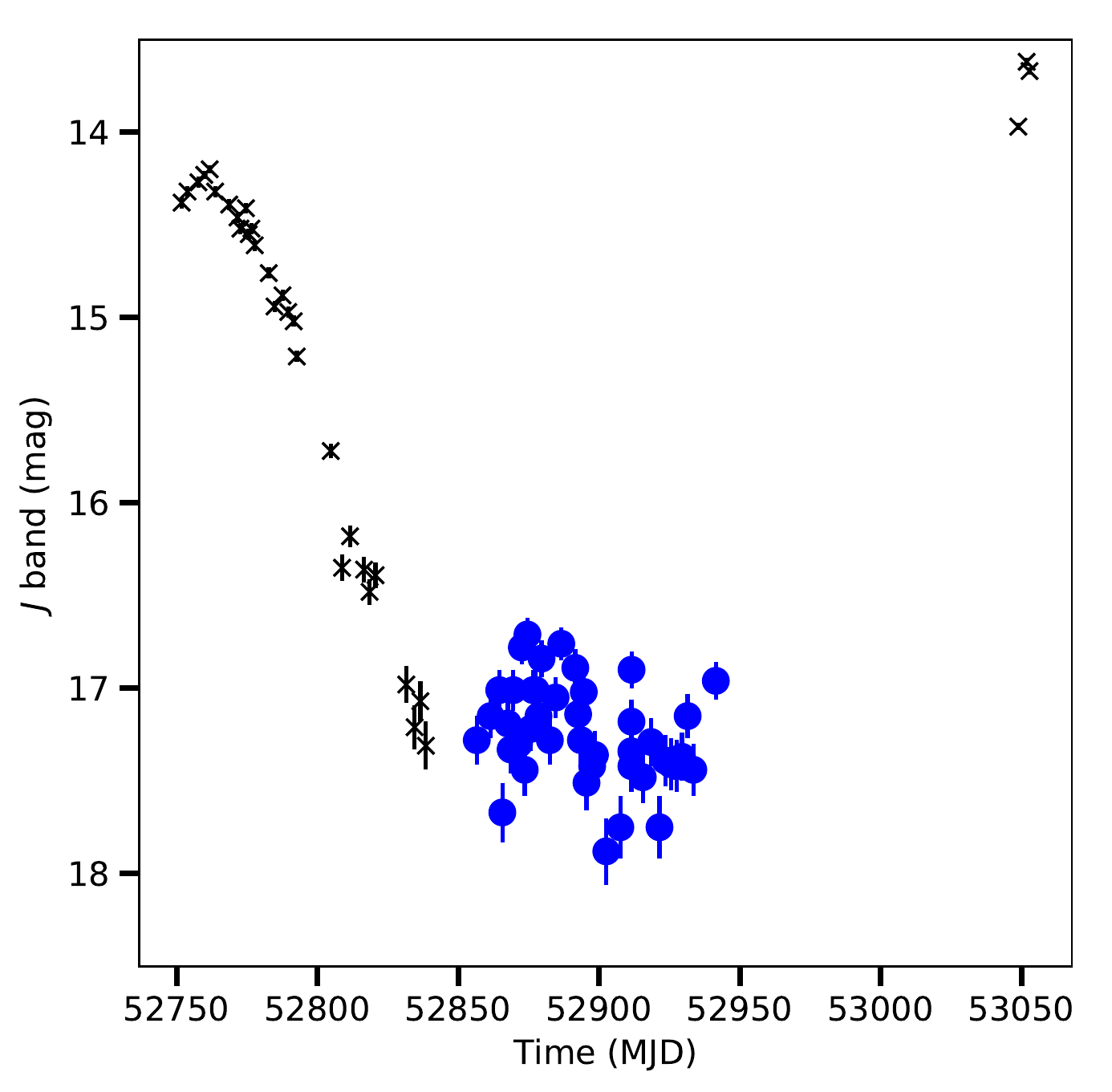}}
\subfloat[2006 outburst]{
\label{f:2006qj}
\includegraphics[width=0.3\textwidth]{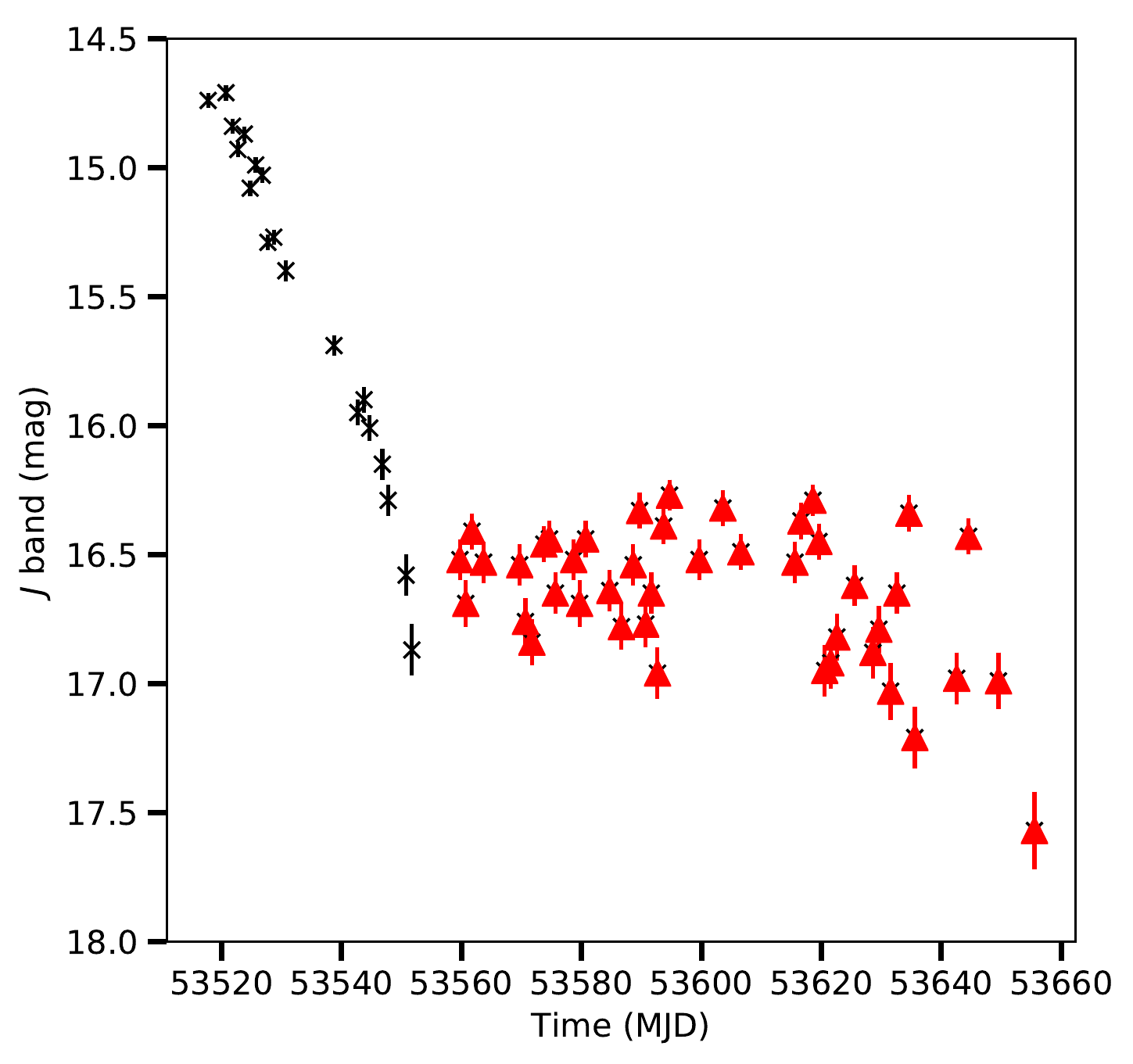}}
\subfloat[2006b outburst]{
\label{f:2006bqj}
\includegraphics[width=0.3\textwidth]{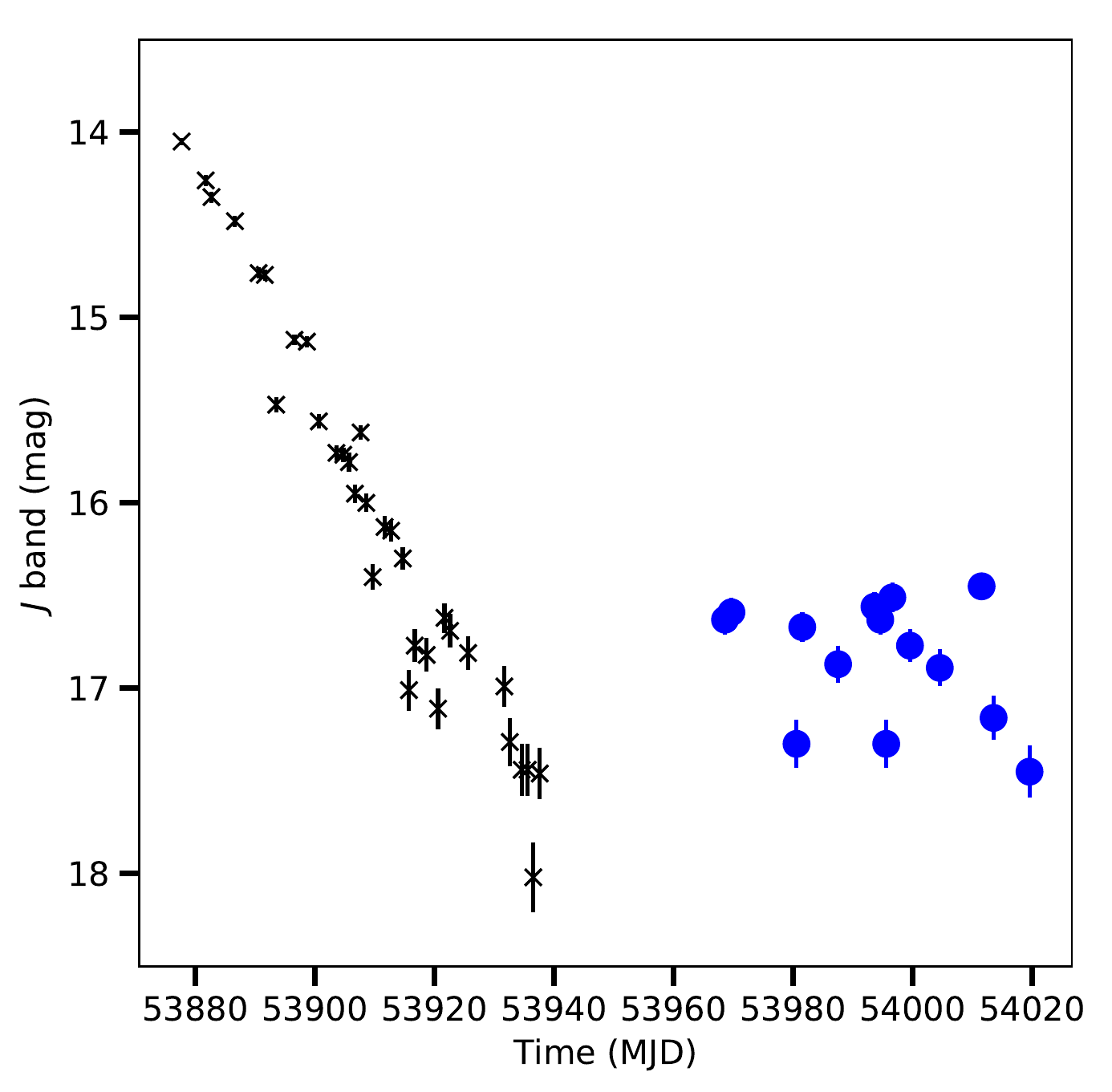}}\\
\subfloat[2008 outburst]{
\label{f:2008qj}
\includegraphics[width=0.3\textwidth]{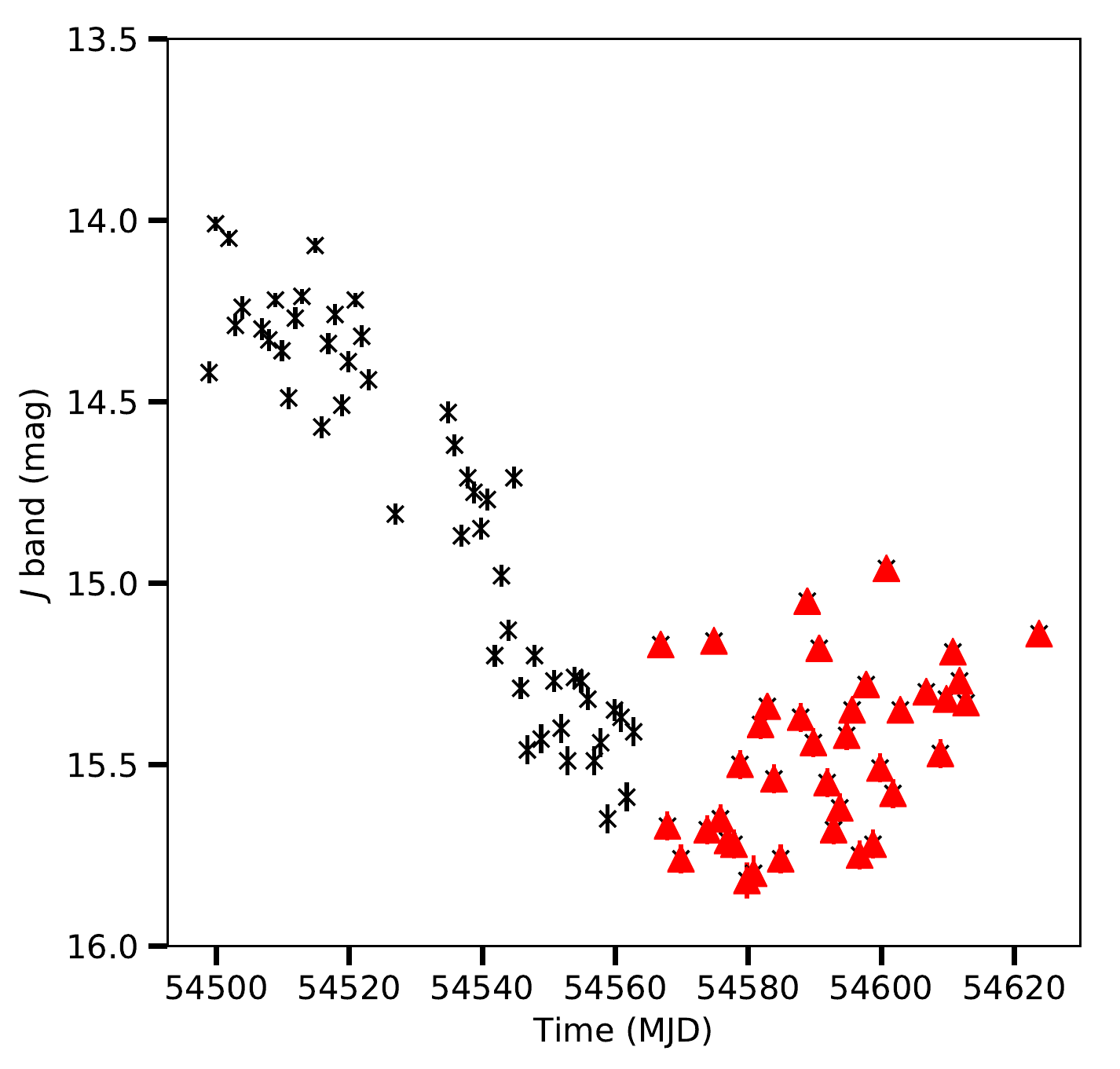}}
\subfloat[2009b outburst]{
\label{f:2009bqj}
\includegraphics[width=0.3\textwidth]{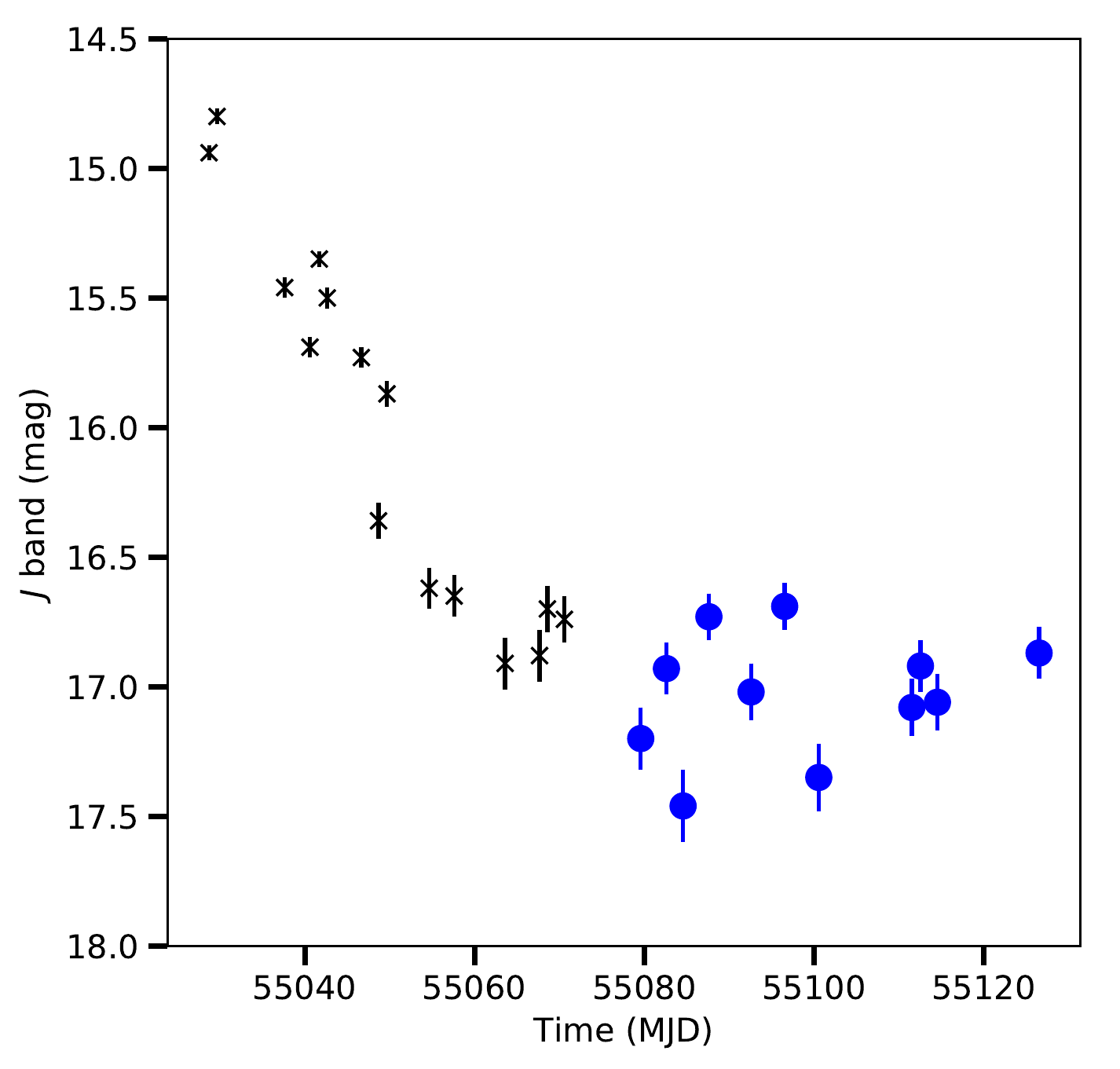}}
\subfloat[2013 outburst]{
\label{f:2013qj}
\includegraphics[width=0.3\textwidth]{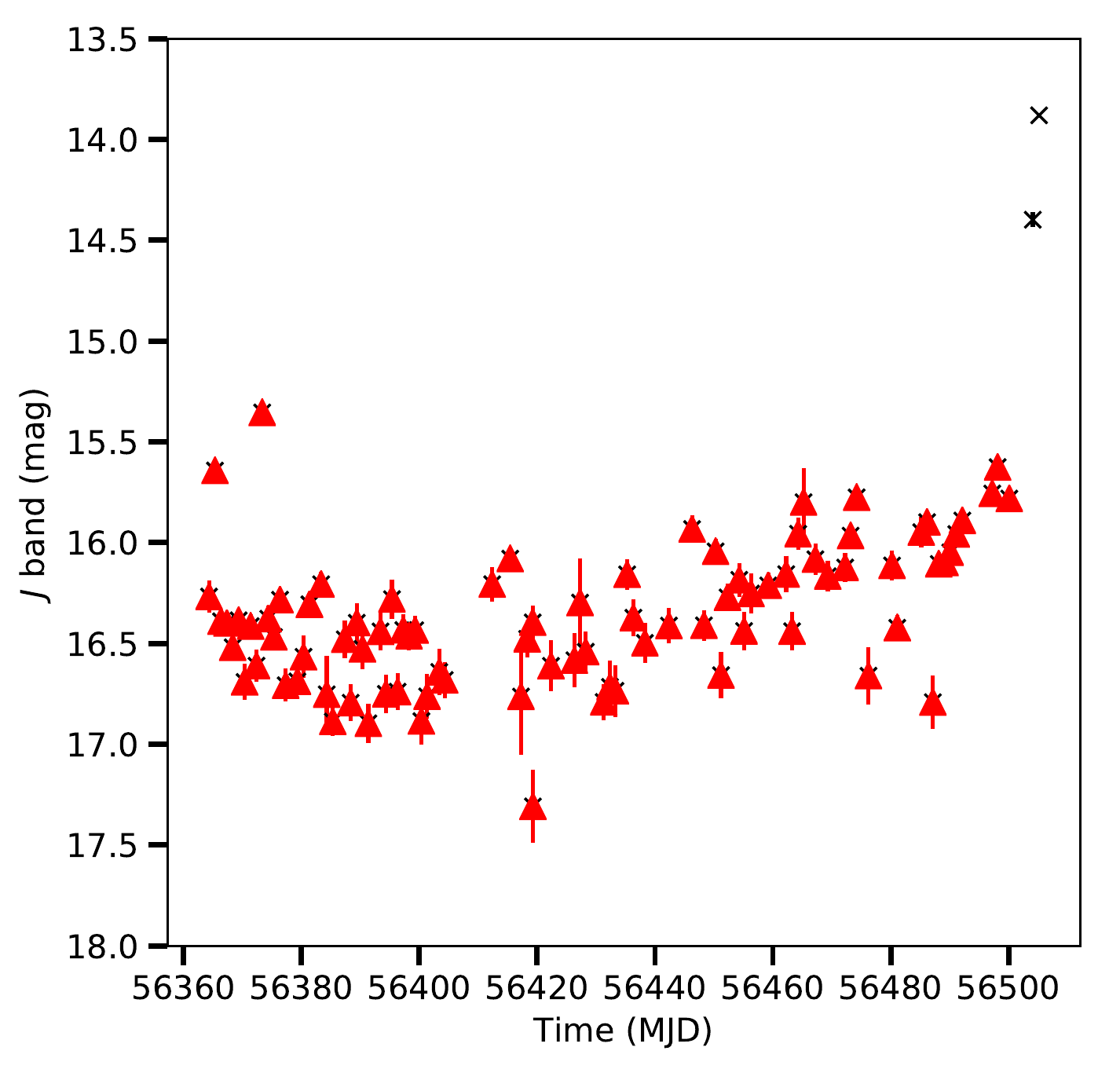}}\\
\subfloat[2014 outburst]{
\label{f:2014qj}
\includegraphics[width=0.3\textwidth]{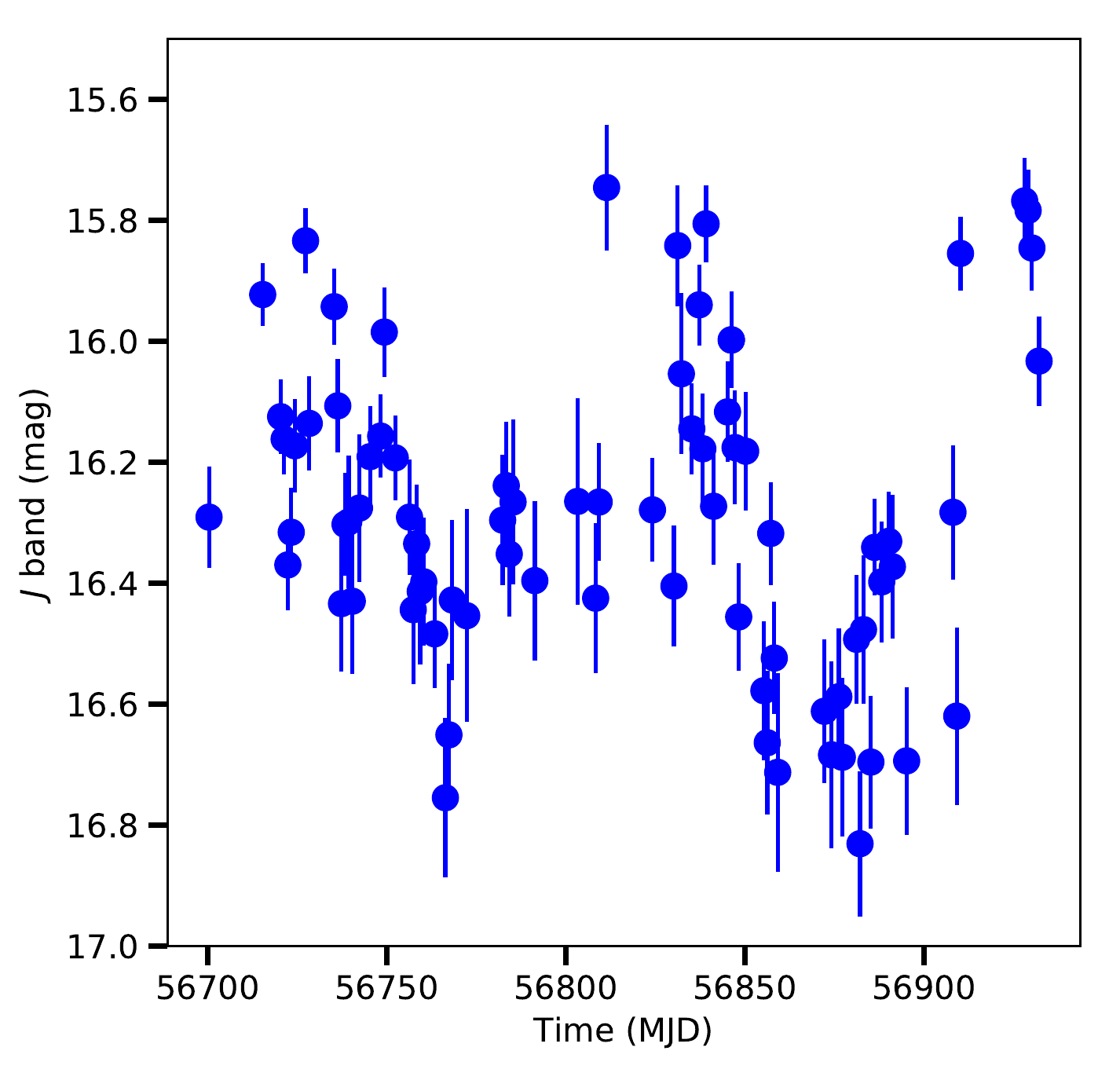}}
\caption{Same as Fig. \ref{fig:priorv} for the J-band.}
\label{fig:priorj}
\end{figure*}

\begin{figure*}
\centering
\subfloat[2004 outburst]{
\label{f:2004qh}
\includegraphics[width=0.3\textwidth]{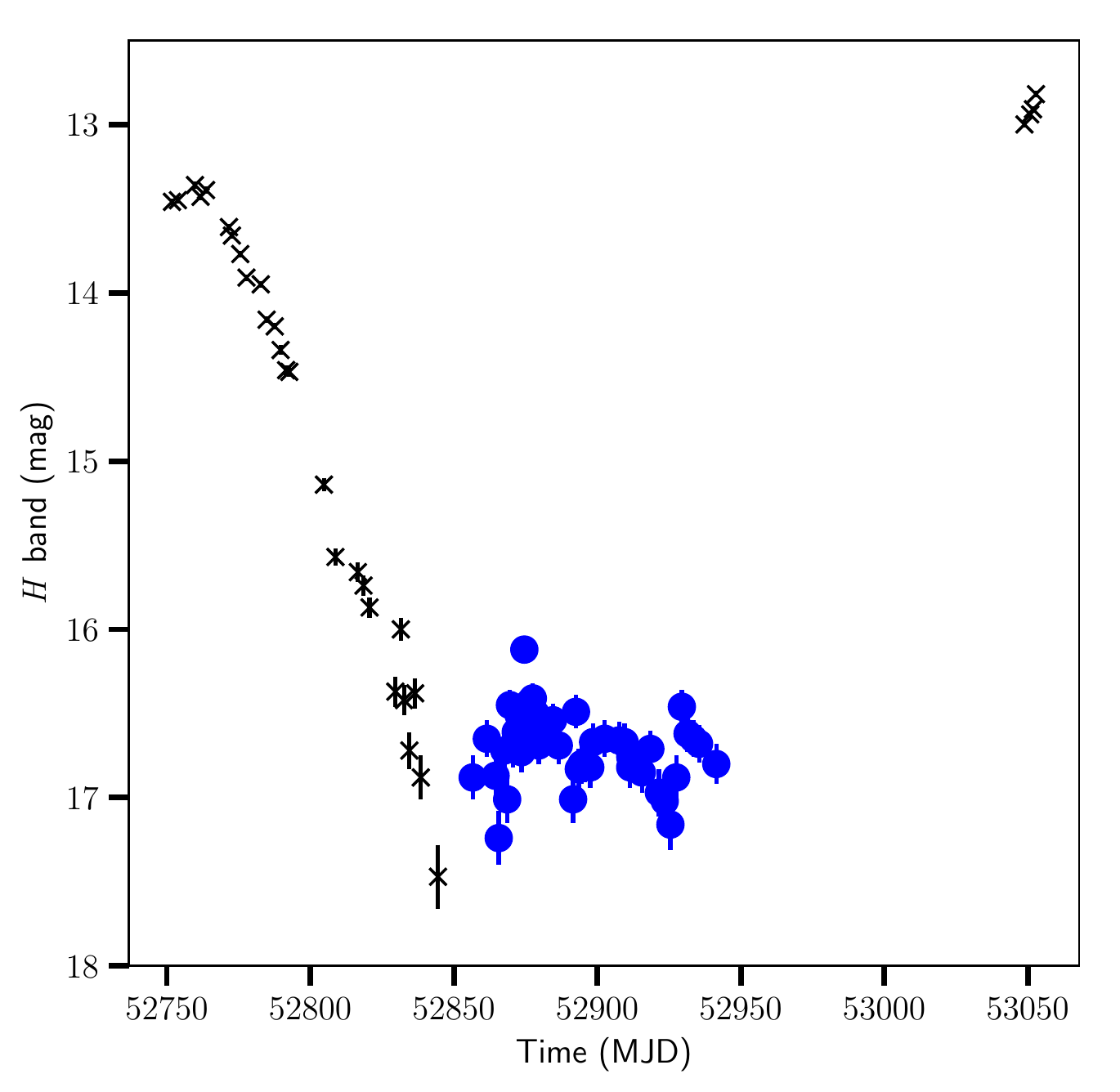}}
\subfloat[2006 outburst]{
\label{f:2006qh}
\includegraphics[width=0.3\textwidth]{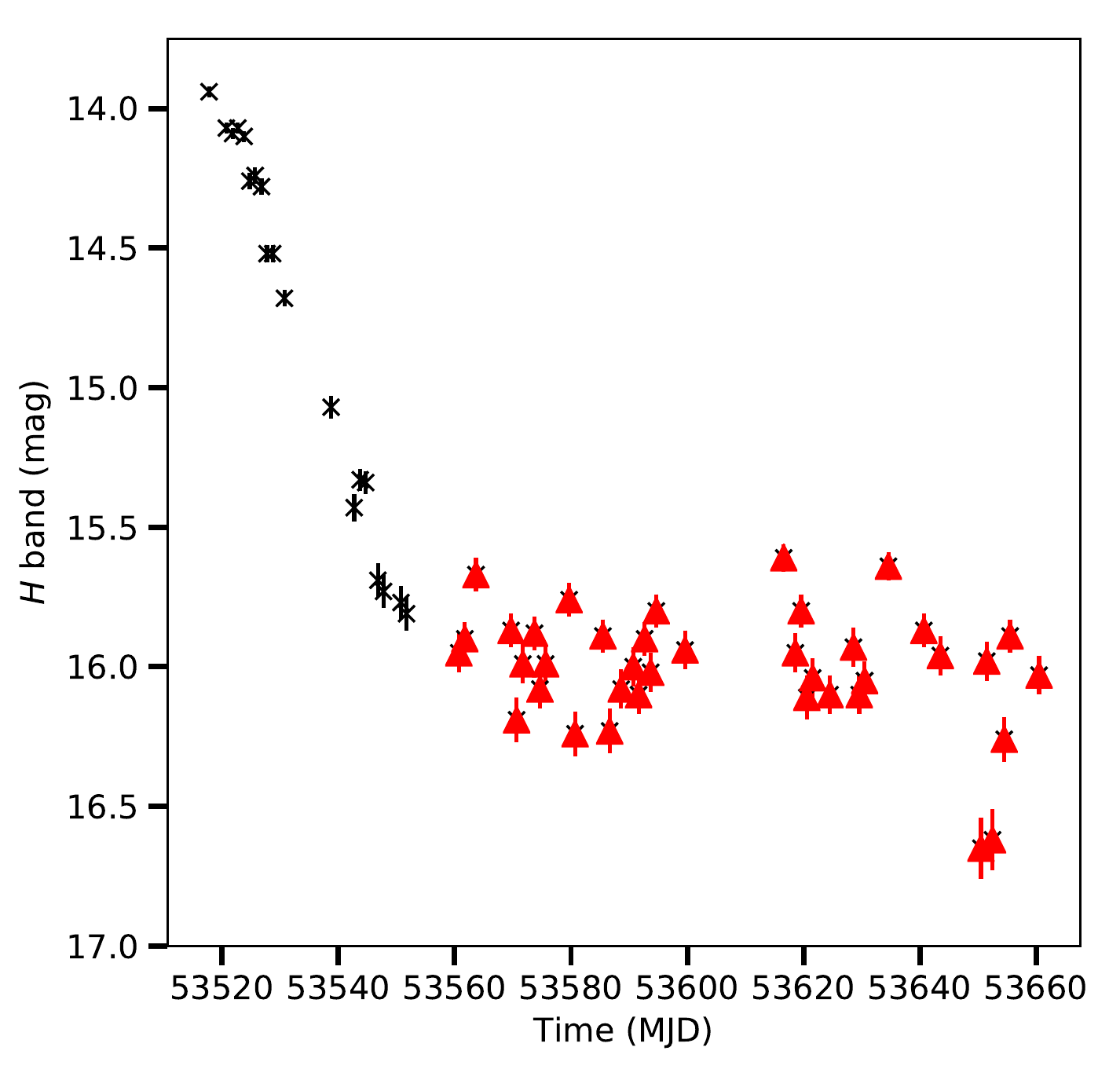}}
\subfloat[2006b outburst]{
\label{f:2006bqh}
\includegraphics[width=0.3\textwidth]{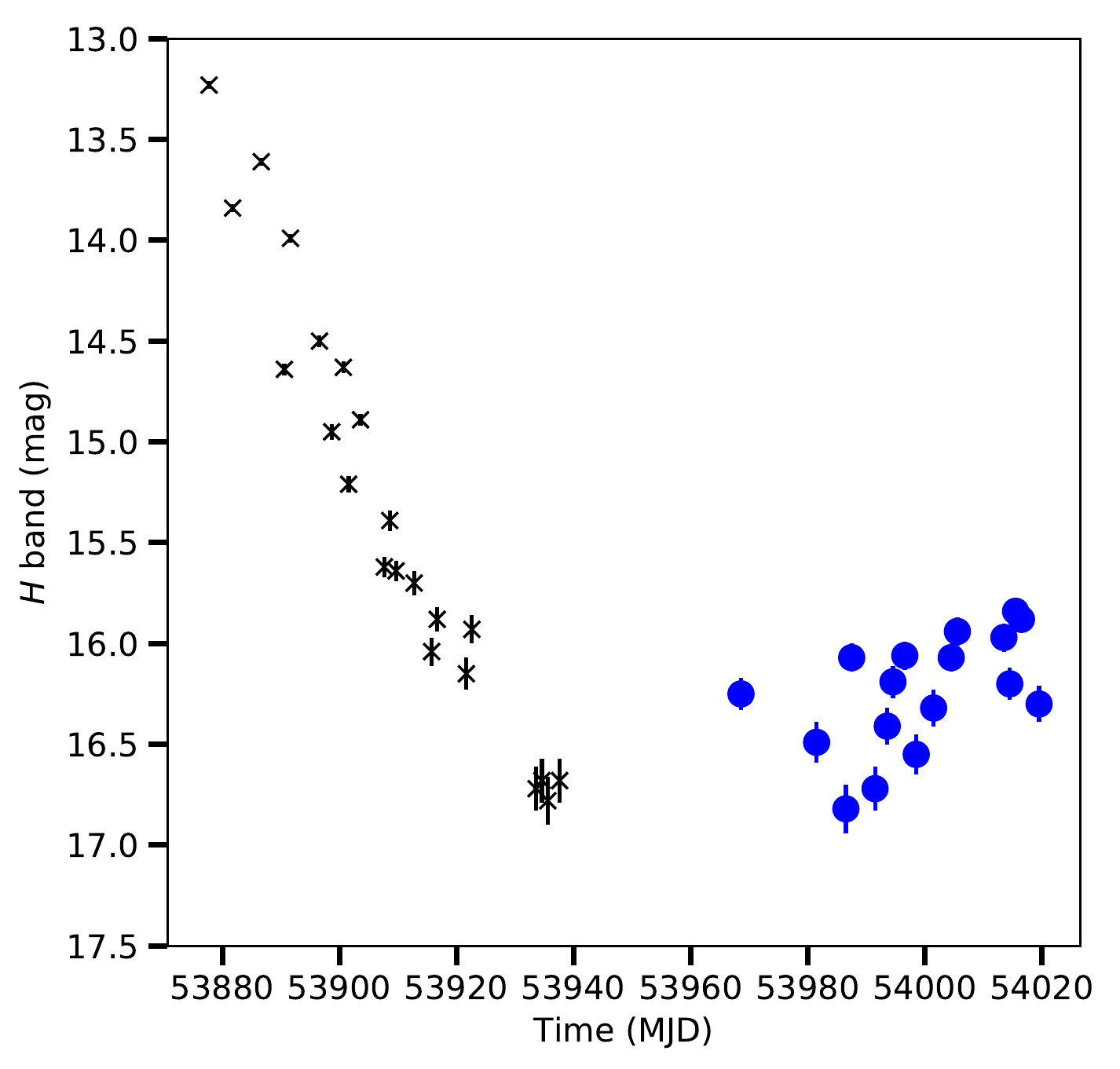}}\\
\subfloat[2008 outburst]{
\label{f:2008qh}
\includegraphics[width=0.3\textwidth]{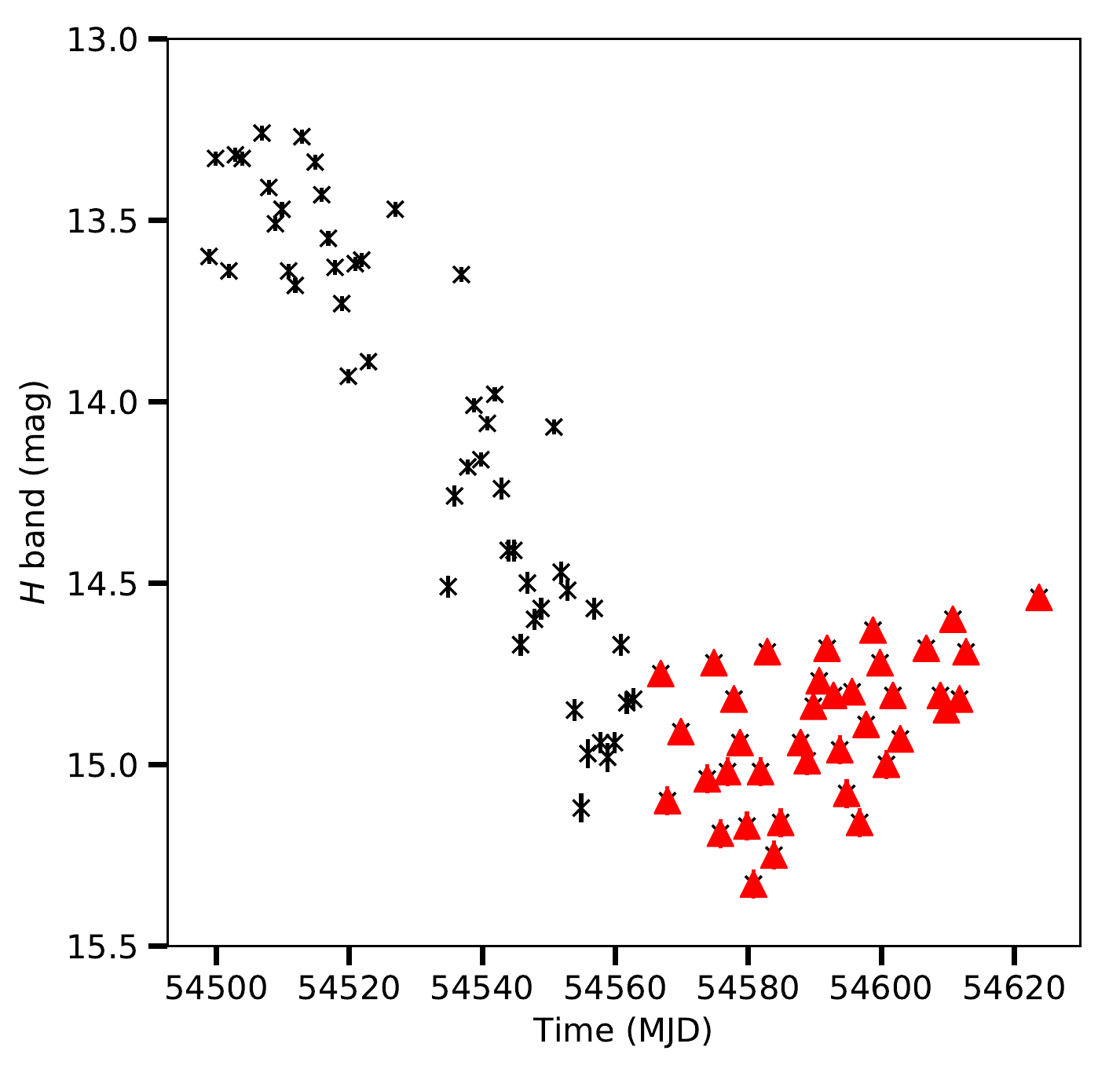}}
\subfloat[2009b outburst]{
\label{f:2009bqh}
\includegraphics[width=0.3\textwidth]{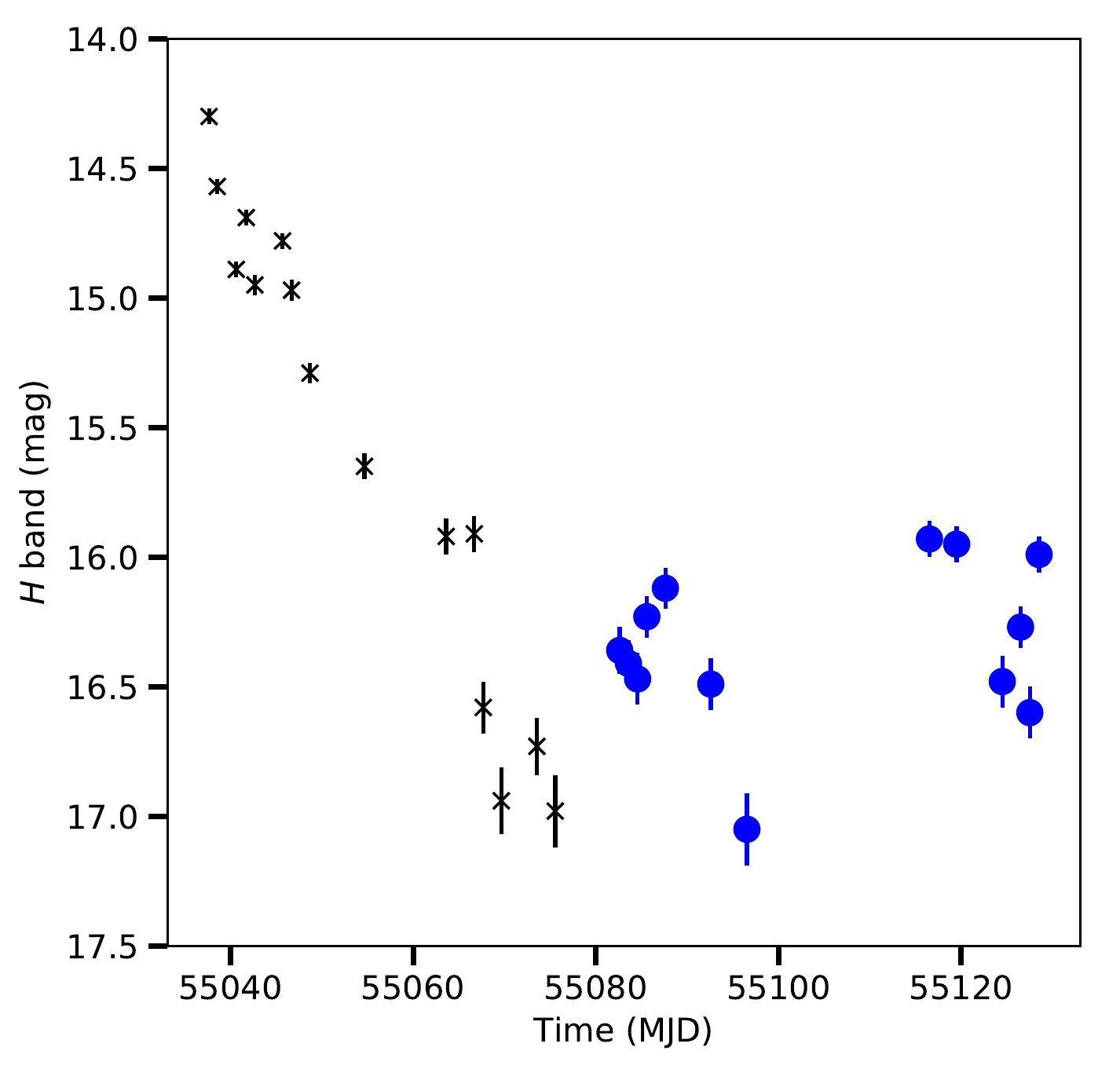}}
\subfloat[2013 outburst]{
\label{f:2013qh}
\includegraphics[width=0.3\textwidth]{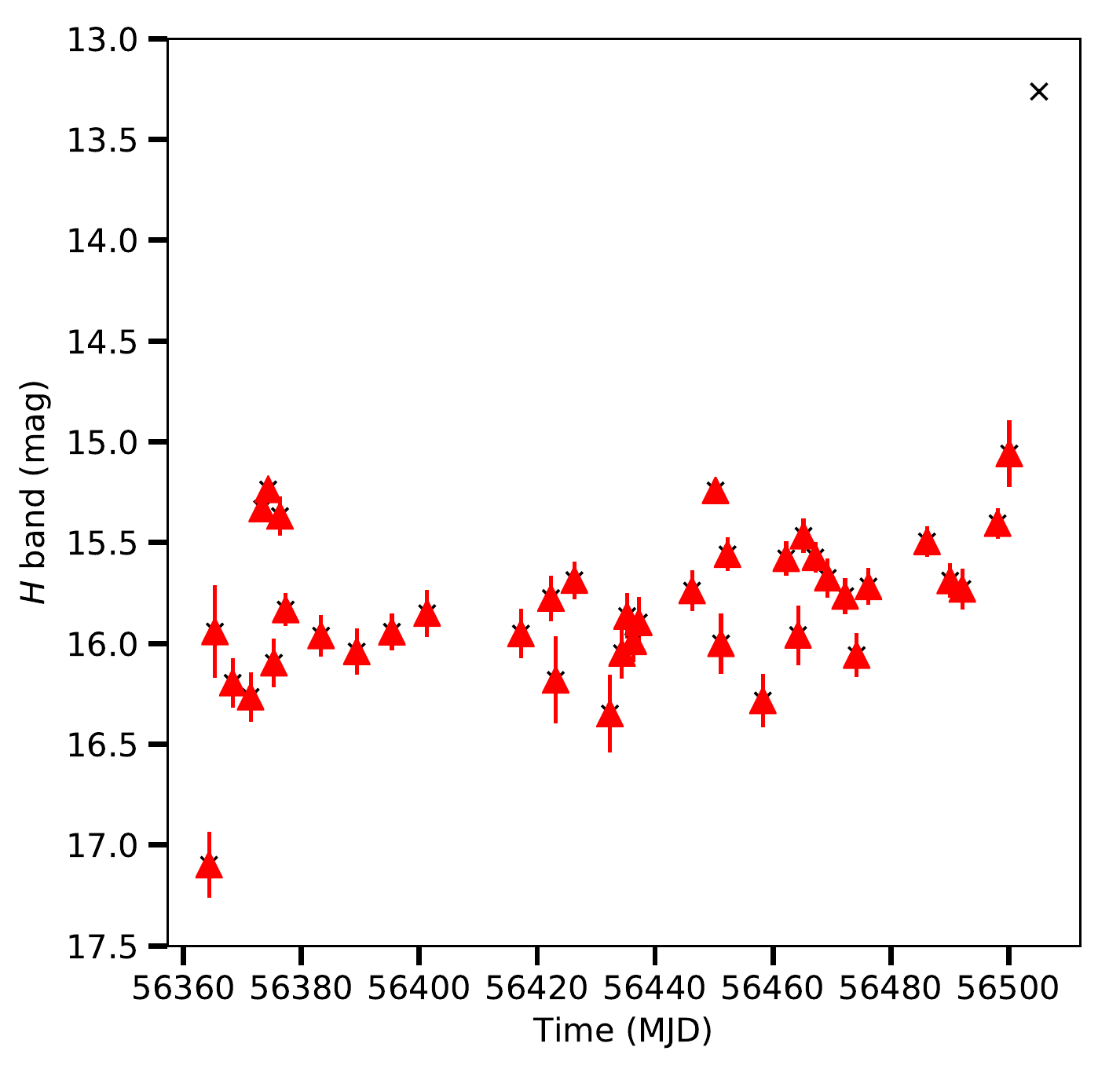}}\\
\subfloat[2014 outburst]{
\label{f:2014qh}
\includegraphics[width=0.3\textwidth]{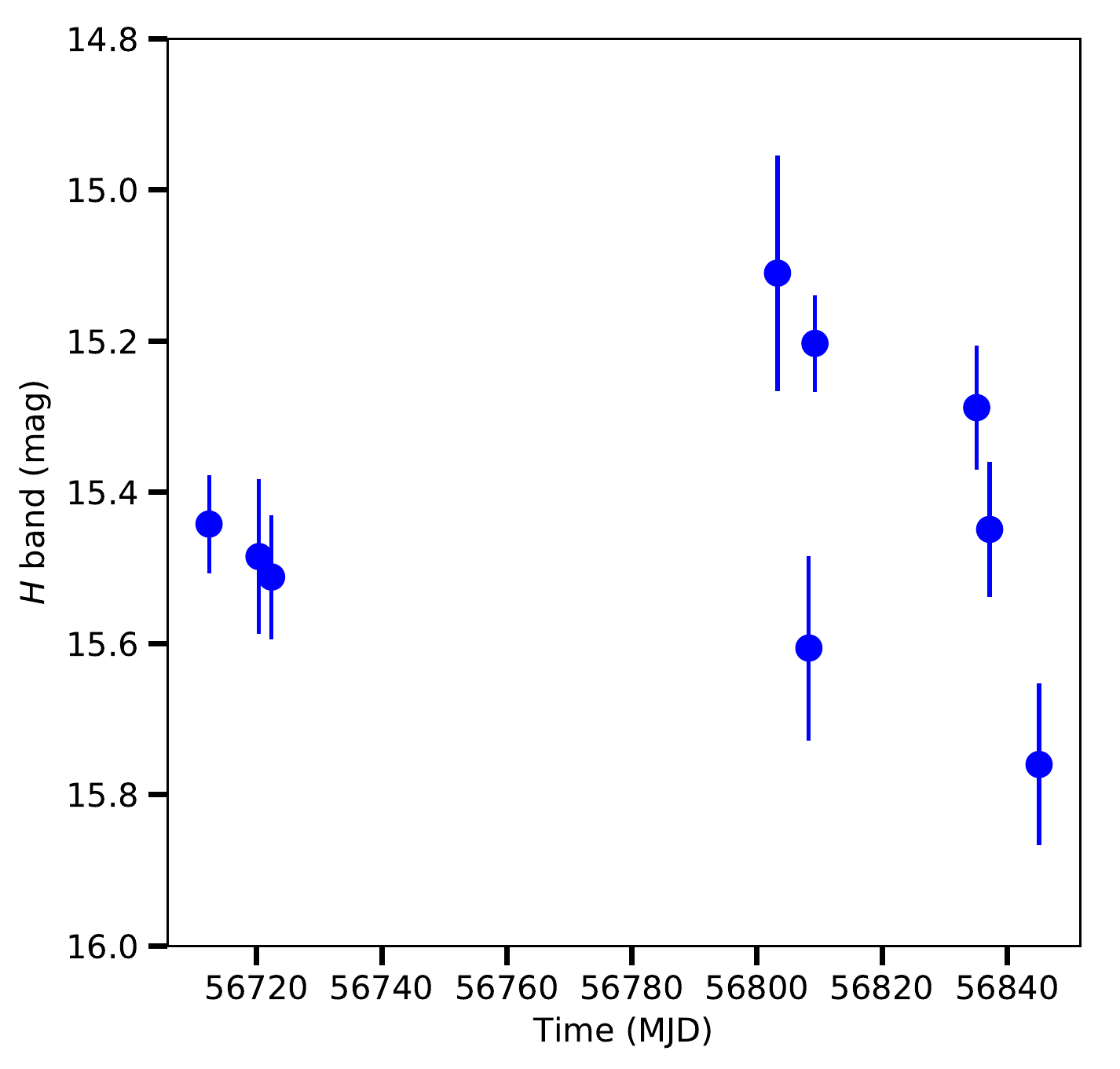}}
\caption{Same as Fig. \ref{fig:priorv} for the H-band.}
\label{fig:priorh}
\end{figure*}


\bsp	
\label{lastpage}

\end{document}